\documentclass[5p,final, authoryear,twocolumn]{elsarticle}

\usepackage{amssymb}
\usepackage{xcolor}
\usepackage{amsmath, float}
\usepackage{txfonts}
\usepackage{fleqn}   
\usepackage{graphicx}
\usepackage{epstopdf}
\usepackage{verbatim}
\usepackage{mathtools}
\usepackage{bm}
\usepackage{hyperref}
\usepackage{threeparttable}
\usepackage{lscape}
\usepackage{epsfig}

\usepackage[T1]{fontenc}

\journal{Journal of High Energy Astrophysics}

\begin{document}

\begin{frontmatter}



\title{How unique are pulsar wind nebulae models? \\
Implementation of a multi-parameter, automatic fitting for time-dependent spectra}


\author[1]{J. Martin\fnref{fn1}}
\author[2,3,4]{D. F. Torres\fnref{fn2}}

\fntext[fn1]{jonatan.martin@inaf.it}
\fntext[fn2]{dtorres@ice.csic.es}

\affiliation[1]{organization={INAF - Osservatorio Astrofisico di Arcetri},
addressline={Largo E. Fermi 5},
postcode={I-50125},
city={Firenze},
country={Italy}}
\affiliation[2]{organization={Institute of Space Sciences (ICE, CSIC)},
addressline={Campus UAB, Carrer de Magrans s/n},
postcode={08193},
city={Barcelona},
country={Spain}}
\affiliation[3]{organization={Institucio Catalana de Recerca i Estudis Avan\c cats (ICREA)},
city={Barcelona},
country={Spain}}
\affiliation[4]{organization={Institut d'Estudis Espacials de Catalunya (IEEC)},
city={Barcelona},
country={Spain}}

\begin{abstract}
Due to the computational cost of calculating a great number of variations of the parameters, detailed radiative models of pulsar wind nebulae (PWNe) do not usually contain fitting algorithms. 
As a consequence, most of the models in the literature are, in fact, qualitative fits based on visual inspection.
This is particularly true when complex, time-dependent models are considered.
Motivated by improvements in the computational efficiency of the current PWN models that were obtained in the last years,  
we here explore the inclusion of automatic fitting algorithms into a fully time-dependent model. 
Incorporating an efficient fitting tool based on the Nelder-Mead algorithm, we blindly find fitting solutions 
for the Crab nebula and 3C 58 with a time-dependent radiation model to compute the spectral and dynamical evolution of young and middle-aged PWNe.
This inclusion allows us, in addition of more faithfully determining the quality of the fit,
to tackle whether there exist degeneracy in the selected PWNe models.
We find both for Crab and 3C58, that the fits are well determined, and that no other significantly different set of model parameters
is able to cope with experimental data equally well.
The code is also able to consider the system's age as a free parameter, recursively determining all other needed magnitudes depending on age accordingly.
We use this feature to consider whether a detailed multi-frequency spectra can constrain the nebula age, finding that in fact this is the case for the two PWNe studied.
\end{abstract}


\end{frontmatter}

\section{Introduction}
\label{sec:intro}

Central to radiative modelling of pulsar wind nebulae (PWN) is the numerically solution the diffusion-loss equation for electron-positron pairs,
\begin{equation}
\label{eq:difflos}
\frac{\partial N(\gamma,t)}{\partial t}=-\frac{\partial}{\partial \gamma}\left[\dot{\gamma}(\gamma)N(\gamma,t) \right]-\frac{N(\gamma,t)}{\tau(\gamma,t)}+Q(\gamma,t),
\end{equation}
where $N(\gamma,t)$ is the electron-positron distribution function with energy $\gamma$ at time $t$.
The full version of the equation is found in \citet{ginzburg1964}, where the radial coordinate and other second order terms are taken into account.
Whereas the lack of radial coordinate $r$ in the equation usually adopted hampers the implementation of a detailed accounting of diffusion of particles and of the magnetic field distribution, it makes the problem more tractable. 
Despite this caveat, this kind of simplified models have been very successful in order to fit the spectra of most of the known PWNe and has taught us as a lot about the underlying PWN physics. Among the most important pieces of information gathered we now know (see e.g., \cite{Aharonian1997,zhang2008, gelfand2009, dejager2009, tanaka2011, martin2012, Bucciantini2011, martin2012, torres2013, vorster2013, Torres2014,torres2018,zhu2021}: 

\begin{itemize}

\item PWNe are particle dominated: magnetization of a few percent are common, although there might be some observational bias embedded into this fact.

\item PWNe are found having very similar injection parameters (high energy slope: 2.2-2.8; low-energy slope: 1.0-1.6)  and break energies (Lorenz factor around $10^{5 \div 6}$), but there is hardly any correlation among parameters.

\item PWNe have high multiplicity (order $10^{4 \div 6}$).

\item Comparing SEDs of the PWNe as observed today mixes pulsars of different spin-down power and age; generates a variety of distributions. A normalized comparison of the SEDs reduces the dispersion leaving no significant outliers. 

\item From the known population of PWNe, only Crab is self-synchrotron Compton (SSC) dominated at high energies (which is understood since we can derive the conditions when this can plausibly happen and prove that essentially only Crab fulfills them). Most others are dominated by the comptonization of the far infrared photon density, with the  
SSC being less relevant. 

\item  The usually adopted GALPROP code may tend to underpredict local values of FIR and NIR needed for PWNe to shine up in TeV as they do. 
GALPROP bins could be coarse for particular PWNe studies.

\item We need to use a more complex dynamics in aged PWNe. Reverberation is very important, but not clearly understood. Reverberation can produce stages of super-efficiency, where, as a result of contraction, e.g., the luminosity in X-rays $L_x > L_{sd}$

\item One key point related to Eq. \ref{eq:difflos} is that, as demonstrated in \citet{martin2012},  all terms of the equation are relevant when studying time evolution, since their relative significance change along time.
Relative differences in flux can reach 100\% between time dependent models once we let them evolve, meaning that the parameters obtained for even a young (worse for a middle-aged PWN) can differ significantly from one model to another depending on the approximations.

\end{itemize}

The recent 3DMHD numerical simulations of the Crab PWN \citep{Porth:2014,Olmi:2016}, despite they are yet limited in time, 
showed that the onset of kink instabilities leads to a large amount of dissipation with respect to that observed at lower dimensionality.
Such dissipation renders the total pressure nearly constant across the whole PWN, as well as to a relatively limited 
magnetic field gradient.
Given that, by definition, one-zone models have zero gradients, this may explain the success of radiative models to explain the spectral energy distribution of PWNe \citep{Gelfand2017}.

In this context, TIDE is a radiative model prepared to compute the spectral and dynamical evolution of young and middle-aged pulsar wind nebulae (PWNe). It solves the diffusion-loss equation for the electron-positron pairs injected into the nebula from the termination shock taking into account synchrotron, inverse Compton (IC), adiabatic, and Bremmstrahlung energy losses, evolving in time the magnetic field, the radius of the PWN
and other dynamical variables, as 
the shock trajectories of the SNR, allowing to follow the evolution further than the beginning of the reverberation phase.
The TIDE code has been used in several occasions to model PWNe observations and study its physics (e.g., see the latest in 
\cite{torres2018,bandiera2020} and references therein. 
Here, we are presenting the latest significant upgrades, including parallelization of some of its components and automatic fitting in a fully time-dependent scenario, and using these upgrades to tackle the issue of possible degeneracy of solutions.
Real fitting of PWNe multi-frequency data is not usually done (despite it is common to find the word {\it fit} in the literature (see the formerly quoted papers, for instance), it rather refer to visually acceptable descriptions when time-dependent, complex models are involved. 
It is thus relevant to briefly clarify what is incorporated in the fitting model and where is room for future improvements.

Using this automatic fitting tool we can more properly address whether degeneracy exists. 
Given a relatively complete set of observational data, how unique are PWN models? 
Can a significantly different set of model parameters lead to similar spectral energy distributions? 
We also use such tool to consider, for the first time to our knowledge, whether a significant set of PWN multi-frequency spectral energy distribution (SED) data can actually imposes a constraint over the age of the system that emits it.
The rest of this work goes directly into presenting the fitting technique and the results obtained, 
leaving a more detailed accounting of the physical components to an Appendix.

\section{Fitting method}
\label{sec:min}

\subsection{Identifying free parameters}

Time-dependent PWNe models, despite all simplifying assumptions,
are still subject to a significant degree of freedom. 
Table \ref{tab:free-para} identifies all (in principle) free parameters that enter into our model.
The meaning and formulae associated with this set of parameters can be read in Appendix~\ref{sec:appa}.
The parameters in Table \ref{tab:free-para} --unless in obvious cases such as the age of the PWN-- are additionally assumed to be 
constant along time, which is in itself an approximation.
There could be occasions (e.g., spin-down rate transitions, glitches, bursts, etc.) where the
injection or the energy distribution or the magnetization could be subject to sudden changes.

\begin{table*}
\scriptsize
\centering
\begin{threeparttable}
\caption{Identification of the whole set of possible free parameters for a time-dependent automatic fitting. All of these parameters can be changed directly in the code, and searched in the minimization procedure when sufficient data warrants it.}
\label{tab:free-para}
\begin{tabular}{lll}
\hline
 & Name & Comment\\
\hline
\hline
$\delta$ & Data systematic uncertainty & Accounts for the systematic uncertainty between data points obtained from different detectors \\
$t_\text{age}$ & Age of the PWN & Relevant for the particle accumulation, dynamics, energetics. \\
$v_\text{psr}$ & Proper motion of the pulsar & Activates/deactivates the injection of new particles depending on the pulsar position with respect to the PWN shell \\
$n$\tnote{$^\dagger$} & Braking index & Affects the evolution of the spin-down luminosity \\
$L_0$\tnote{$^\dagger$} & Initial spin-down luminosity & Affects the evolution of the spin-down luminosity \\
$\tau_0$\tnote{$^\dagger$} & Initial spin-down age & Affects the evolution of the spin-down luminosity \\
$d$ & Distance & Relevant for the flux normalization \\
$\gamma_\text{min}$ & Minimum energy at injection & Relevant for the spectrum normalization and can affect to the luminosity of the spectrum at radio frequencies \\
$\gamma_\text{b}$ & Energy break & Relevant for the spectrum population and the position of the spectrum break \\
$\alpha_l$ & low-energy index & Relevant for the spectral definition at low energies.  \\
$\alpha_h$ & high-energy index  & Relevant for the spectral definition at high energies. \\
$\varepsilon$ & Containment factor & Relevant for fixing the synchrotron cut-off \\
$E_\text{sn}$ & Supernova explosion energy & Relevant for the evolution of the PWN radius \\
$M_\text{ej}$ & Ejected mass & Relevant for the evolution of the PWN radius \\
$\rho_\text{ism}$ & Interstellar medium density & Relevant for the evolution of the PWN radiu \\
$\delta_\text{core}$ & SNR core density index & Relevant for the evolution of the PWN radius \\
$\omega$ & SNR envelope density index & Relevant for the evolution of the PWN radius \\
$\eta_\text{B}$ & magnetic fraction & Controls synchrotron radiation and magnetic field. \\
$\eta_\text{other}$ & Other energy releases fraction & Controls spectrum normalization and magnetic field. \\
$w_\text{i}$ & Photon background energy density & Affects inverse Compton luminosity. Can come from observations or environmental modelling. \\
$T_\text{i}$ & Photon background energy density & Affects inverse Compton luminosity. Can come from observations or environmental modelling. \\
$n_\text{He}/n_\text{H}$ & He/H density ratio & Relevant for Bremsstrahlung radiation spectrum. \\
\hline
\end{tabular}
\begin{tablenotes}
\item [$\dagger$] Their values are taken consistently 
with the age of the PWN is the latter is let to vary. 
\end{tablenotes}
\end{threeparttable}
\end{table*}

\subsection{Fitting }

Our model now incorporates a new routine package in order to perform fits of multi-wavelength PWNe 
spectra from radio to VHE, in a multi-dimensional space, along time. We refer to this as TIDEfit.
It uses the Downhill Simplex Method (also called Nelder-Mead method, see \cite{numerical_recipes}) where the initial set of parameters is represented in a polygon (called simplex) in a $N$-dimensional parameter space, being $N$ the number of free parameters, with $N+1$ vertices.
The Nelder-Mead method prompts a sort of transformations to this polygon, which provide new solutions that are tested, and accepted or rejected according to its nearness to the data, in such a way that the polygon surrounds the final solution up to the desired accuracy for final convergence.
It is known that such a method suffer of a great loss of efficiency when the number of parameters is high enough \citep{gao2012}.
However, these authors proposed a modified version of the method to improve the efficiency in these situations, which is already implemented in the {\it minimize} function of the {\it Scipy} library for Python.

Getting a good fit via methods such as the Monte Carlo Markov Chain implies realizing a high number of iterations  (of the order of tens of thousands or even more) in which the diffusion-loss equation must be solved at every time. 
Due to the complexity that implies solving this equation in detail and the potentially high number of free parameters, other typical methods (e.g., minimum least squared) which require the calculations of first partial derivatives (and probably second partial derivatives too) with respect of all the free parameters are not preferred either, due to their time-consumption.
Instead, the simplex method avoid the use of first partial derivatives and can typically converge in $\sim 1000$ iterations for 6-8 free parameters if the initial choice of the simplex is not extremely far from the fitting one.
See Table 1 for typical search ranges.

\subsection{Data treatment}

The multi-wavelength data for a generic PWN come from different detectors, and from different observation times (sometimes decades apart). Thus, in addition of the observational uncertainty of each detector per se and the complexities arising from treating data with vastly different relative error bars, there exist a cross-calibration uncertainty that must be taken into account.
This issue --as well as some misconceptions arising when fitting data in astrophysics-- are discussed by \cite{hogg2010}.
We pose that the most convenient prescription is to introduce an extra free parameter in our fit such that the real uncertainty for each point is systematically enlarged in proportion to the flux calculated from the model. 
That is (see \cite{hogg2010}),
\begin{equation}
s_i^2 = \sigma_i^2 + \delta^2 F_i^2,
\end{equation}
where $\sigma_i$ is the observational uncertainty, $\delta$ is the free parameter that accounts for the systematic uncertainty and $F_i$ is the spectral flux obtained from the model.
As discussed in \cite{hogg2010}, the $\chi^2$ function typically used in astrophysics is only a part of the full formula for the likelihood $\chi^2$ function, because the coefficient depending on the uncertainty is neglected. This approximation commonly done in the literature is right if we do not consider systematic errors in our data, so they are constant independently of the values of the model parameters. Here, instead of minimizing the typical $\chi^2$ function, we maximize the likelihood function \citep{hogg2010}
\begin{eqnarray}
\label{eq:likefunc}
\ln p(y | \nu, \sigma, \vec{x}, \delta) = - \frac{1}{2} 
\sum_i \left \{ \frac{[y_i(\nu) - F(\nu,\vec{x})]^2}{s_i^2(\nu,\sigma_i,\delta)} - \right.  \nonumber \\  \left. 
\ln 2 \pi s_i^2(\nu,\sigma_i,\delta) \right \},
\end{eqnarray}
where $\nu$ is the observed frequency and $\vec{x}$ is our set of free parameters.
Note that Eq.~\ref{eq:likefunc} uses the logarithm of $p$
to have smoother control over the function's gradient and the variation tolerance before consider convergence of the numerical scheme.

If the likelihood function has a complicated shape in the parameter phase-space, it is possible --in the context of the Nelder-Mead method-- to fall into local minima of the likelihood function 
depending on the first guess of the parameters. 
To avoid this, we noticed that a good practise is to fix first the systematic uncertainty at an a-priori seemingly reasonable value (10 - 20\%), 
then fit the rest of parameters, 
in order to use the results as an initial guess for a second refined fit now leaving the systematic uncertainty as a free parameter too.
All of this is automatically implemented in our code (see Appendix B for some further notes on implementation and numerical cost).

\section{Applications to two canonical PWNe with varying data quality: How
unique are their fits?}
\label{sec:results}

We now fit the spectra of two canonical PWNe with TIDE: the Crab Nebula and 3C 58. Fortunately, most of the parameters can be fixed for observations in both nebulae, as we can see in Table \ref{tab:fixed}. However, the number of data points and their uncertainty varies.
For instance, whereas the age of Crab is certain, the age of 3C 58 is assumed. 
A total of 8 parameters has been left free to fit both PWNe.
Apart the systematic uncertainty, these are the energy break and the spectral slopes of the injection function, the ejected mass of the SNR, the magnetic fraction and the FIR and NIR energy densities.

A special note goes for the containment factor. 
The value taken for Crab comes from a previous fit where this parameter is free, but since it only affects the synchrotron cut-off where the number of data points is small, the uncertainty in this parameter is high because the $\chi^2$ changes little when we look for the confidence level. 
Thus we have taken it fixed to minimize the number of parameters.
We come back to this in Appendix~\ref{sec:appa}, and show in detail how different containment factors affects the spectra.
The fitting ranges and the results obtained for each nebula are compiled in Table \ref{tab:results}.

\begin{table}
\scriptsize
\centering
\caption{Data from observations -first panel-, computed from the first panel -second panel-, or assumed -third panel- in the models of the Crab nebula and 3C 58.}
\label{tab:fixed}
\begin{tabular}{llcc}
\hline
\hline
Period (ms) & $P$ & 33.4 & 65.7\\
Period derivative (s s$^{-1}$) & $\dot{P}$ & $4.2 \times 10^{-13}$ & $1.93 \times 10^{-13}$\\
Braking index & $n$ & 2.509 & 3\\
Observed radius (pc) & $R_\text{pwn}$ & 1.8 & 2.3\\
Distance (kpc) & $d$ & 2 & 2\\
Pulsed radiation fraction & $\eta_\text{other}$ & 0.0142 & 0.0018\\
\hline
Characteristic age (yr) & $\tau_\text{c}$ & 1296 & 5398\\
Age (yr) & $t_\text{age}$ & 968 & 2500\\
Spin-down  (erg s$^{-1}$) & $L$ & $4.5 \times 10^{38}$ & $2.7 \times 10^{37}$\\
Initial spin-down  (erg s$^{-1}$) & $L_0$ & $3.1 \times 10^{39}$ & $9.3 \times 10^{37}$\\
Initial spin-down age (yr) & $\tau_0$ & 750 & 2878\\
\hline
SN energy explosion (erg) & $E_\text{sn}$ & $10^{51}$ & $10^{51}$\\
ISM density (cm$^{-3})$ & $\rho_\text{ism}$ & 0.5 & 0.1\\
FIR temperature (K) & $T_\text{cmb}$ & 70 & 25\\
NIR temperature (K) & $T_\text{cmb}$ & 5000 & 2900\\
Containment factor & $\varepsilon$ & 0.3 & 0.5\\
\hline
\end{tabular}
\end{table}

\subsection{Crab Nebula (with LHASSO data)}
\label{sec:crab}

\begin{figure*}
\centering
\includegraphics[width=0.45\textwidth]{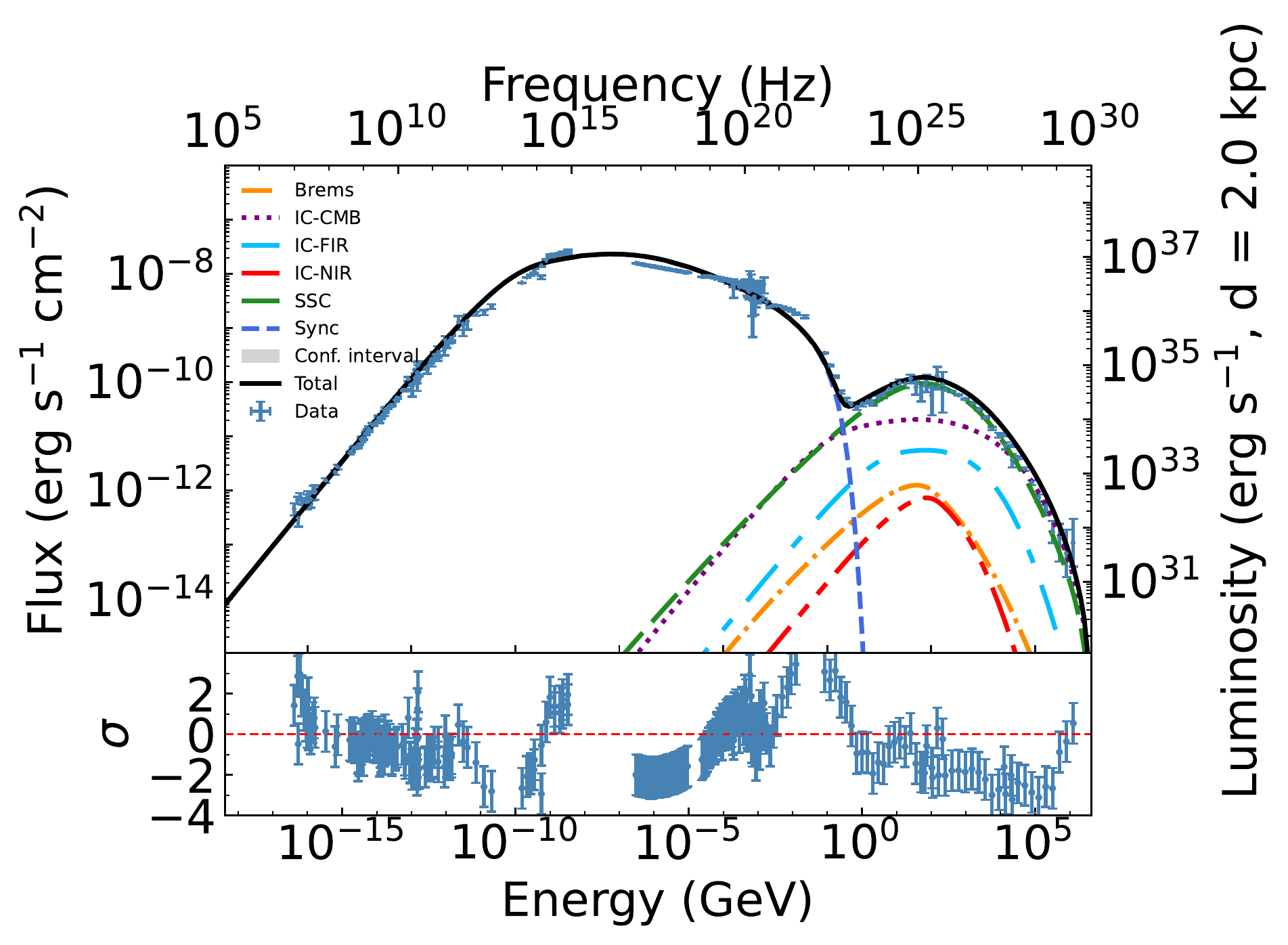}
\includegraphics[width=0.45\textwidth]{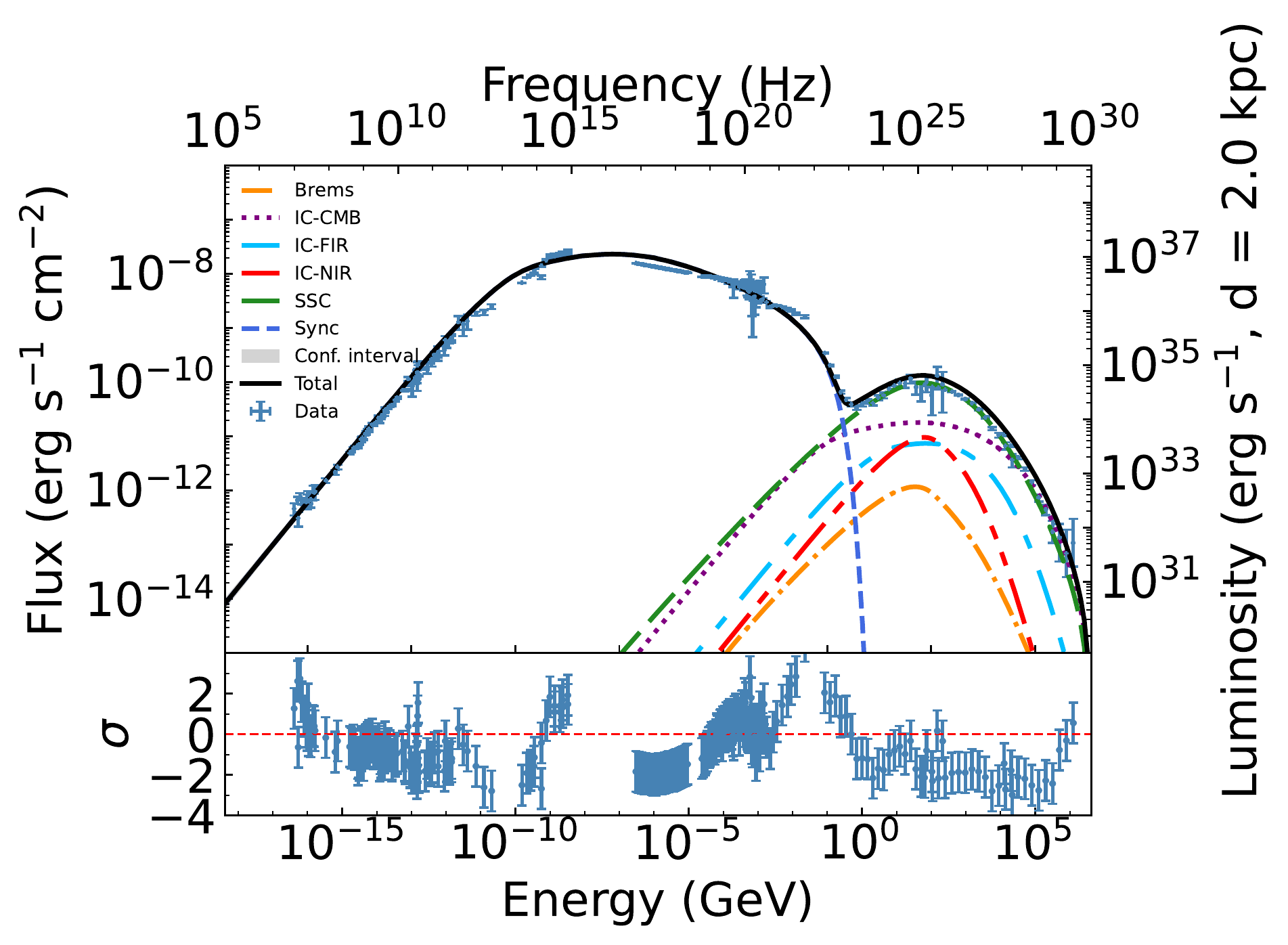}
\\
\includegraphics[width=0.45\textwidth]{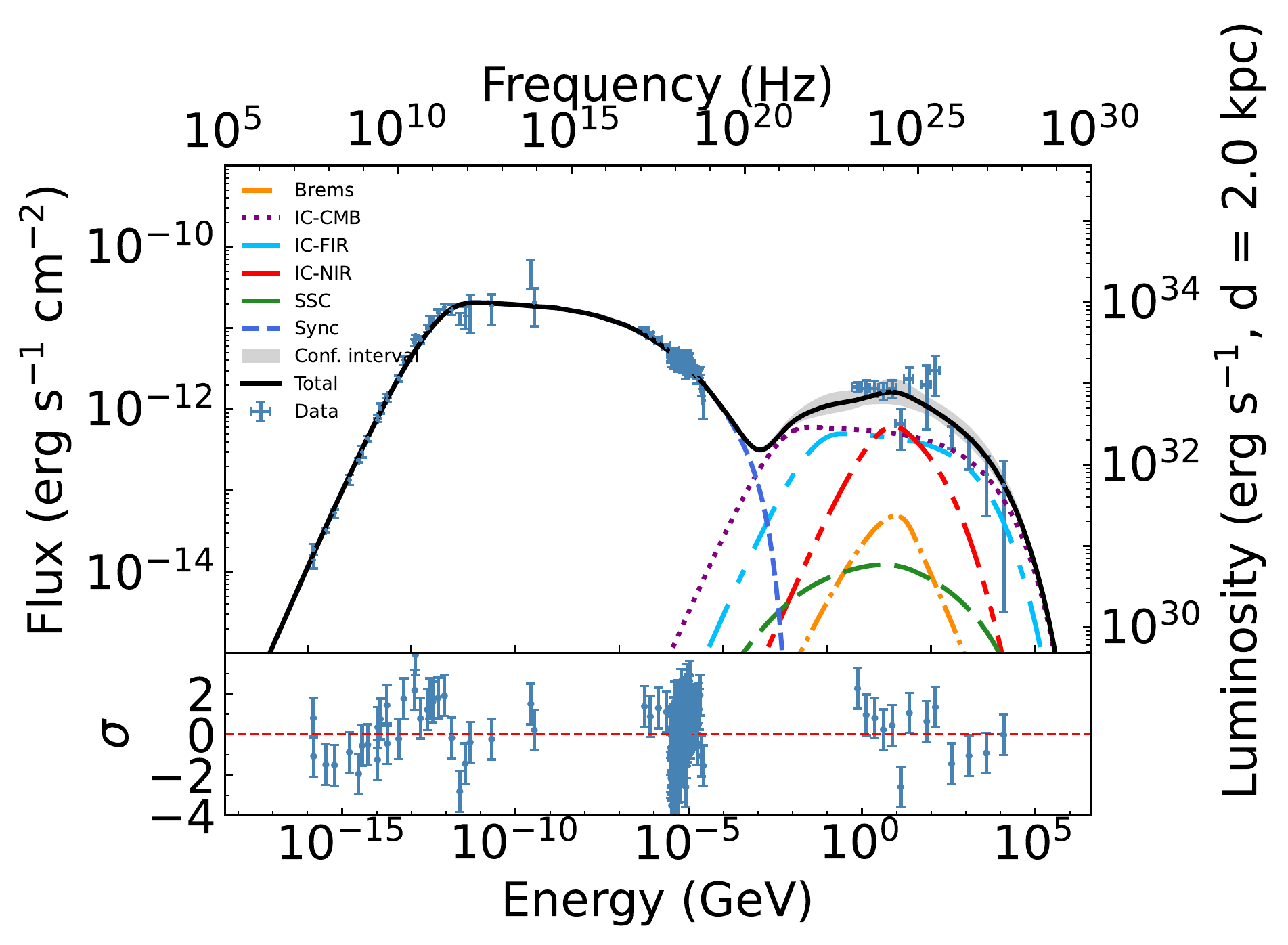}
\includegraphics[width=0.45\textwidth]{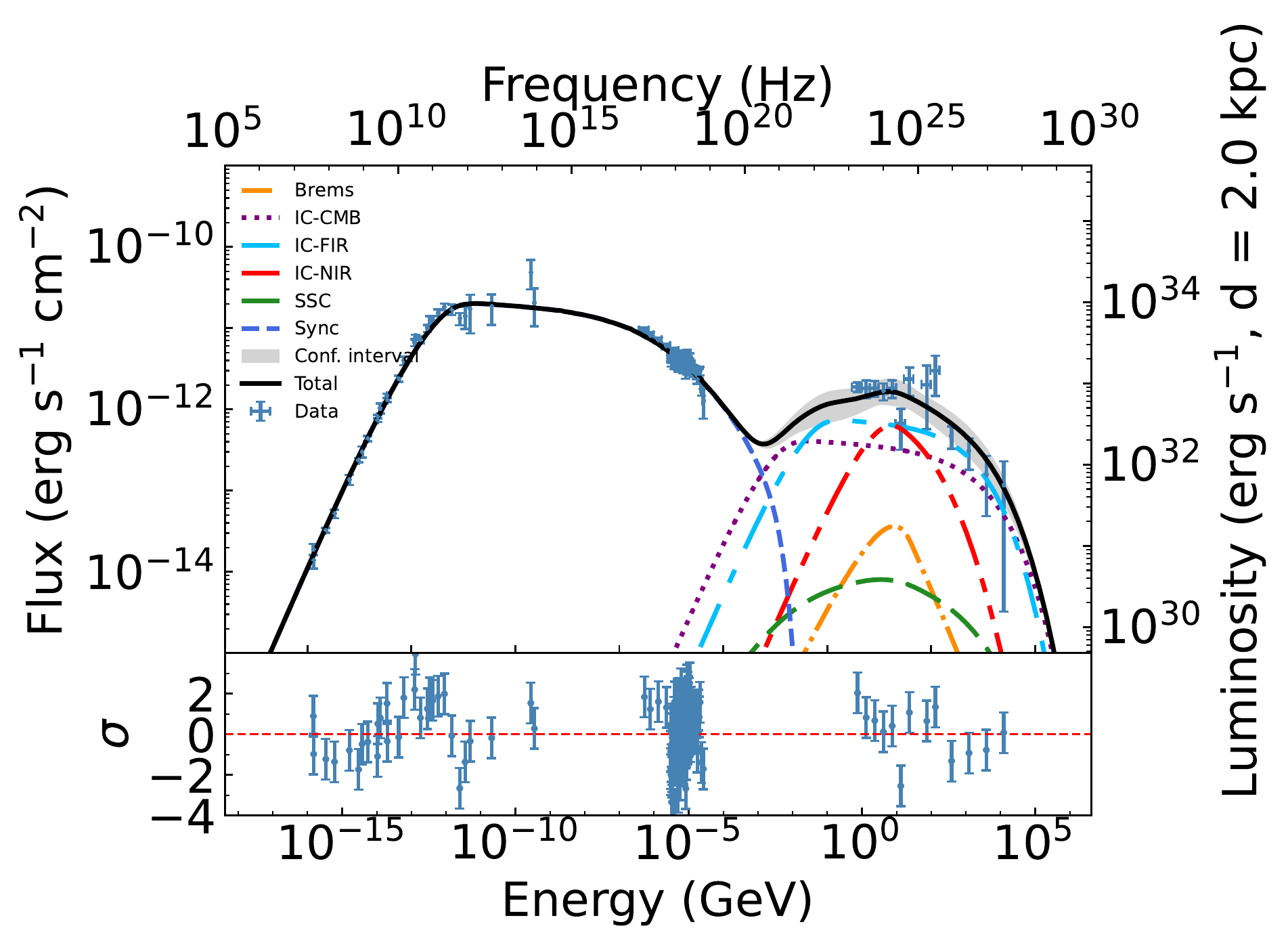}
\caption{Models fitted for the Crab Nebula (top) and 3C 58 (bottom). On the left column, the age is fixed and on the right, it is a free parameter (see results in Table \ref{tab:results}) The data points for Crab are taken from \cite{baldwin1971,maciasperez2010}, (compilations of radio observations), \cite{buehler2014} (from IR to $\gamma$-rays and VHE) and the new points offered by the LHAASO collaboration \citep{lhaaso2021}. For 3C 58, they come from \cite{green1986,morsi1987,salter1989} (radio), \cite{torii2000} (infrared), \cite{green1994,slane2008} (X-rays), \cite{abdo2013,ackermann2013,li2018} (gamma-rays) and \cite{aleksic2014} (VHE).}
\label{fig:sed}
\end{figure*}

The Crab Nebula is the most studied PWN, and has precise measurements of its flux at all frequencies from radio to VHE.
Regarding the flux data, we included the compilation shown in \cite{buehler2014} and the recently VHE data points by LHAASO \citep{lhaaso2021}.
The resulting spectral energy distribution (SED) from the automatic fitting process is shown in Figure \ref{fig:sed}.
The automatic fit has reduced $\chi^2$ of 1.013, and the 1$\sigma$ band (depicted in grey in the plot) is almost invisible at all energies in this scale (compare with the companion panel for 3C 58, where this band is more prominently seen).
The fit reproduces the data, particularly in critical regions such as in the synchrotron decay or in the gamma-ray peak.
The Crab PWN is SSC dominated at the gamma-ray peak, with a significant contribution of comptonized CMB photons at larger energies.
Despite the generous range of the parameters search, there are no surprises in the fit, neither in the radiative process dominance nor in the free parameters values.
There is no alternative set of solutions outside the range quoted for the free parameters providing a similarly good fit. 
We have tested this thoroughly, starting from dozens of very different initial conditions in the Nelder-Mead simplex:
We performed $\sim 150$ fits with randomly selected initial simplex 
obtaining convergence to the solution quoted in more than 73\% of the cases. 
In the remaining ones, visual inspection or the value of the reduced $\chi^2$ would immediately qualify the fit as non-compliant.
These non-convergent cases happen when the values of the energy break $\gamma_b$ and/or the magnetic fraction $\eta_\text{B}$, and/or the initial systematic uncertainty
are close to one of the extreme values inside its range of variation.
The Crab nebula spectra is thus uniquely fit in the surrounding of the parameters described, and in the surrounding of no others. 
There seems to be little room for degeneracy in our PWNe models when such significant set of data is considered.

\begin{figure}
\centering
\includegraphics[width=0.5\textwidth]{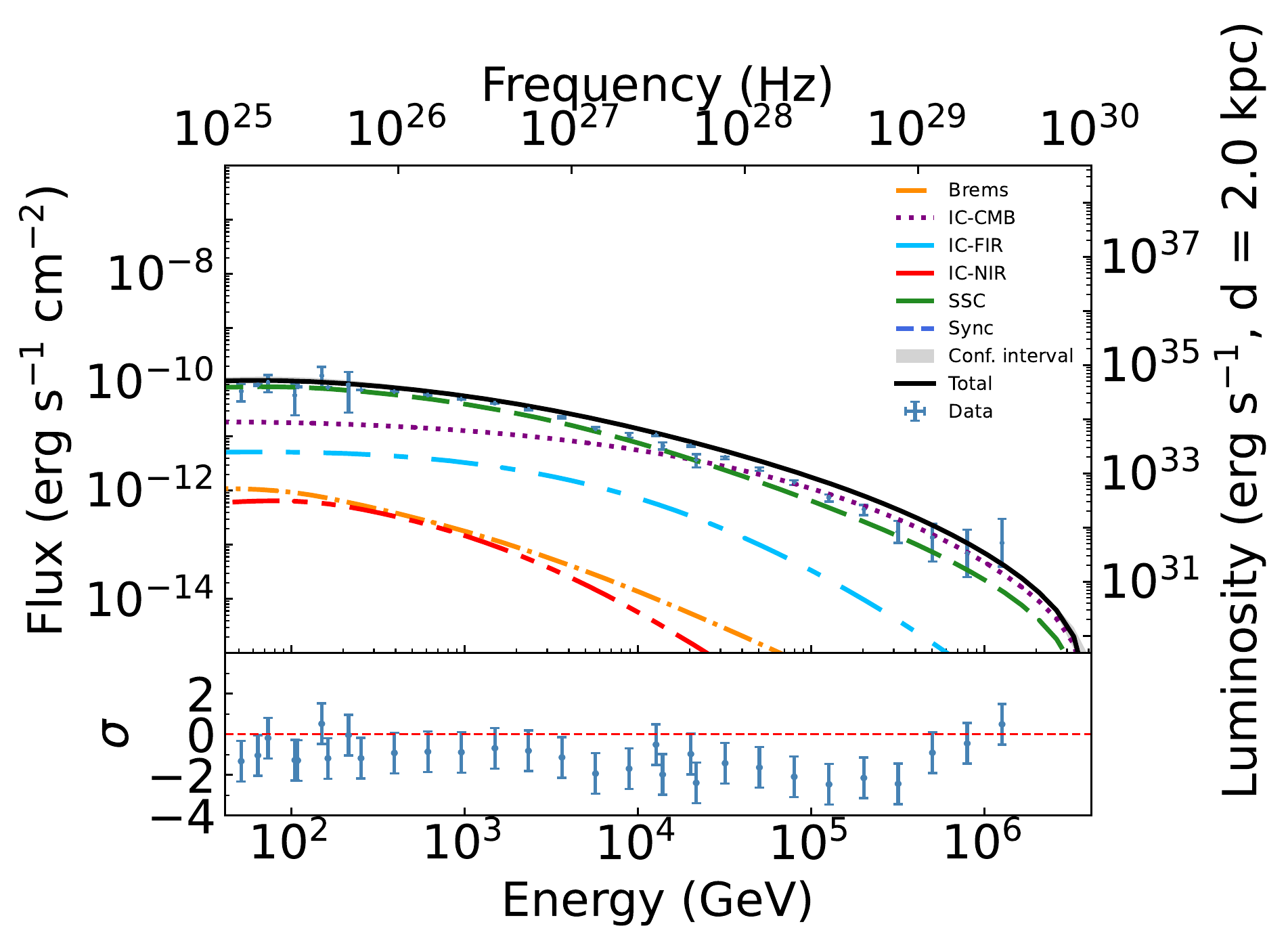}
\caption{Zoom of the Crab Nebula spectrum shown in Figure \ref{fig:sed} between 100 GeV and 1 PeV.}
\label{fig:sedzoom}
\end{figure}

It is interesting to remark that the best model actually slightly overpredicts the flux at TeV energies, especially around 100 TeV (see Figure \ref{fig:sedzoom} to zoom in the details). 

Given that the PWN radius in the model is just an average value representing a more complex physical situation, one can think that the inverse Compton yield at high energies could be reduced by if the radius of the nebula is larger, at least within the corresponding uncertainty.
If so, the magnetic energy density would be lower and, in consequence the magnetic field.
In turn, the magnetic fraction could be larger in order to power up the magnetic field sufficiently, decreasing the normalization of the particle distribution function to compensate.
Despite of the reduction of the normalization, the synchrotron luminosity could be fitted because it is proportional to $\sim B^2$.
A lower number of pairs would decrease the IC flux helping to fit the IC flux in both components (IC-CMB and SSC).
This strategy may however be considered disadvantageous since it would require an ejected mass of about $6M_{\odot}$, which seems too low for current models of supernovae \cite{Smartt:2009}. 
\cite{Owen2015} state
that the total mass of gas plus dust in the Crab Nebula is $7.2 \pm 0.5 M_\odot$, consistent with a progenitor star mass of $\sim
9 M_\odot$.
However, 
we note on the side that disregarding this fact, and allowing for smaller ejecta masses than the range we adopt in Table 
\ref{tab:fixed}, we obtain even better fits than the ones shown.

Another option to decrease the inverse Compton yield at high energies is that part of the energy invested in the acceleration of electron-positron pairs, $(1-\eta_\text{B})$, is used to accelerate an hadronic component, or to emit pulsed emission, or other multi-messenger emission such as neutrinos or gravitational waves.
That is, that part of the energy is simply not available due to it being spent in other processes that are or may be, in fact, naturally occuring.
We have taken this into account. We take the fraction of pulsed emission integrating the data from \cite{Torres2019}, which corresponds to 1.42\% of the spin-down luminosity, and we have actually discounted such a quantity from our contribution to particles (as explained in the Appendix~\ref{sec:appa}). However,  Figure \ref{fig:sed} shows that this imply little change.

\begin{figure*}
\raggedleft
\includegraphics[width=0.16\textwidth]{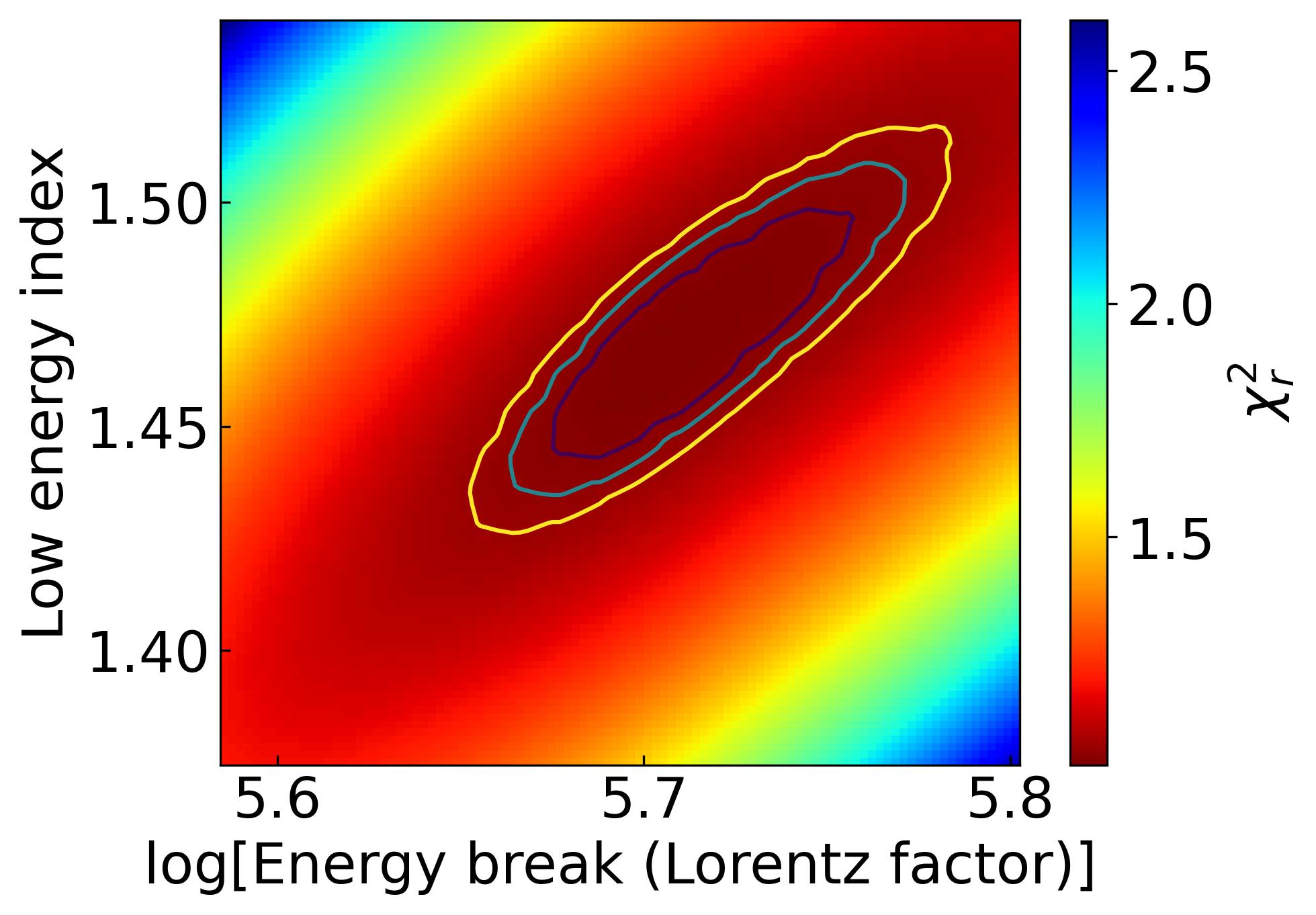}
\includegraphics[width=0.16\textwidth]{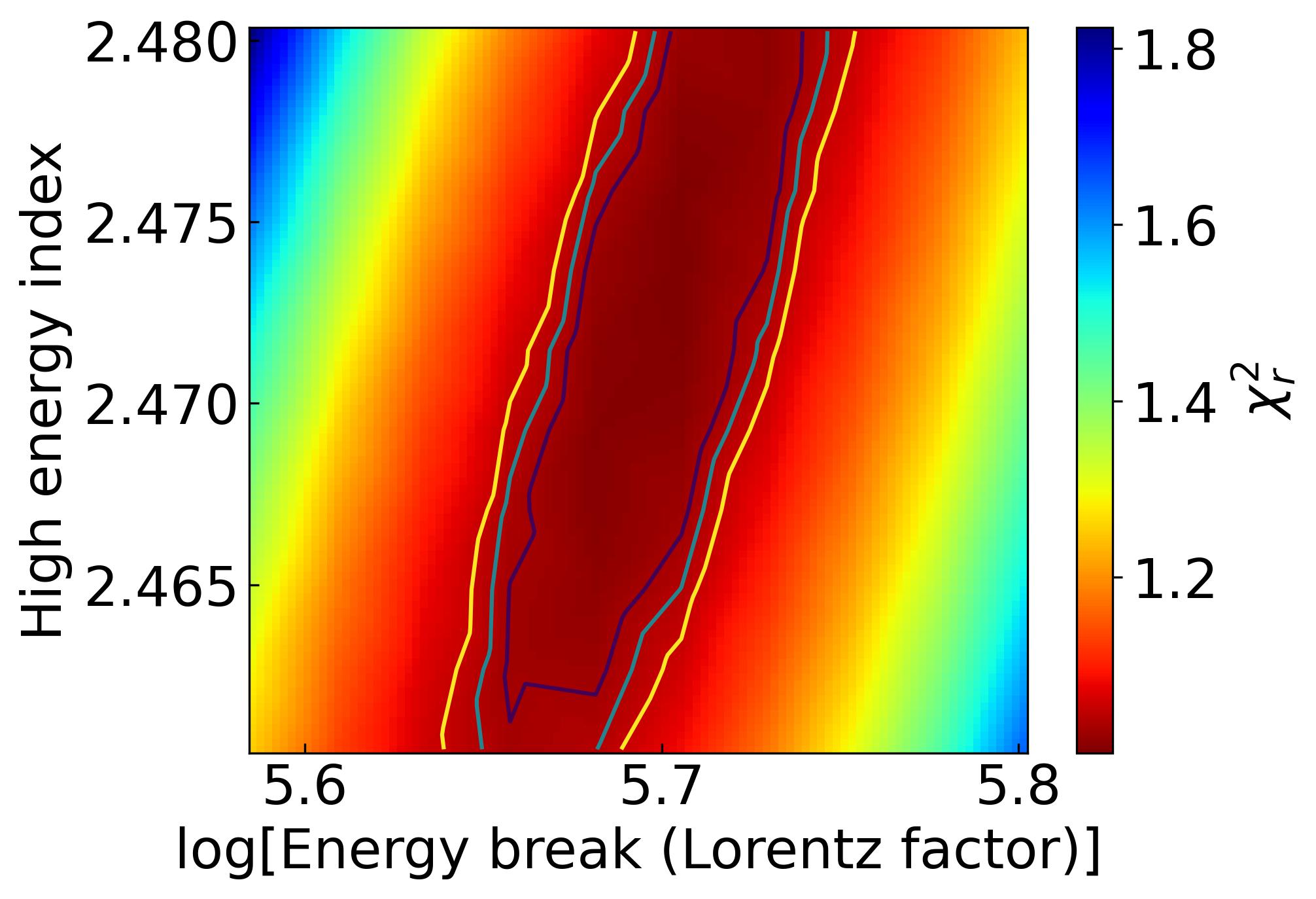}
\includegraphics[width=0.16\textwidth]{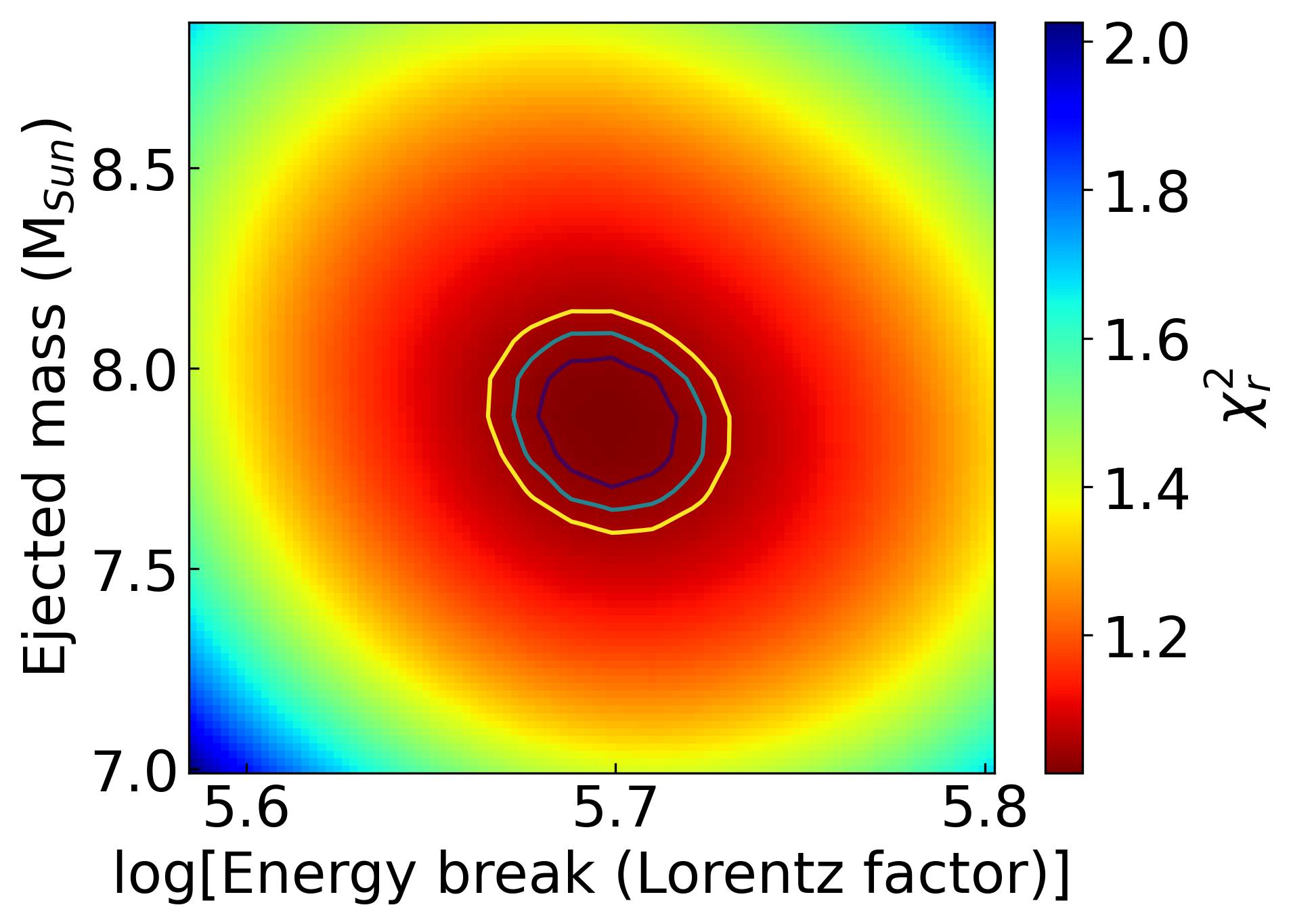}
\includegraphics[width=0.16\textwidth]{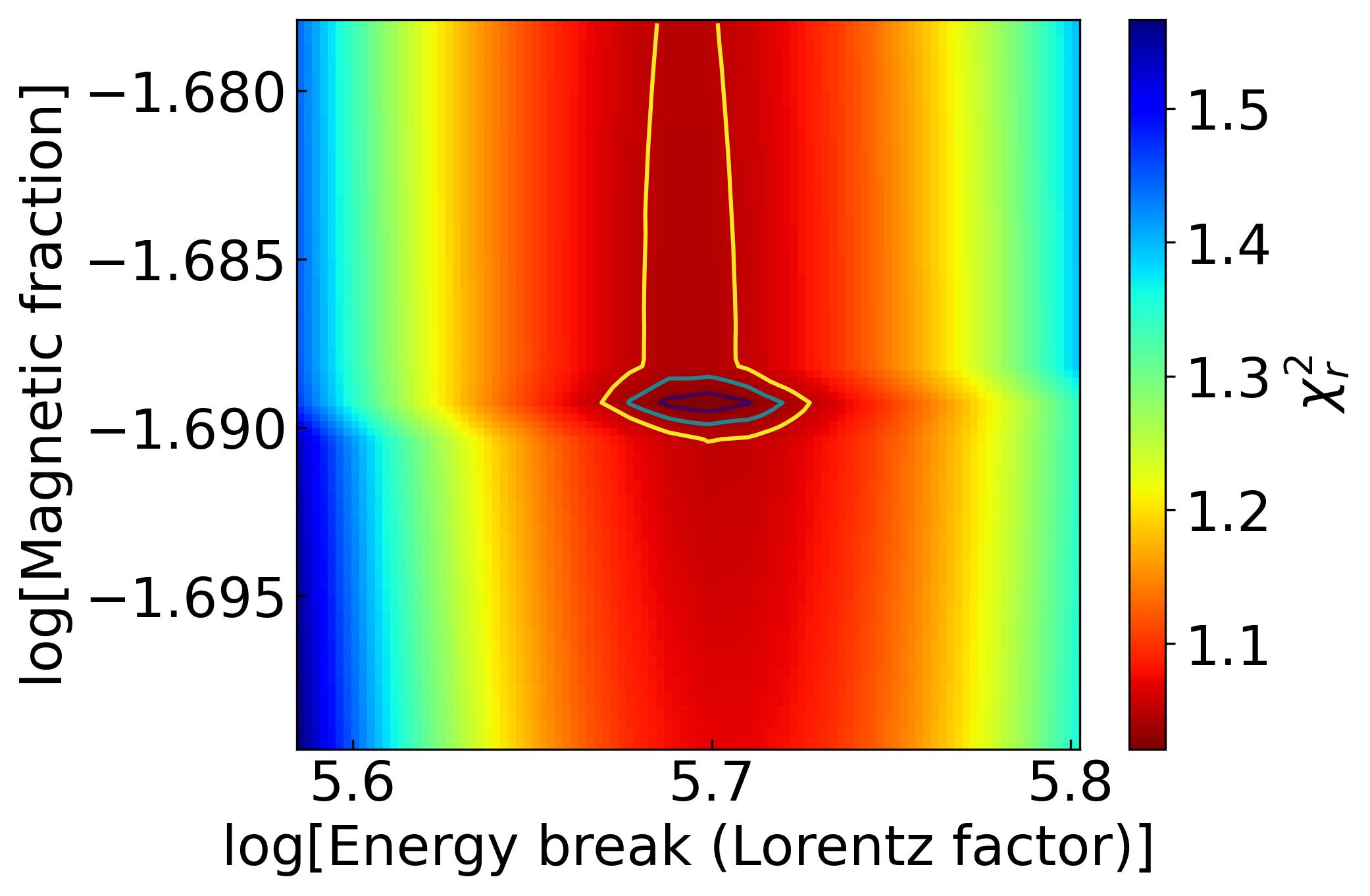}
\includegraphics[width=0.16\textwidth]{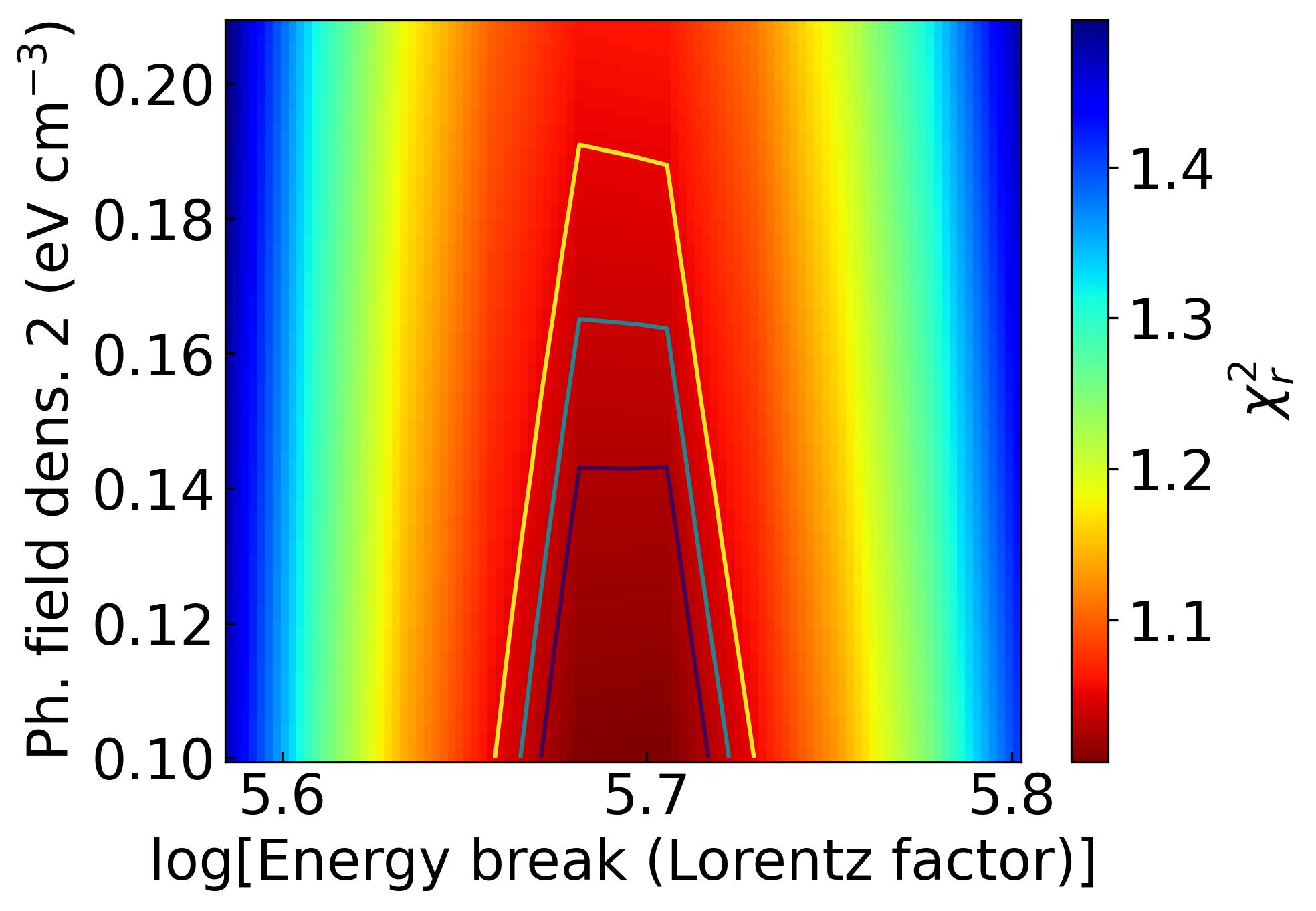}
\includegraphics[width=0.16\textwidth]{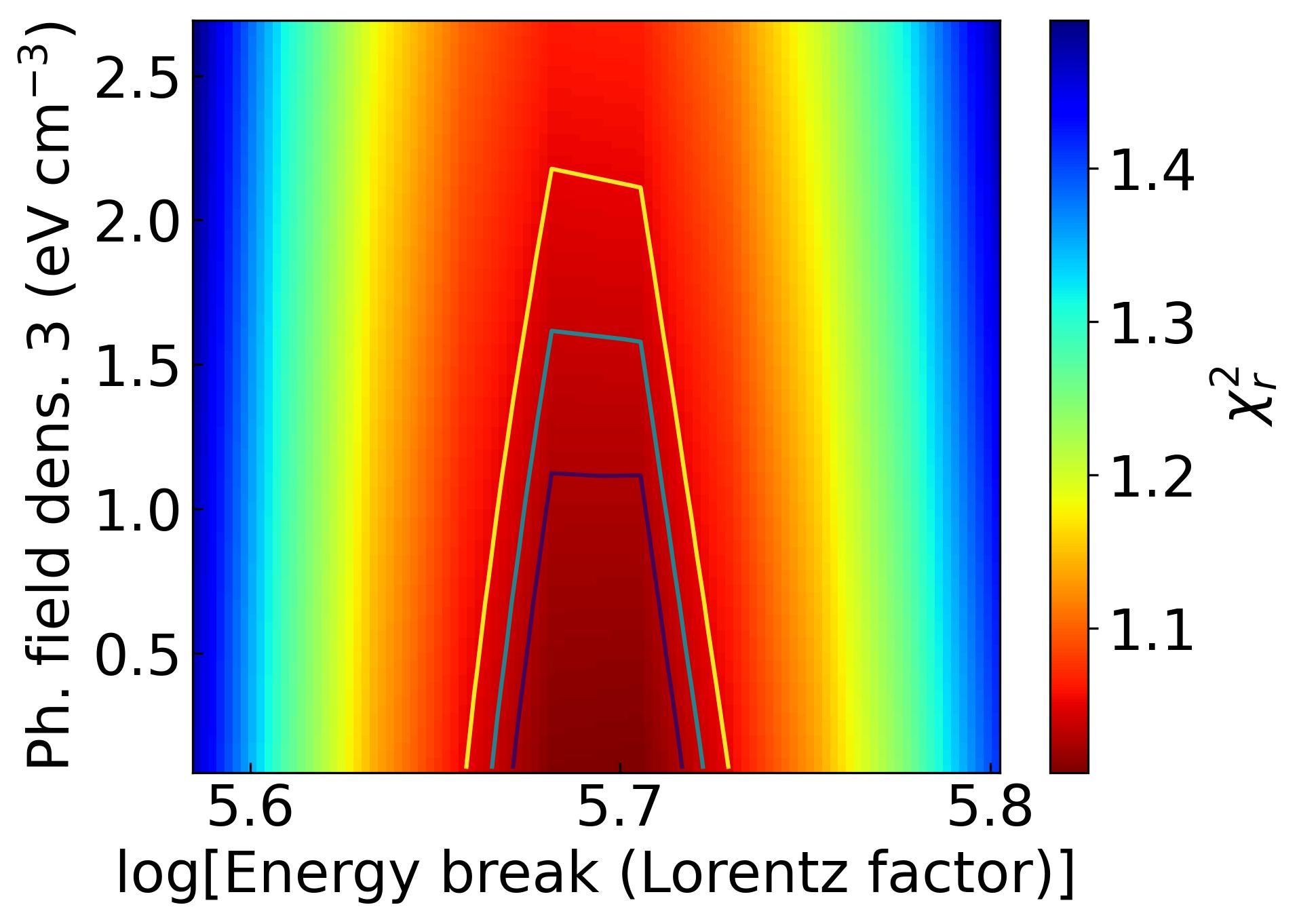}
\\
\includegraphics[width=0.16\textwidth]{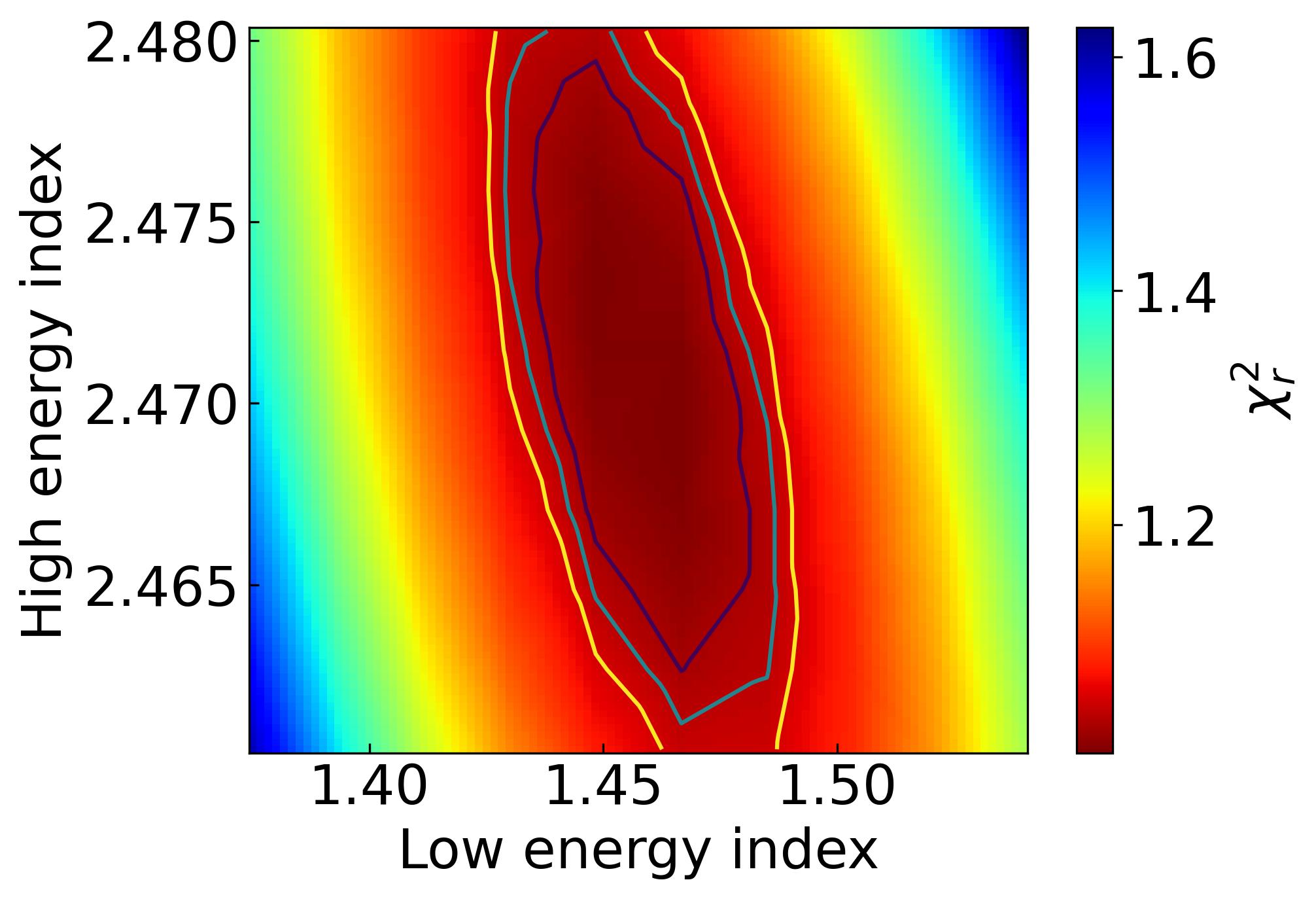}
\includegraphics[width=0.16\textwidth]{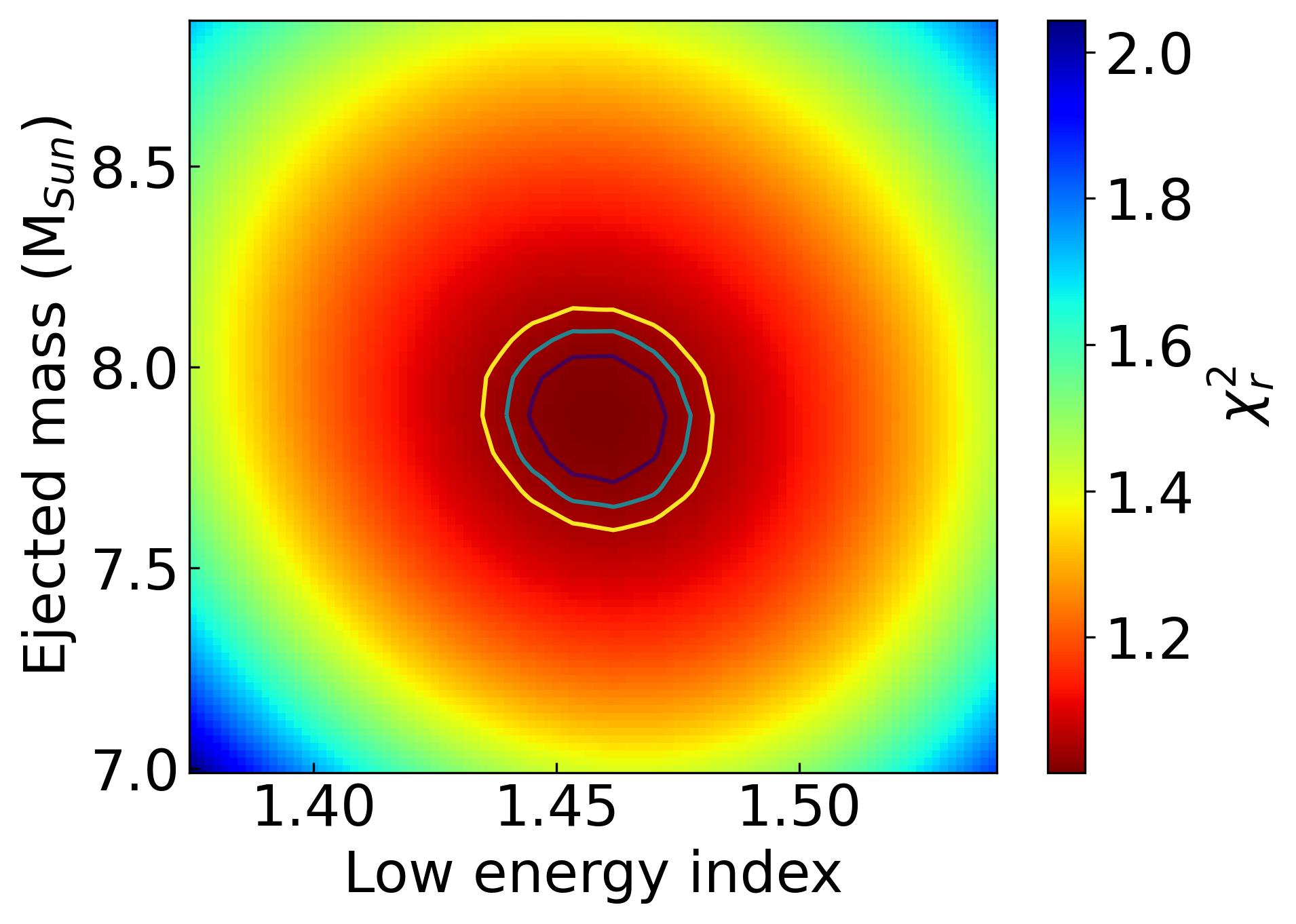}
\includegraphics[width=0.16\textwidth]{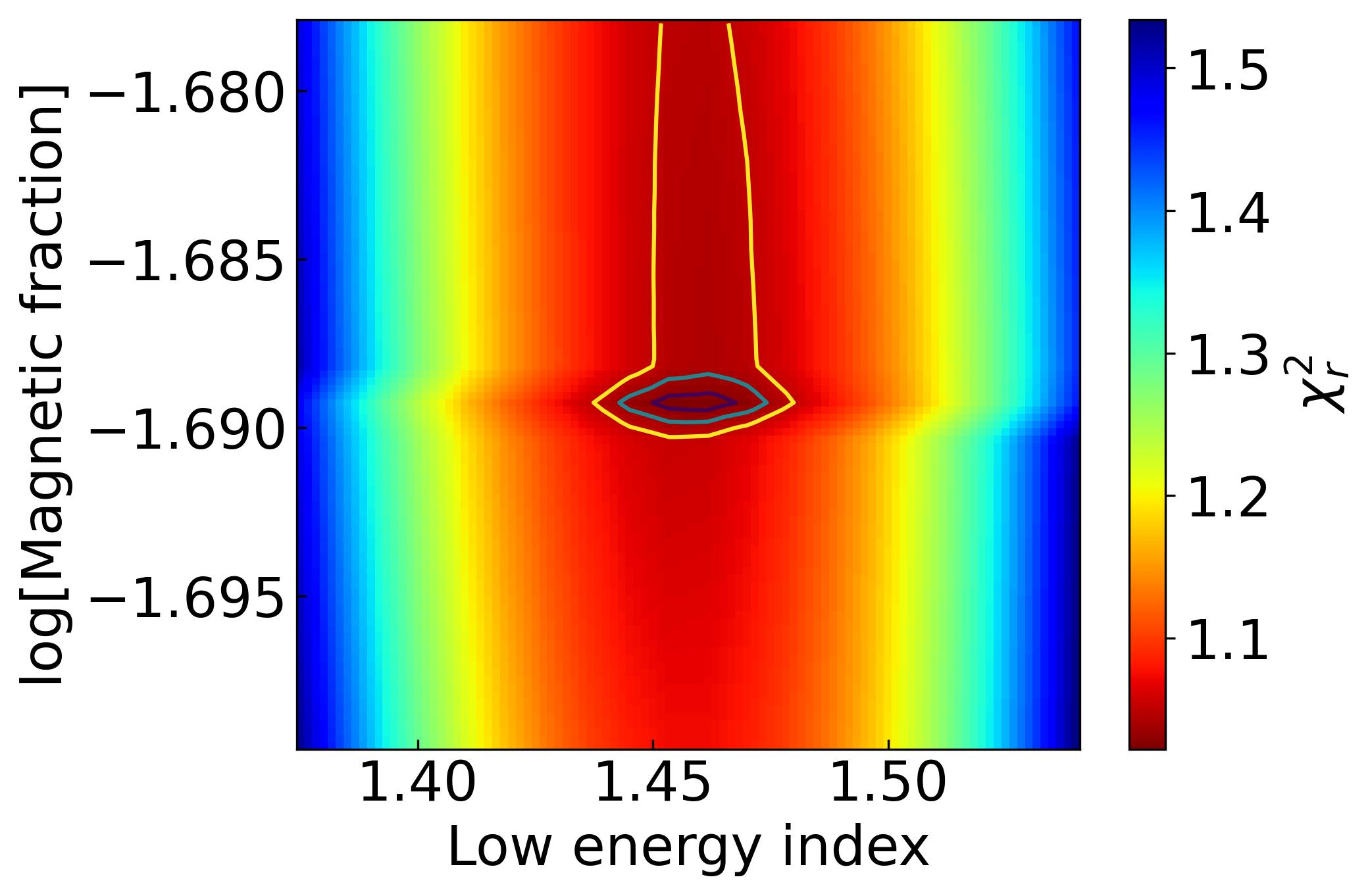}
\includegraphics[width=0.16\textwidth]{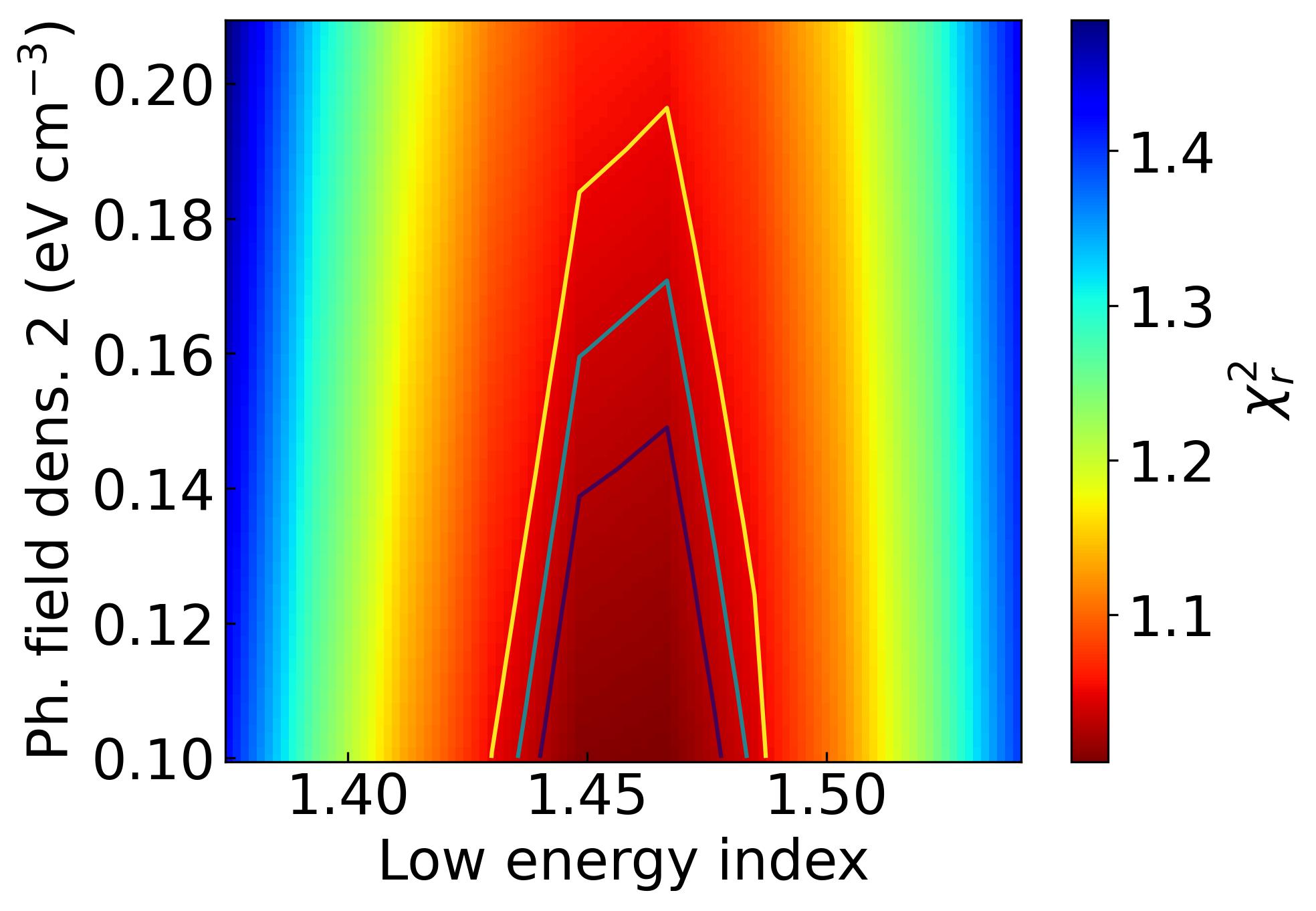}
\includegraphics[width=0.16\textwidth]{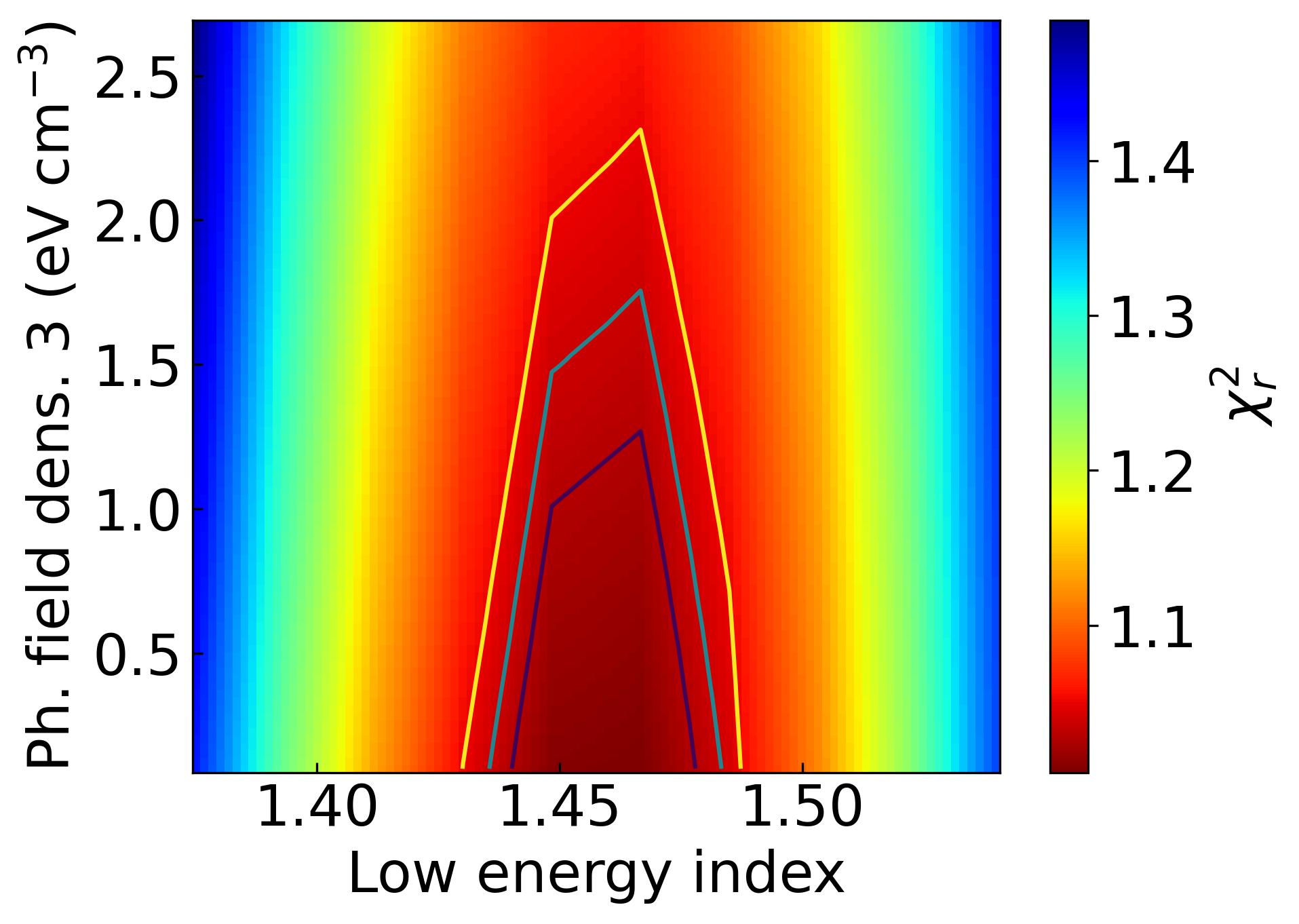}
\\
\includegraphics[width=0.16\textwidth]{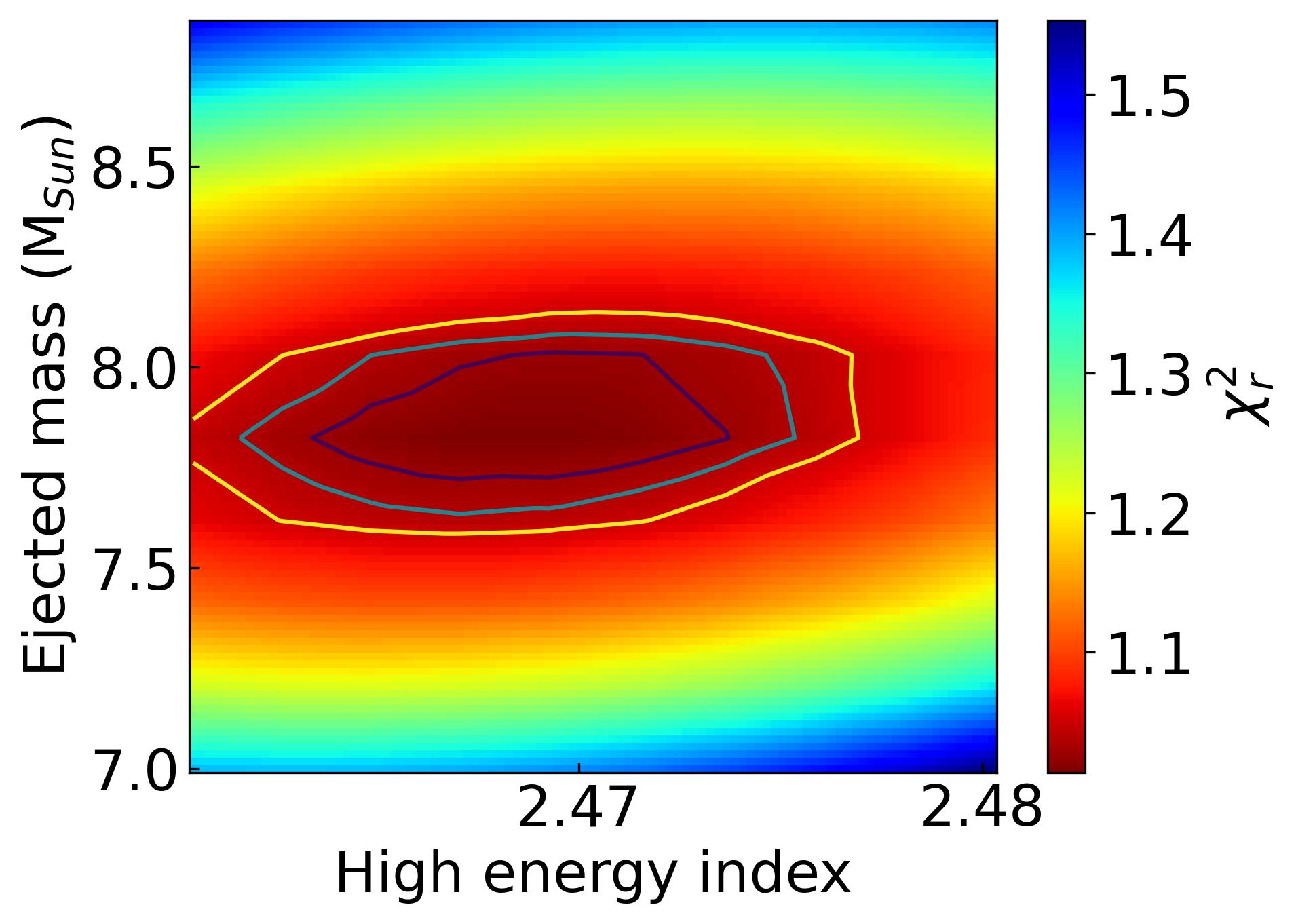}
\includegraphics[width=0.16\textwidth]{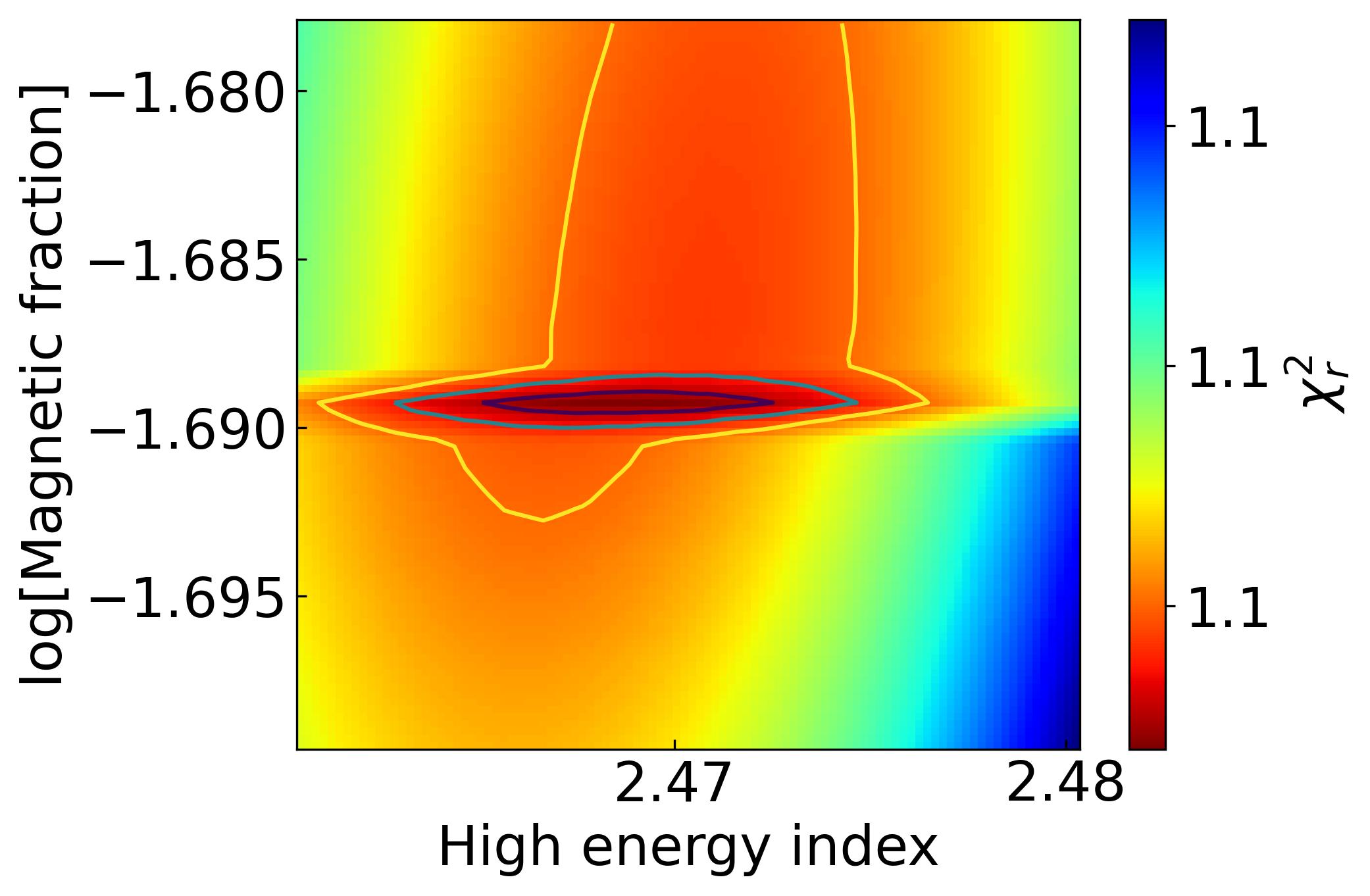}
\includegraphics[width=0.16\textwidth]{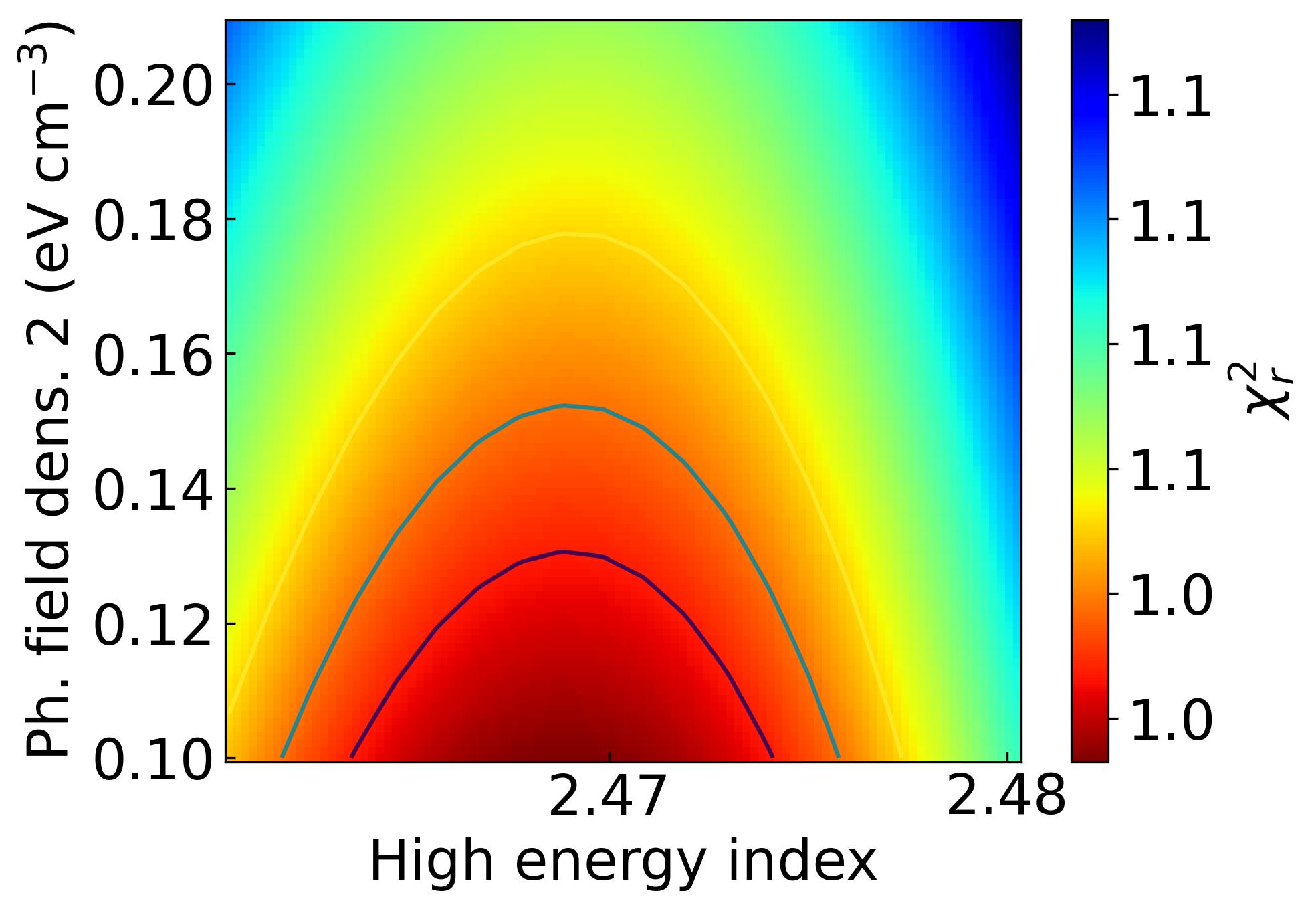}
\includegraphics[width=0.16\textwidth]{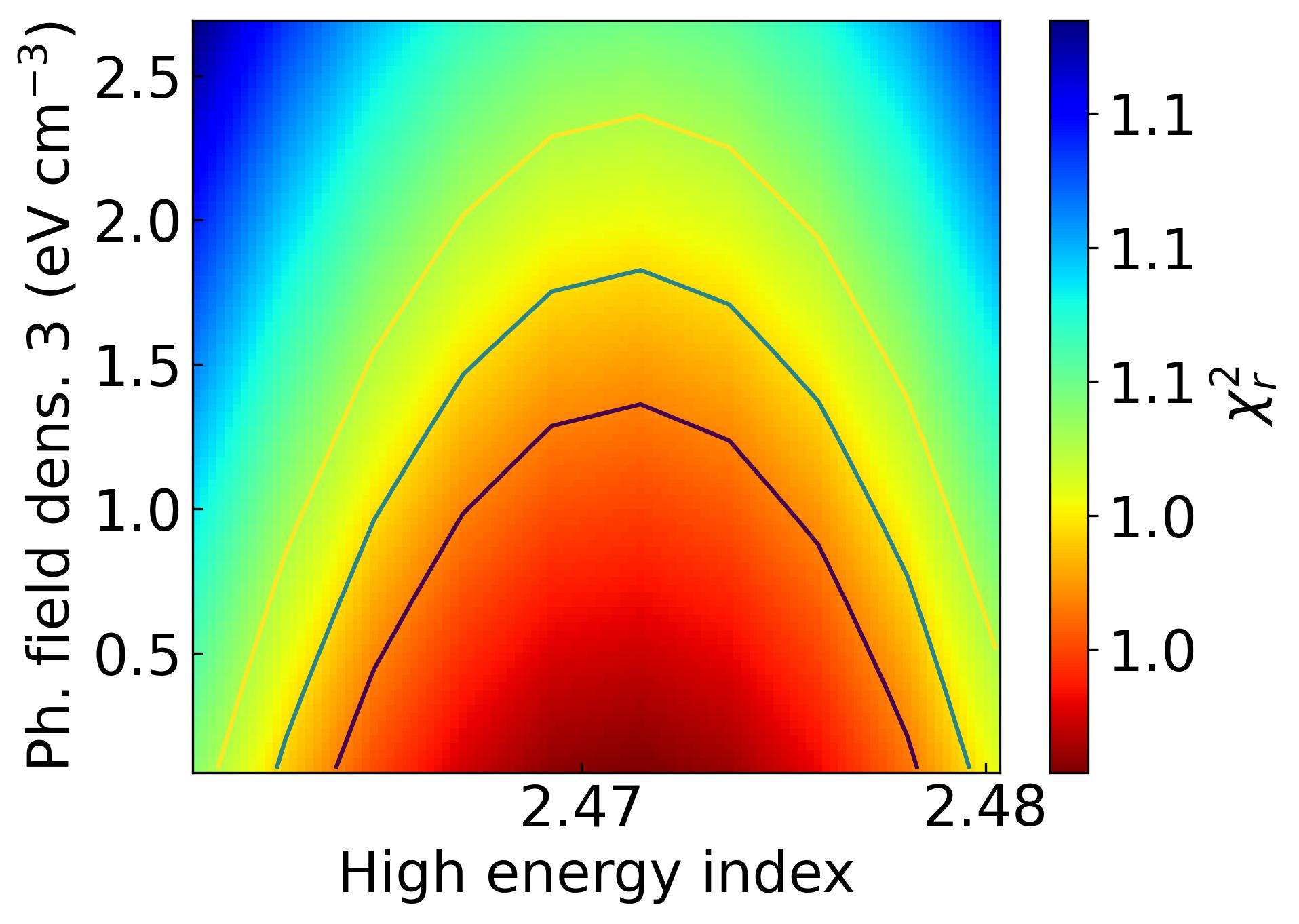}
\\
\includegraphics[width=0.16\textwidth]{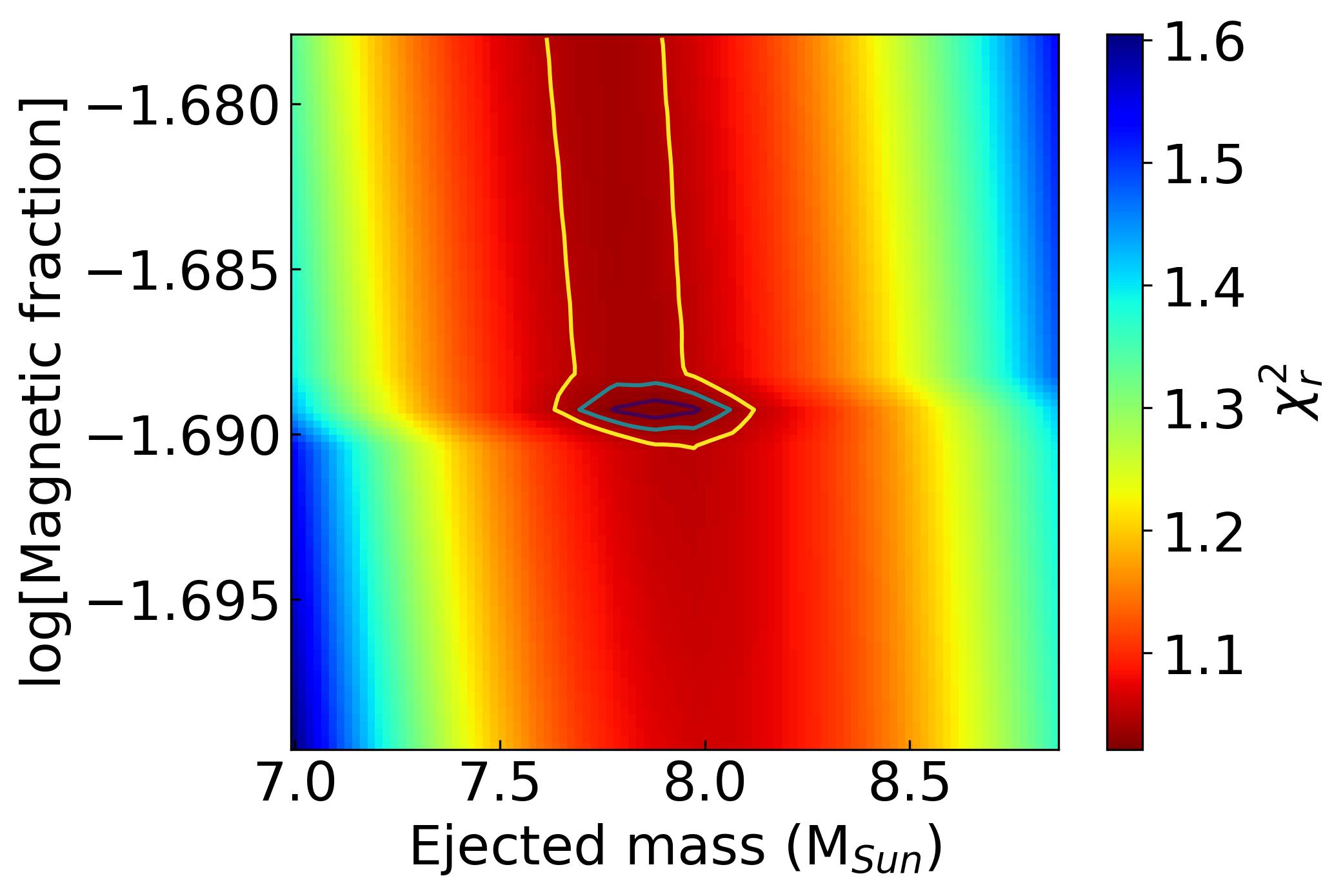}
\includegraphics[width=0.16\textwidth]{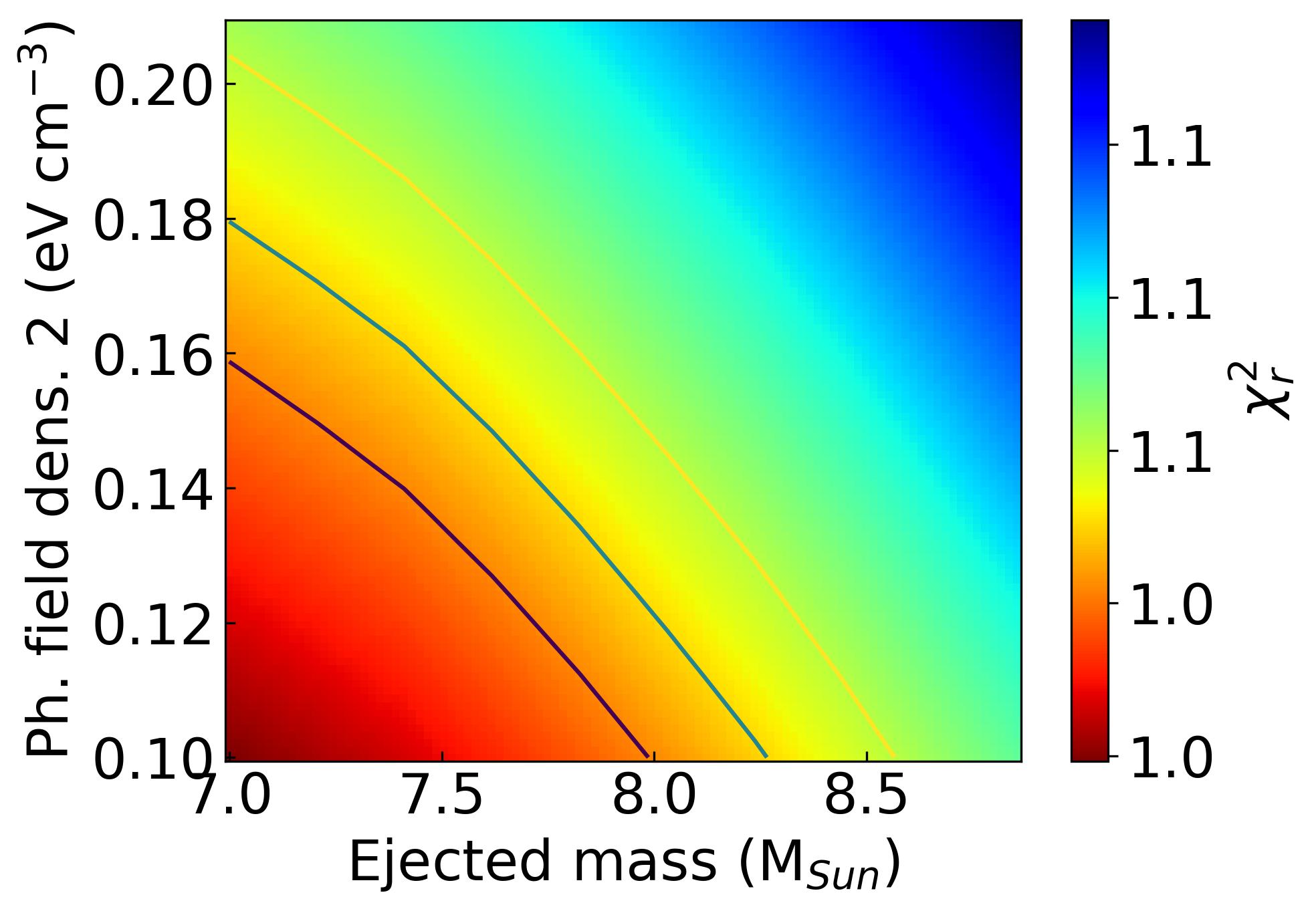}
\includegraphics[width=0.16\textwidth]{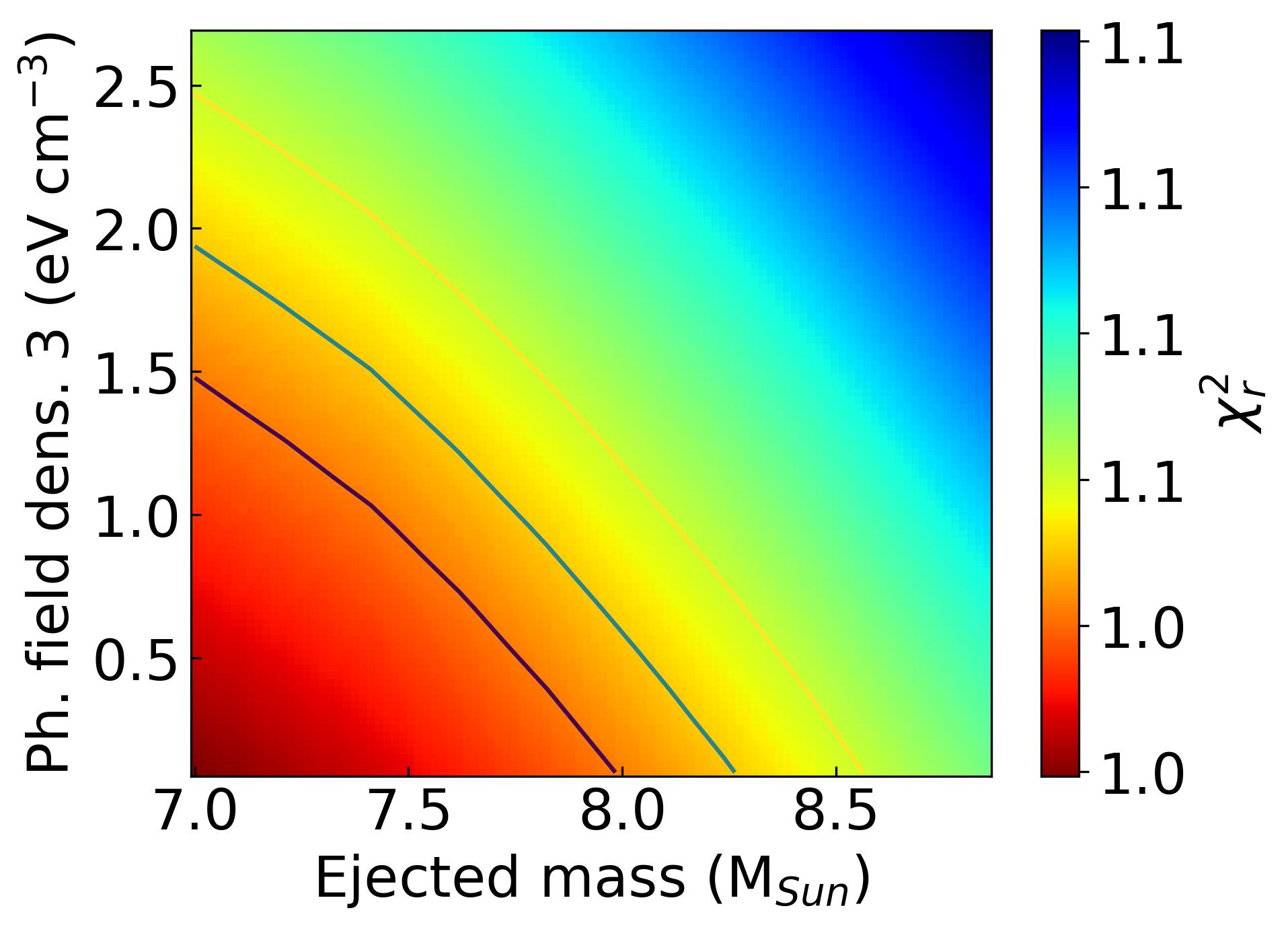}
\\
\includegraphics[width=0.16\textwidth]{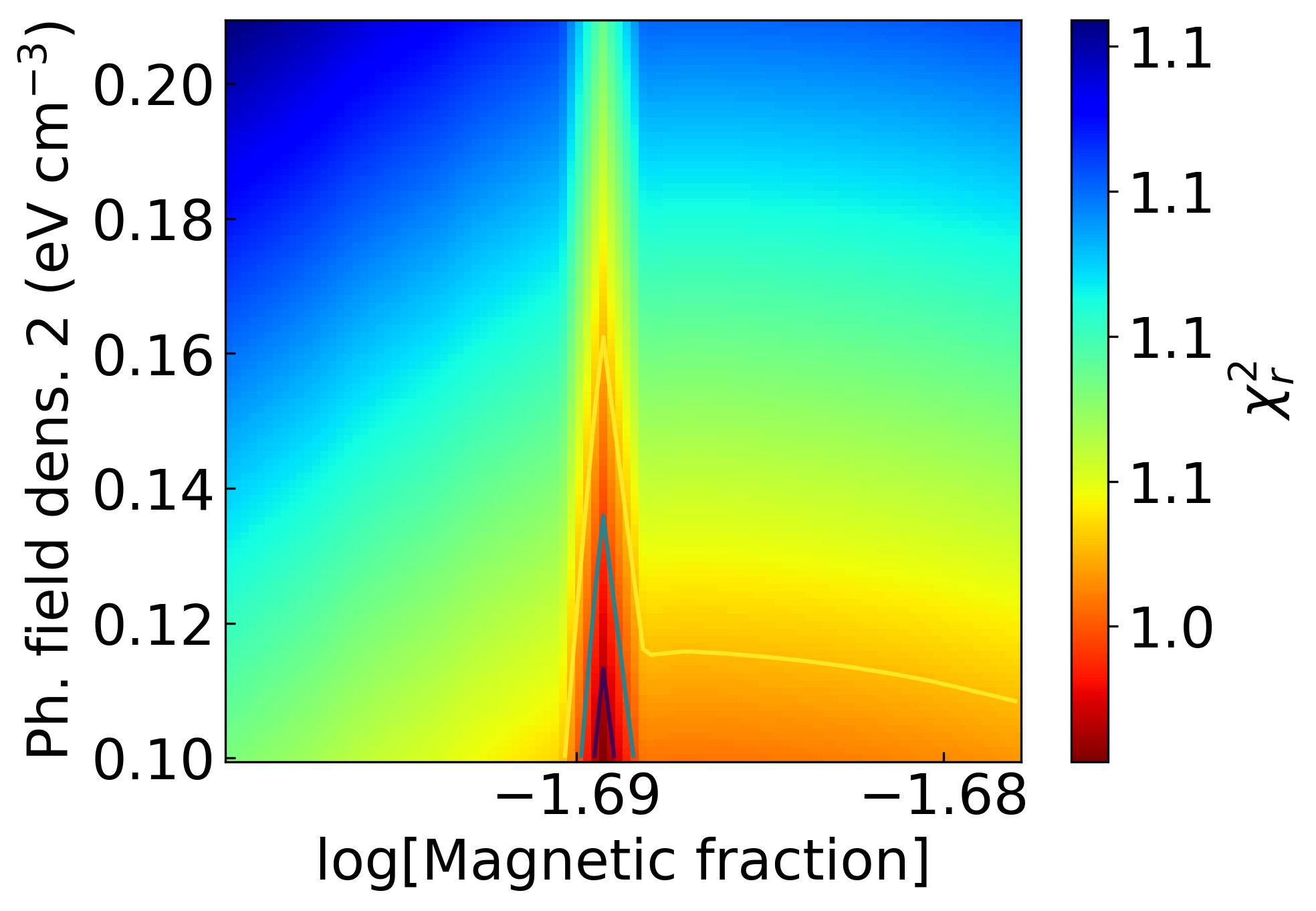}
\includegraphics[width=0.16\textwidth]{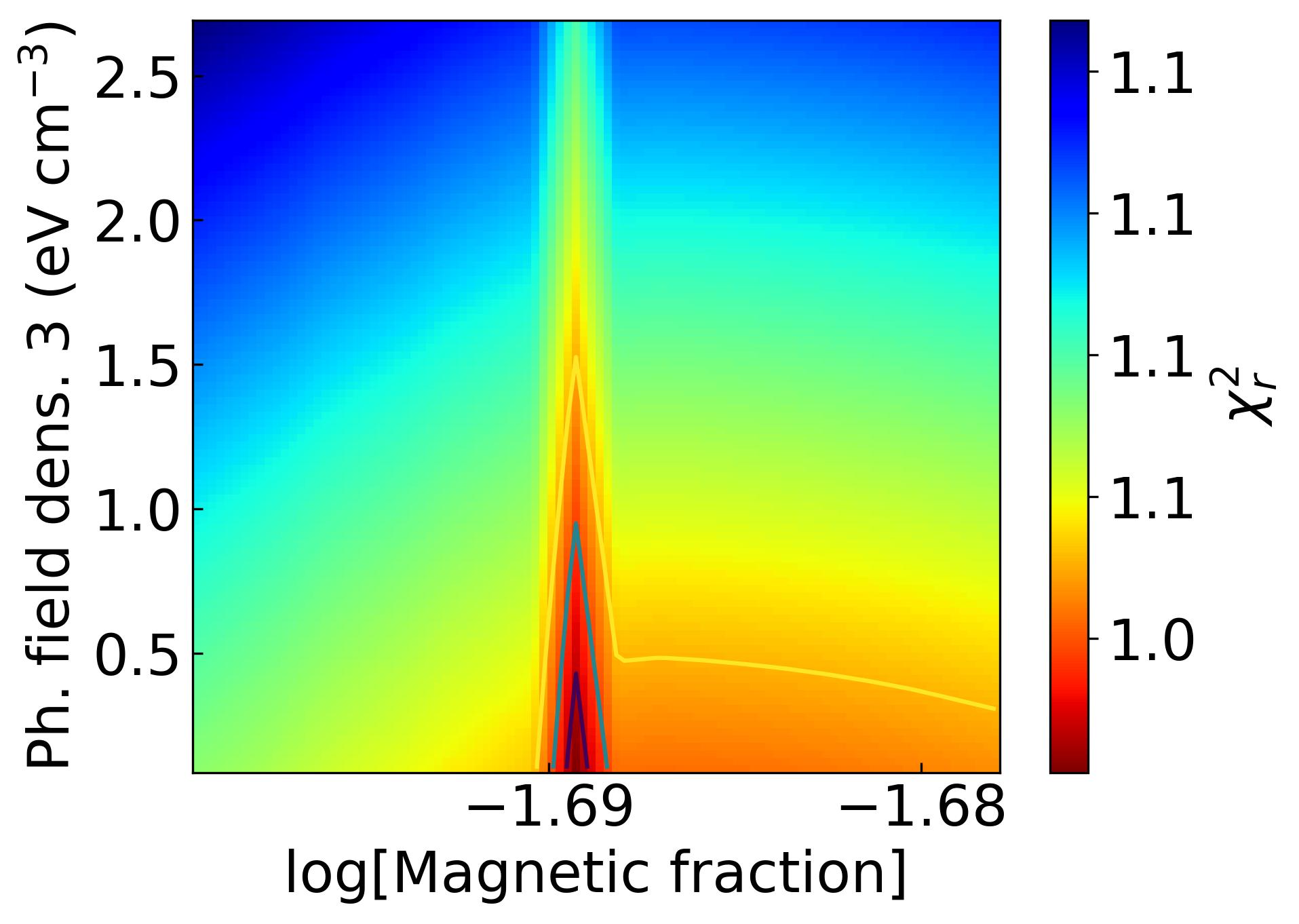}
\\
\includegraphics[width=0.16\textwidth]{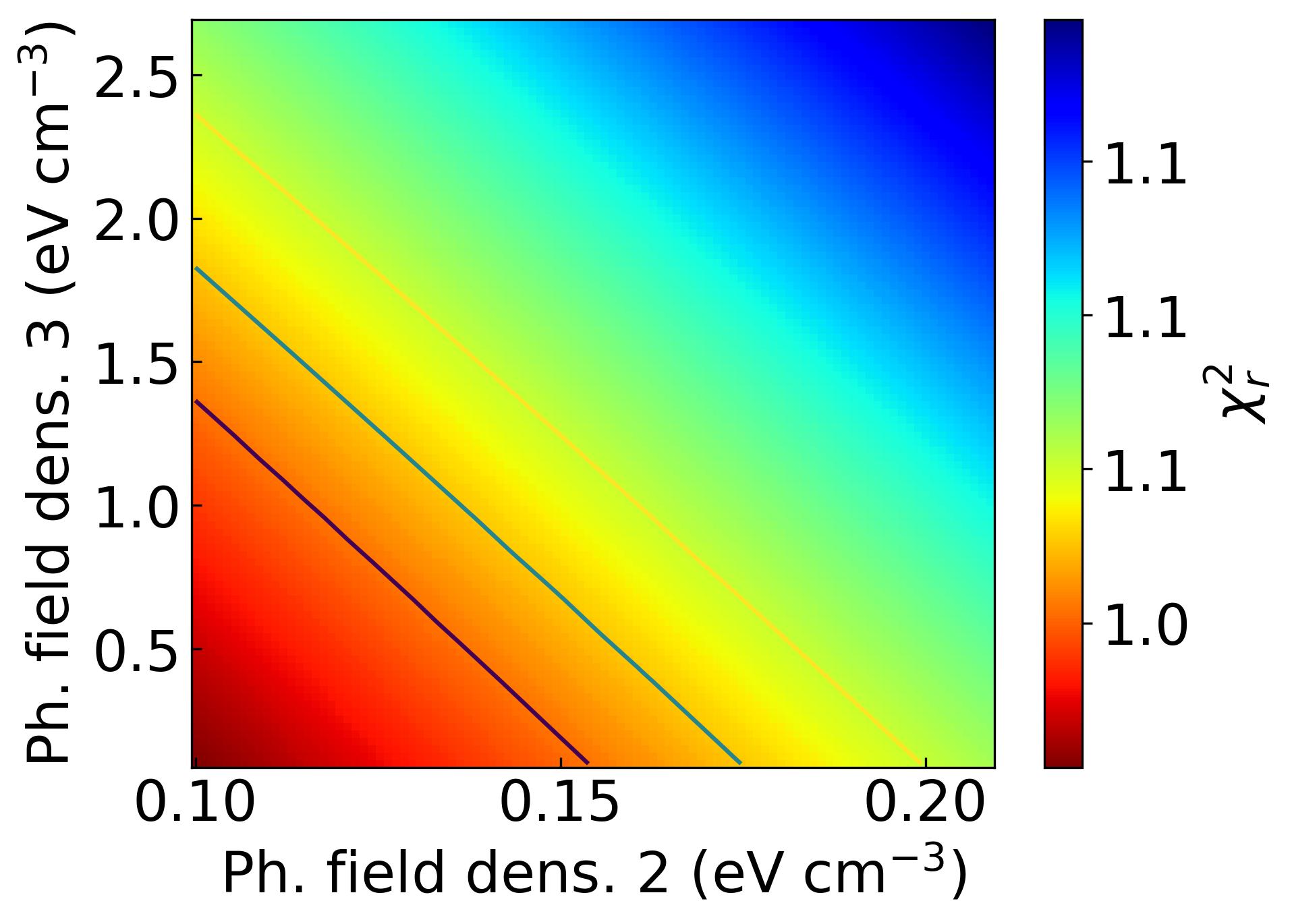}
\caption{Reduced $\chi^2$ maps for Crab Nebula. The dark blue, turquoise and yellow contours correspond to the 1, 2 and 3$\sigma$ confidence levels.}
\label{fig:crabchi}
\end{figure*}

Figure \ref{fig:crabchi} shows the shape of the reduced chi-squared function 
around the solution given by TIDE for each parameter pair. 
Most of the uncertainty in the model is introduced by the subdominant FIR and NIR energy densities: the 1, 2 and 3$\sigma$ ellipsoids are elongated and cover the entire possible value parameter range. This occurs as a result of the fact that none of them is actually relevant for the fit. Interestingly, however, the automatically selected FIR inverse Compton contribution is larger than the NIR one. 
The rest of the parameters, those related with the injection function (energy break and low and high energy injection slopes) together with the magnetic fraction and the ejected mass are well determined in the $\chi^2_r$ maps, confirming that the degeneracy is minimal.

\subsubsection{The automatic fit in the context of other models}

The automatically chosen parameters are compatible with earlier studies using previous incarnations of the code (e.g., \cite{Torres2014, aleksic2015}). 
Differences introduced by the improved treatment of the dynamics explain 10-20\% changes in the resulting radius and ejected mass (we now discount magnetospheric pulsed emission, use the new SNR/PWN shock positions as in \cite{bandiera2021}, time-dependence of the spin-down luminosity in the radius evolution equations as in \cite{martin2016}, and an improved treatment for the energy conservation in the dynamical evolution as in \cite{bandiera2020}).
The small variation in radius also reflects in the magnetic fraction, while the magnetic field is similar.
Regarding the FIR and NIR energy densities, the fitting algorithm confirms that the IC-FIR and IC-NIR contributions are negligible in order to explain the IC flux.

Apart from TIDE, other models use the new data from LHAASO to update their corresponding fit, but it is difficult to compare them directly due to the features of each model being different.
Here, we will compare our results with three different models. One of them is previous to the LHAASO data, and based on the model of \cite{meyer2010} and published with the MAGIC data points released in \cite{aleksic2015}.
The other two models include the new data: the paper of the LHAASO collaboration itself \citep{lhaaso2021} and the one proposed in \cite{nie2022}.
\cite{meyer2010} (hereinafter MEY) fits the Crab spectrum with a steady one zone model composed by two pair populations: one which accounts for the radio emitting pairs and another one for the wind accelerated particles. These two populations are not calculated by evolving the diffusion-loss equation but  as a fit of the particle spectrum observed today. 
The model uses 15 parameters to fit the spectrum.
These are thus significant differences, which makes a direct comparison unfeasible.
The model used in \citep{lhaaso2021} (hereinafter LHA) is also a steady one zone model, but more simplified in comparison with MEY, needing only three parameters, in particular,  the spectral index at high energies, the magnetic field, and the super exponential energy cut-off.
Finally, TIDE and the model of \cite{nie2022} (hereinafter NIE) assume a broken power-law injection of particles and evolve them in time taking into account adiabatic and only synchrotron radiative energy losses.
From the right-hand side of Equation \ref{eq:difflos}, NIE neglects the second term and radiative losses are included in the escape term (third term of Equation \ref{eq:difflos}).
As it was shown in \citealt{martin2012}, this kind of approximations has a significant impact in the results along the time evolution, obtaining differences in the calculated flux of more than a factor 2 by using the same parameters.
Qualitatively bearing in mind account all said differences, selected model values are not far from one another.
For instance, the maximum energy at injection for the models MEY, LHA and NIE is, respectively, 3.7, 2.15 (we take the super-exponential energy cut-off as a reference) and 5.649 PeV (this is the initial maximum energy at injection, so we should take it as an upper limit).
In the case of TIDE, the maximum energy gives 4.7 PeV (see Table \ref{tab:results}) which is an intermediate value of those obtained by these models.
We have seen in \cite{lhaaso2021,nie2022} some trials to include an hadronic component into the spectrum to explain the enhance of the VHE flux at $~$1 PeV in the Crab Nebula.
Introducing this component is certainly possible, but for the current set of data, there is not enough data to really favour a lepto-hadronic model over a leptonic one: it would introduce more free parameters (at least two) to fit only one additional data point.
However, the appearance of such enhancement may require this inclusion in the future.

\subsubsection{A lower-energy explosion}
\label{crab-low}

It has been earlier suggested that SN 1054 was an atypical, low-energy explosion (e.g., \cite{Frail1995,Yang2015}).
Here, we shall also explore this possibility by considering, on the one hand, a model with $E_{sn} = 10^{50}$ erg, and on the other
by additionally consider $E_{sn}$ as a free parameter within a range $10^{49}$ --  $10^{51}$ erg. 

For the $E_\text{sn}$ case, the model clearly worsens and requires an ejected mass lower than 7M$_\odot$ to be competitive. 
If we prevent this, then we cannot fit the radius of the nebula correctly. 
For the latter case, when we allow $E_\text{sn}$ to vary, we obtain a very similar result as show in Table~\ref{tab:results}: $E_\text{sn}=8.3^{+0.5}_{-0.4} \times 10^{50}$ erg, which is close to the a priori canonical value of $10^{51}$ erg as initially assumed.

\subsection{3C 58}
\label{sec:3c58}

3C 58 is a PWN associated with the pulsar PSR J0205+6449 \citep{murray2002}.
Regarding the morphology, this PWN shows a faint jet and a toroidal structure perpendicular to it.
3C 58 has been detected in radio, IR, X-rays, and gamma-rays (see the references in the caption of Figure \ref{fig:sed}).
This makes 3C 58 another good example to test the automatic fitting code, albeit
there is still an ongoing discussion on its age and distance in comparison to the more certain values of the Crab nebula. 
Regarding the age, \cite{stephenson2002} associate the nebula to the supernova SN 1181, resulting a current age of 841 yr, but other estimates give an age of up to 7 kyr (see e.g., \citealt{fesen2008} for a compilation).
For our model, we use the dynamical age of 2.5 kyr given by \cite{chevalier2004,chevalier2005} and come back to this discussion below, with a new approach.
We also note that the required ejected mass in order to fit the radius using an age of 841 yr would also be low ($< 8M_\odot$).
On the other hand, the distance ranges from 2 kpc \citep{kothes2010} to 3.2 kpc \citep{roberts1993}.
We take the newest determination of 2 kpc, where new H I data is used.

\begin{figure*}
\raggedleft
\includegraphics[width=0.16\textwidth]{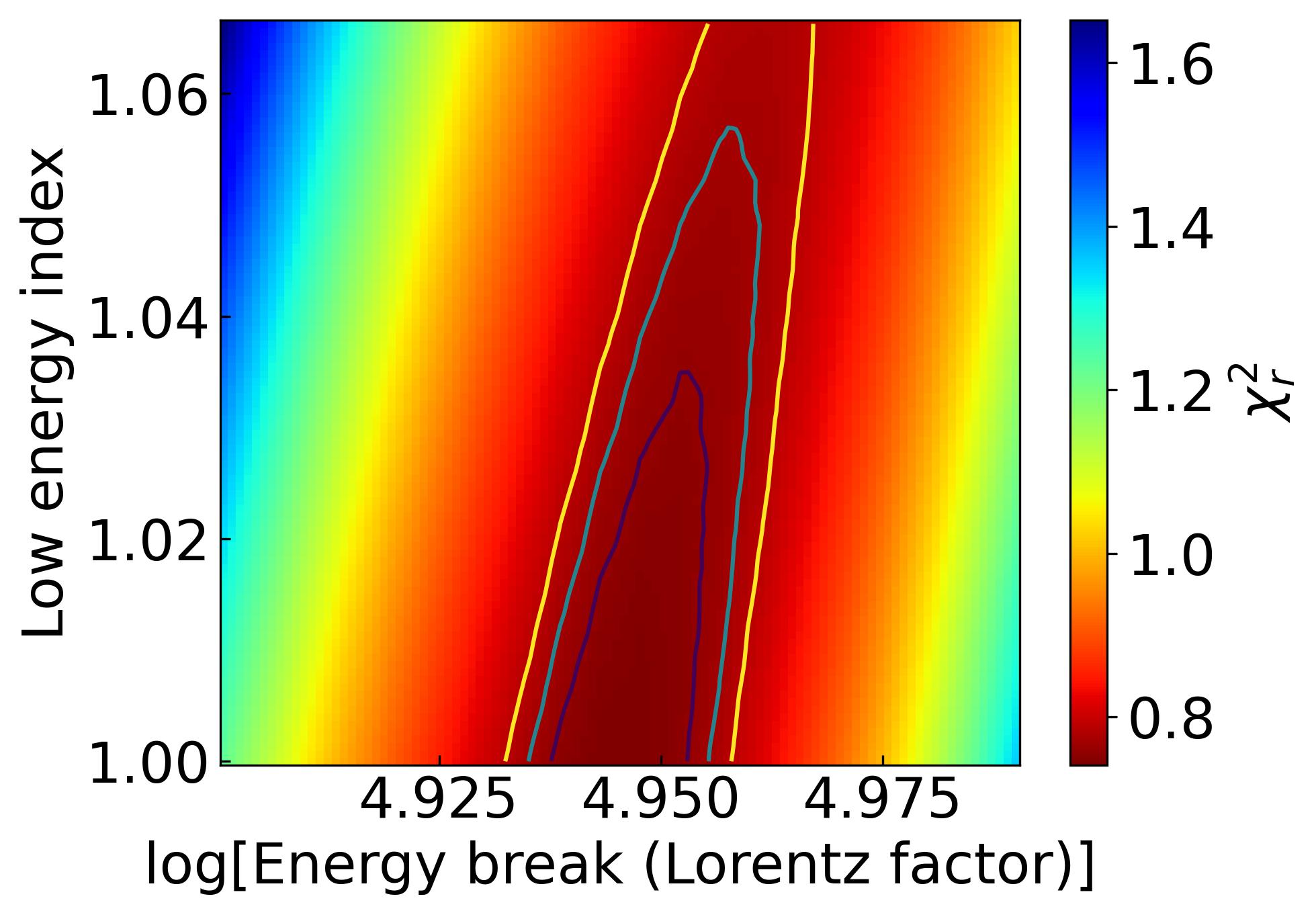}
\includegraphics[width=0.16\textwidth]{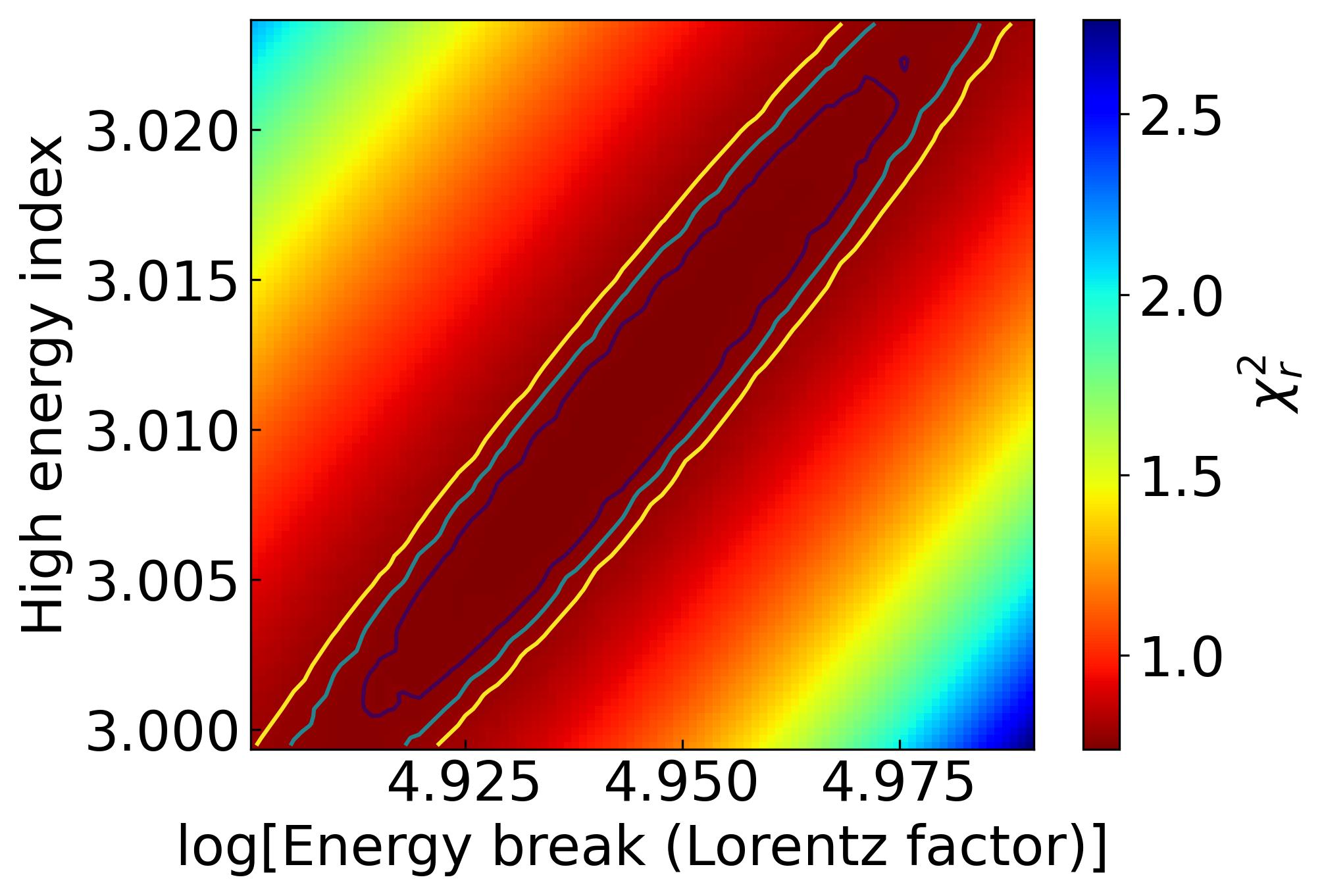}
\includegraphics[width=0.16\textwidth]{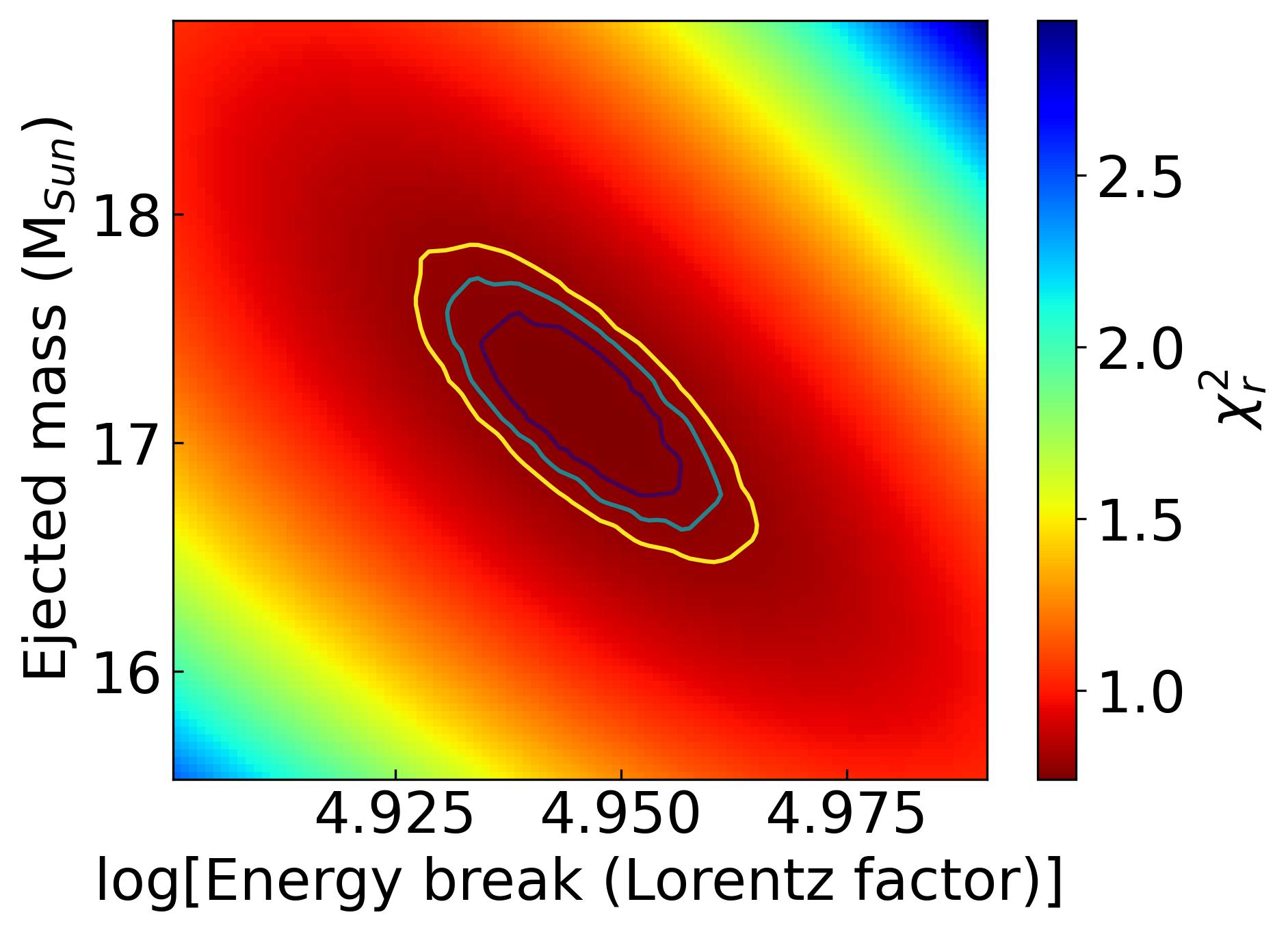}
\includegraphics[width=0.16\textwidth]{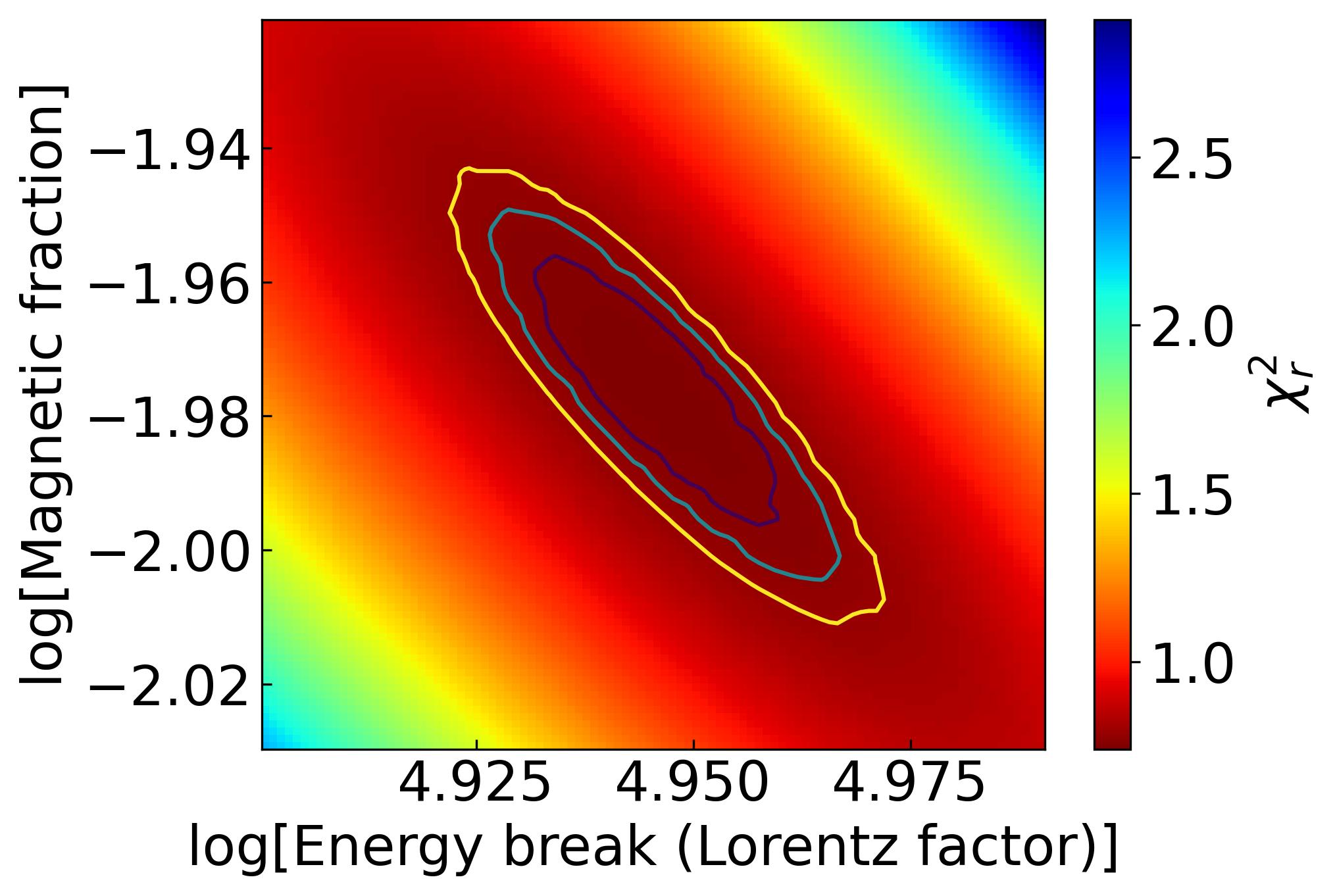}
\includegraphics[width=0.16\textwidth]{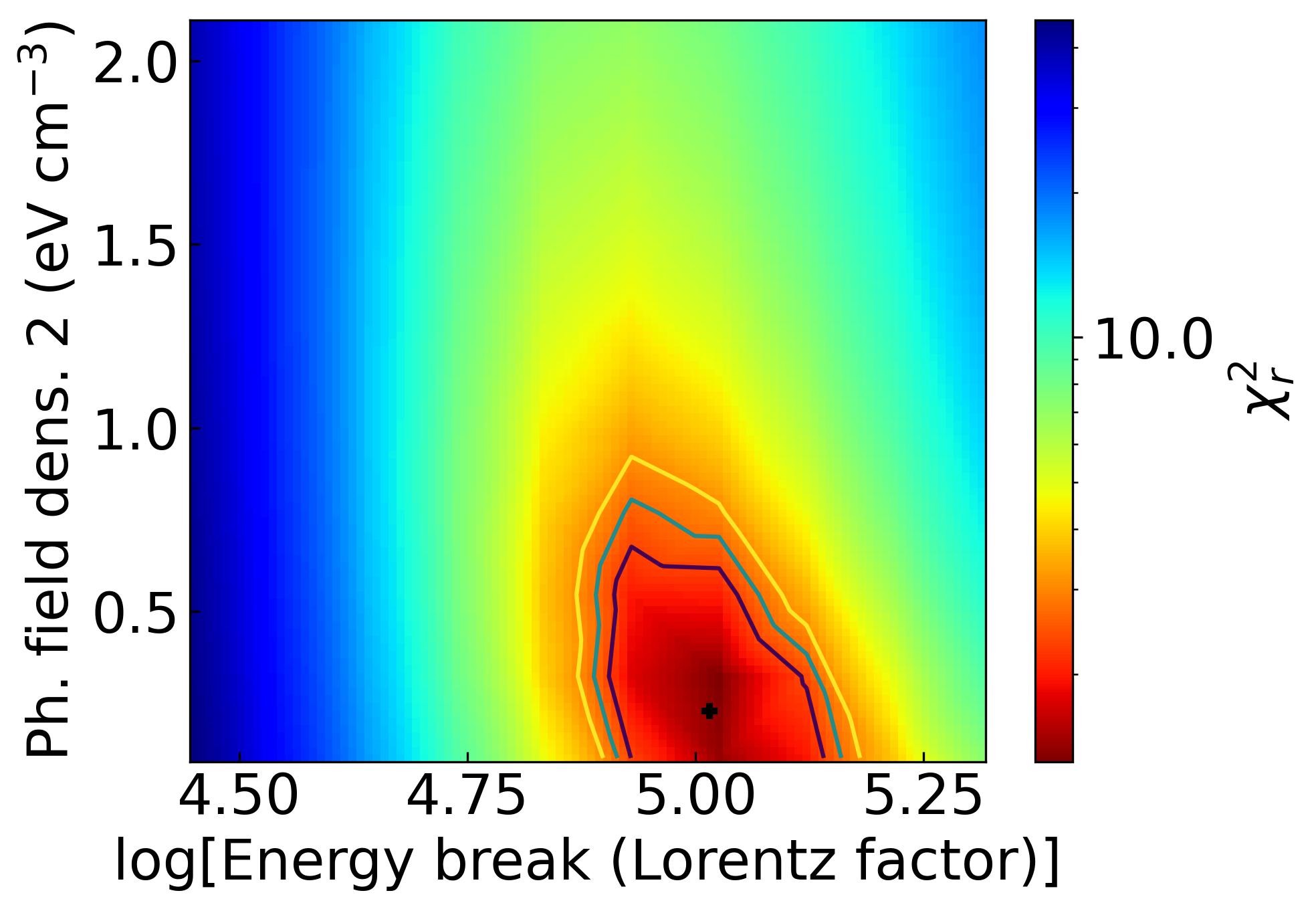}
\includegraphics[width=0.16\textwidth]{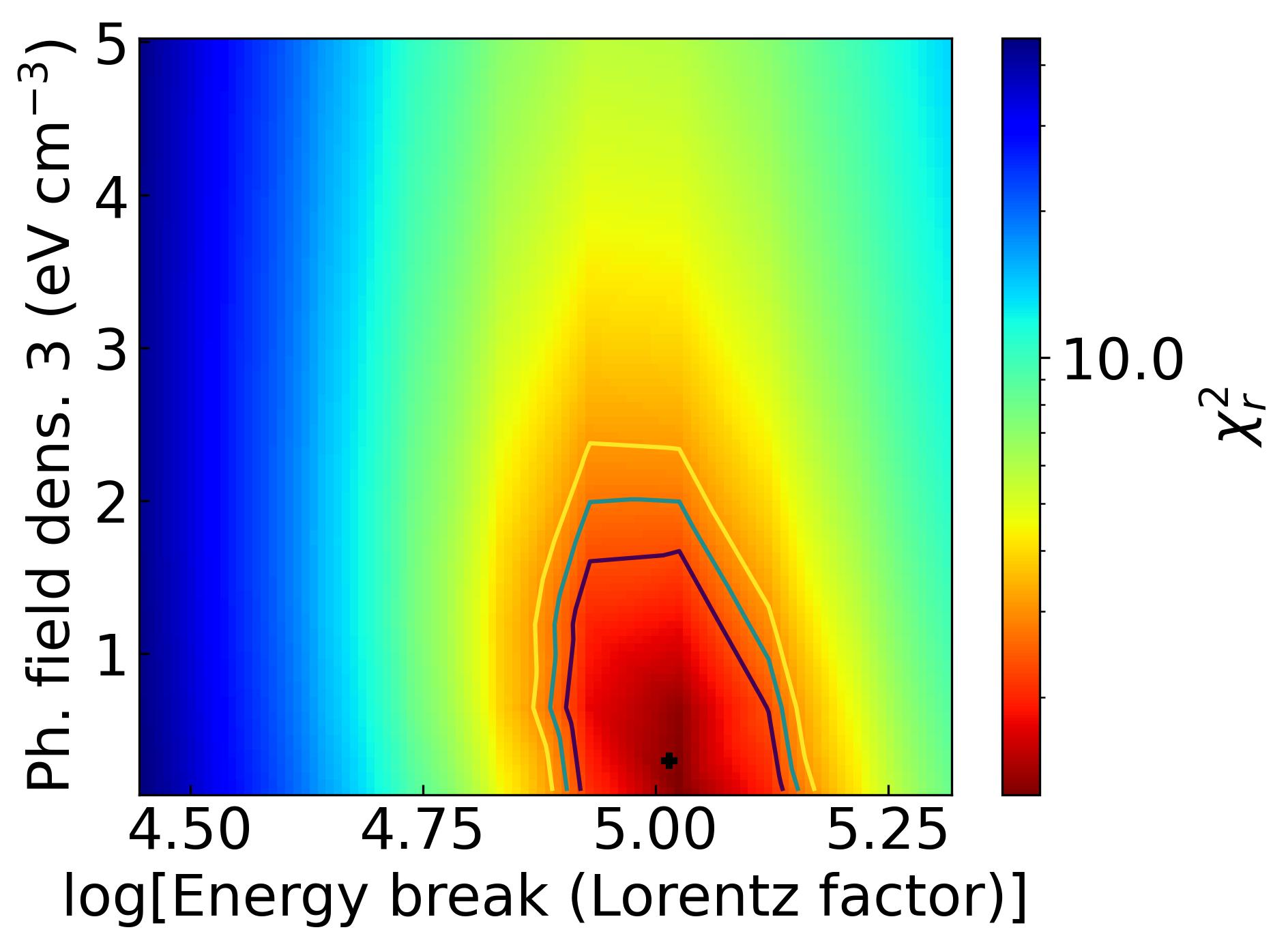}
\\
\includegraphics[width=0.16\textwidth]{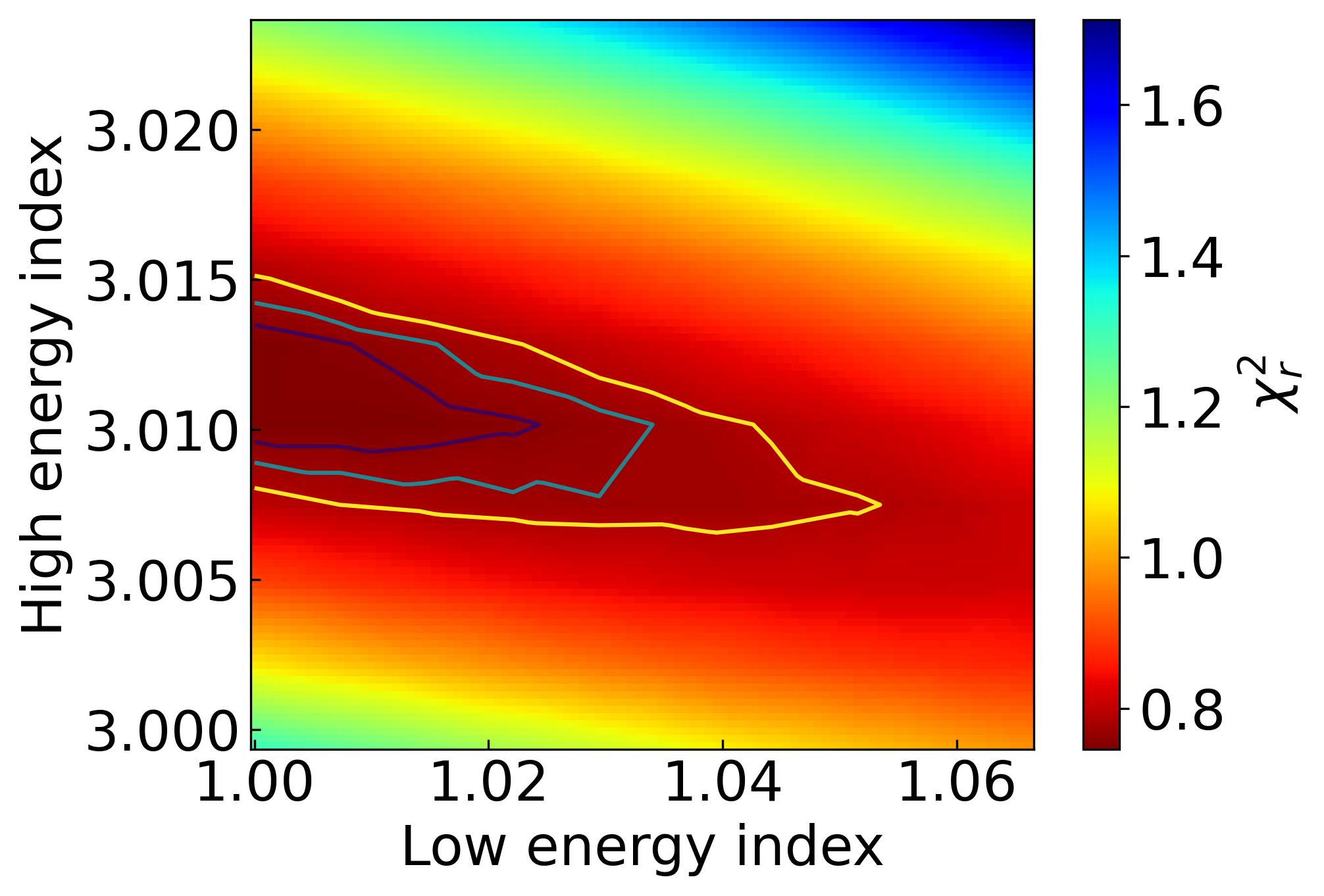}
\includegraphics[width=0.16\textwidth]{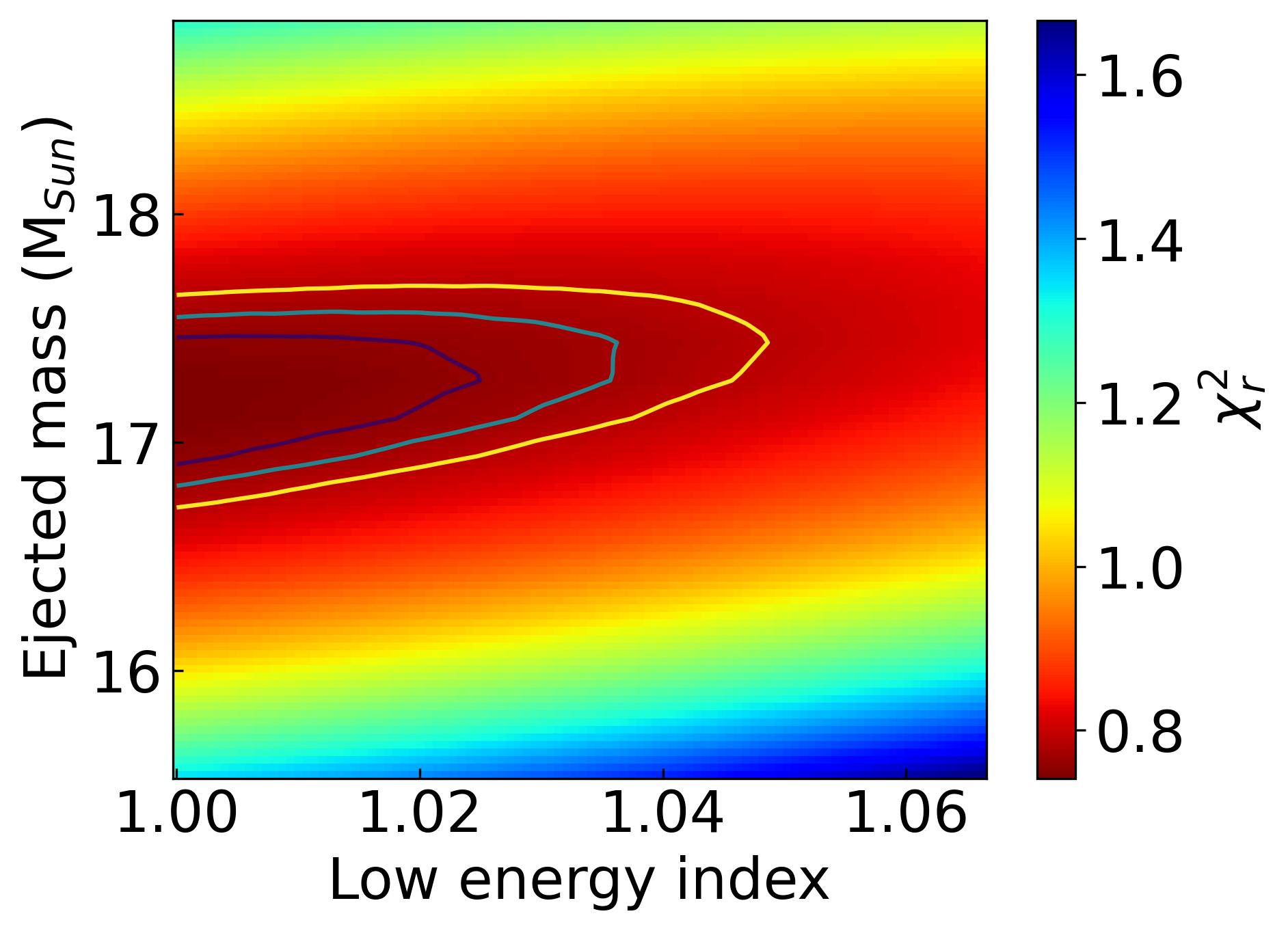}
\includegraphics[width=0.16\textwidth]{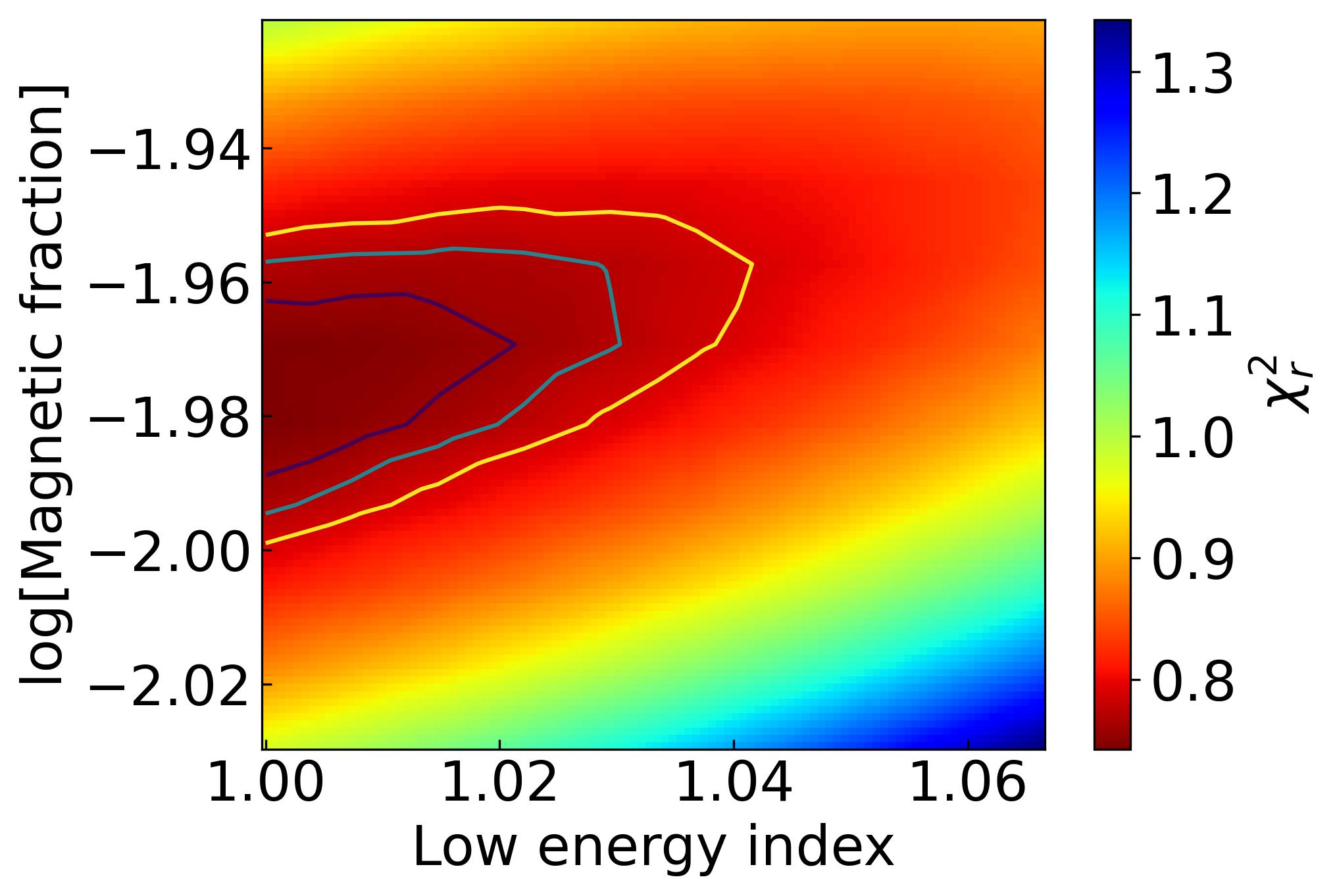}
\includegraphics[width=0.16\textwidth]{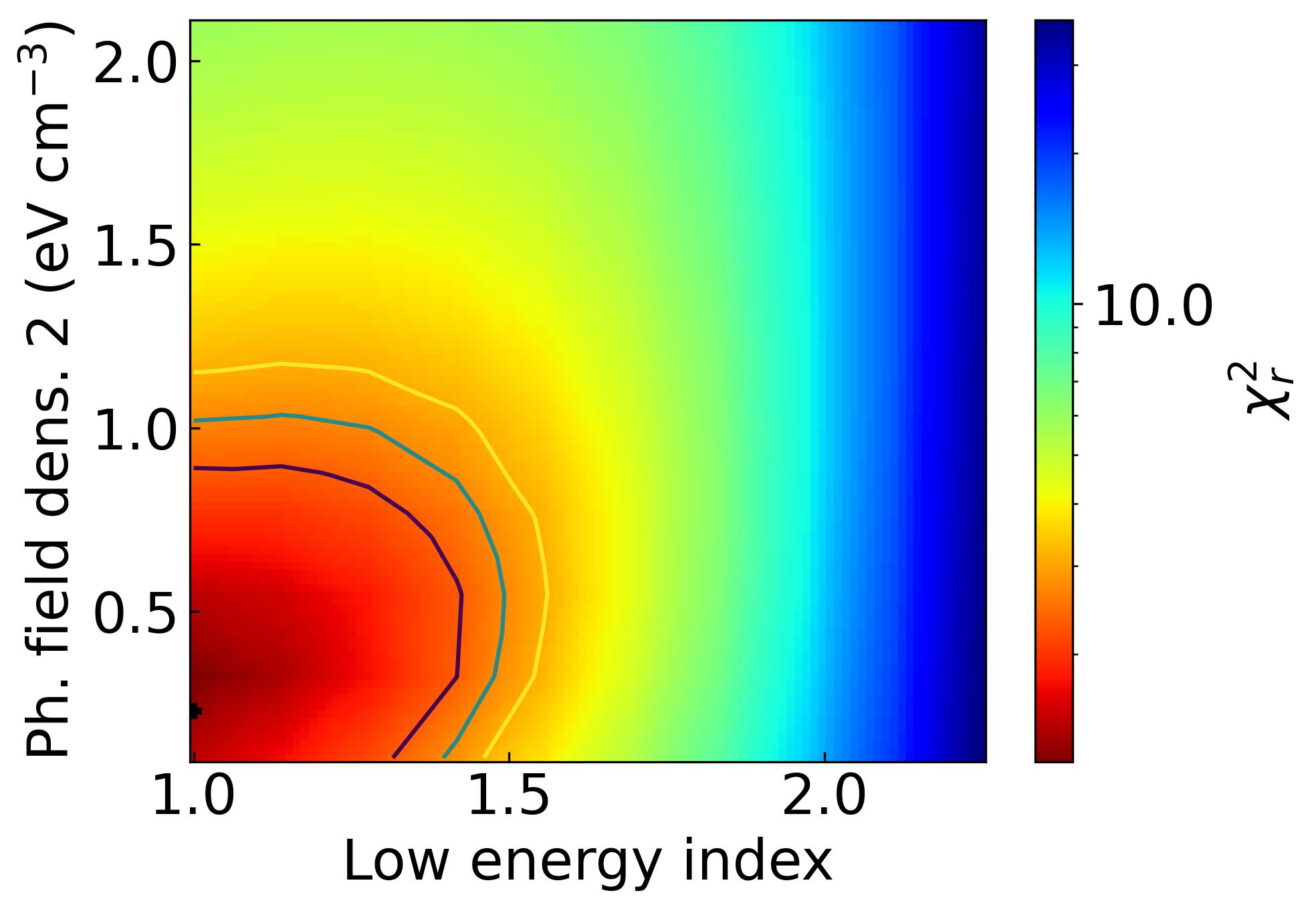}
\includegraphics[width=0.16\textwidth]{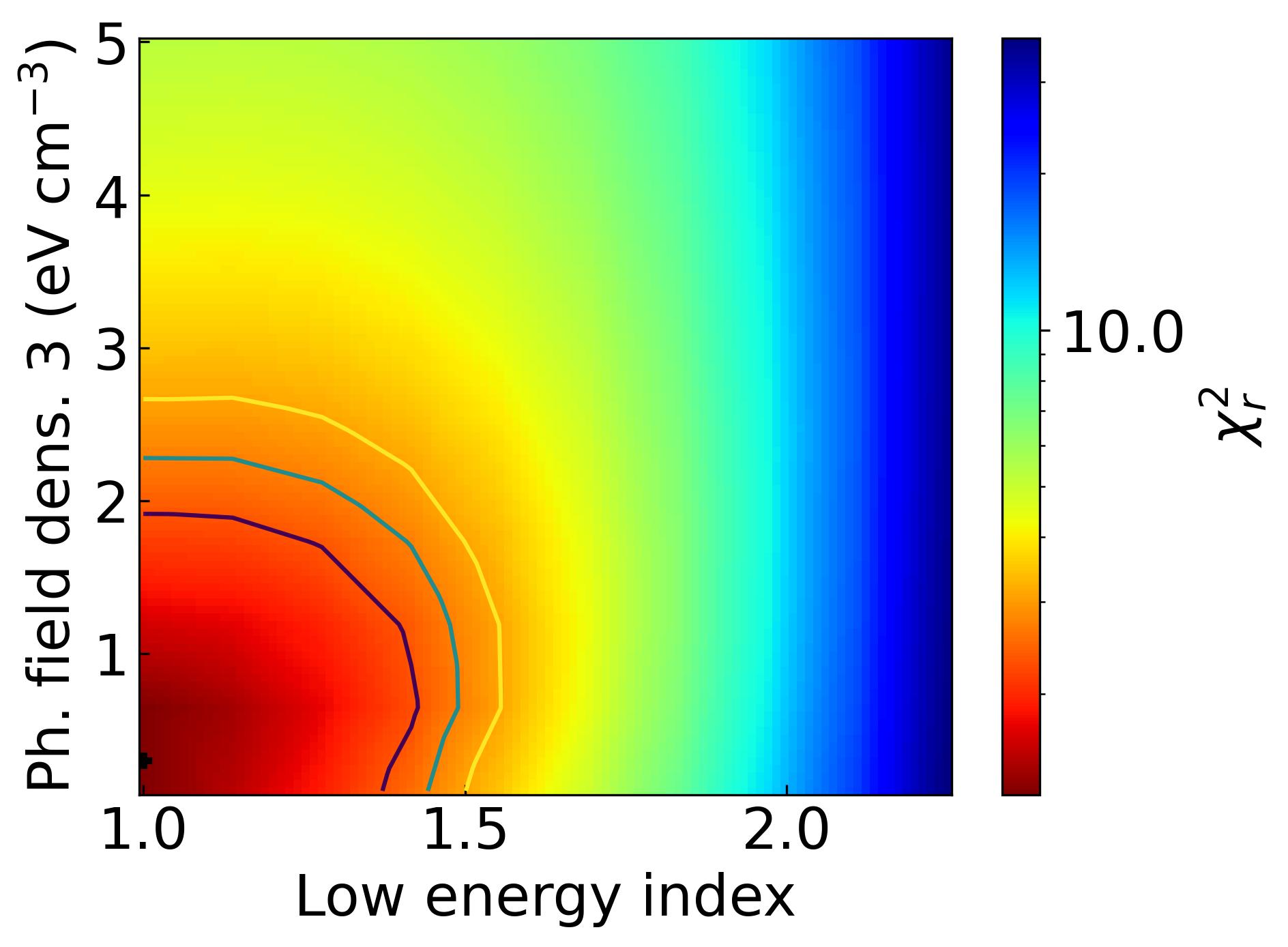}
\\
\includegraphics[width=0.16\textwidth]{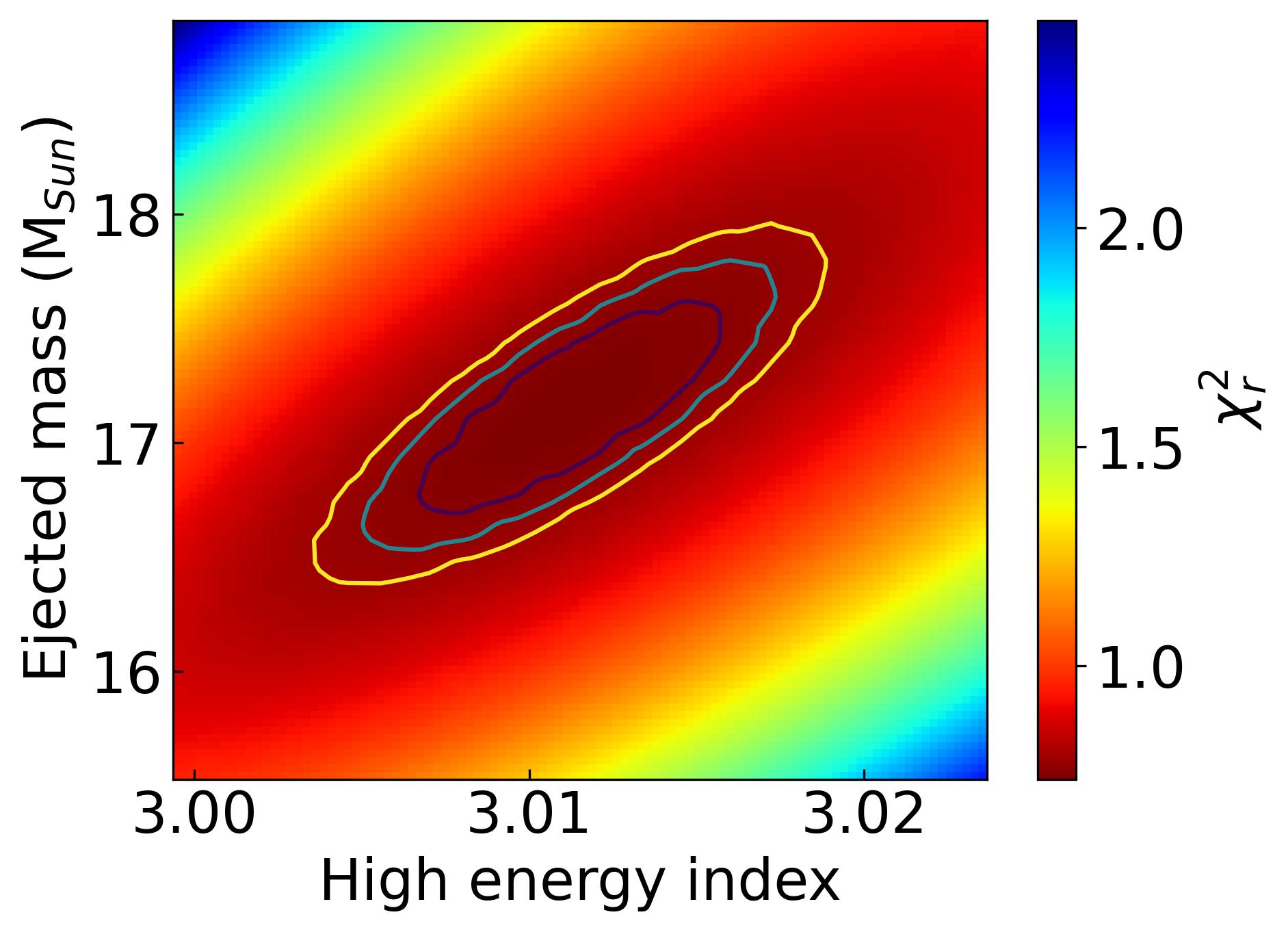}
\includegraphics[width=0.16\textwidth]{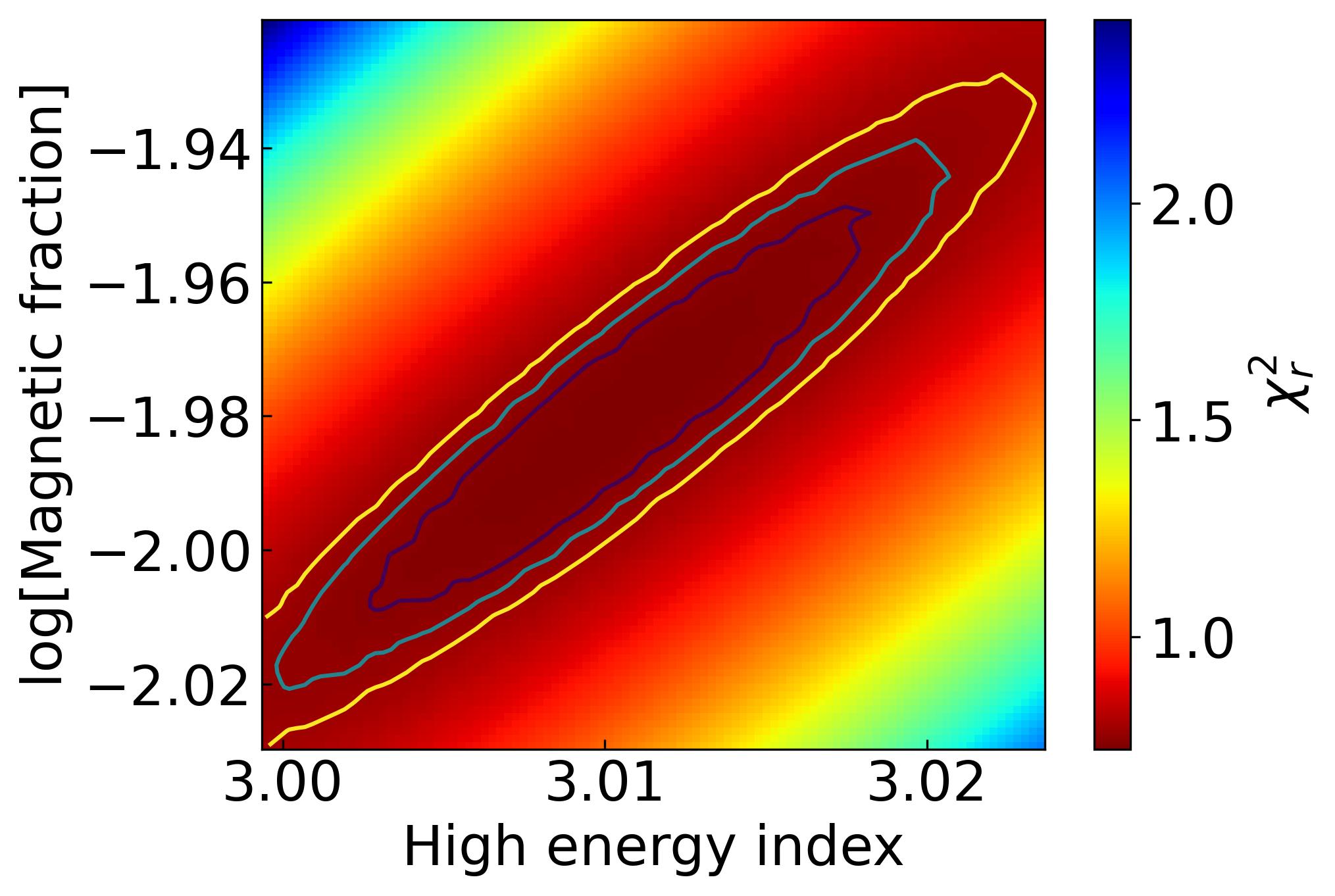}
\includegraphics[width=0.16\textwidth]{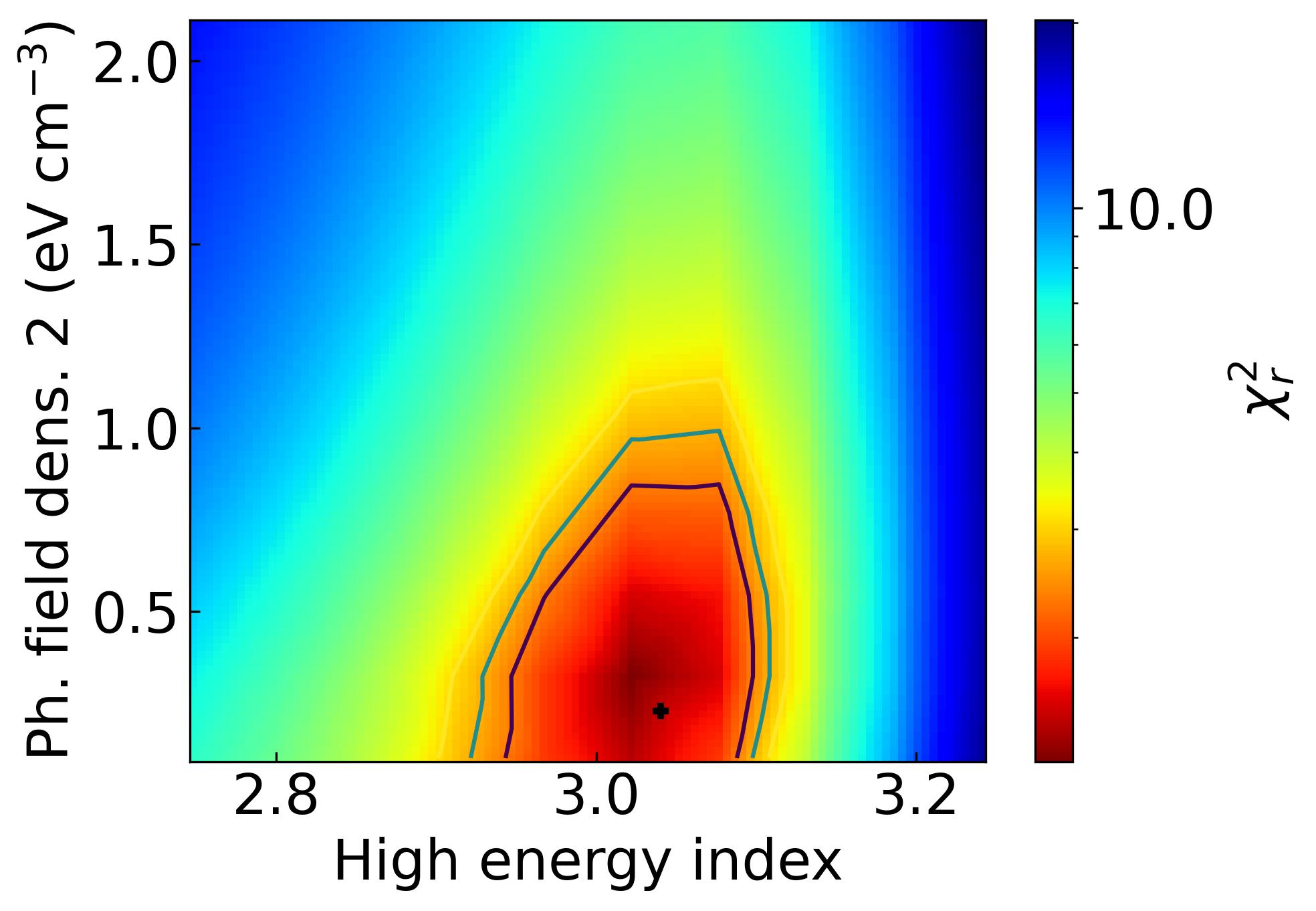}
\includegraphics[width=0.16\textwidth]{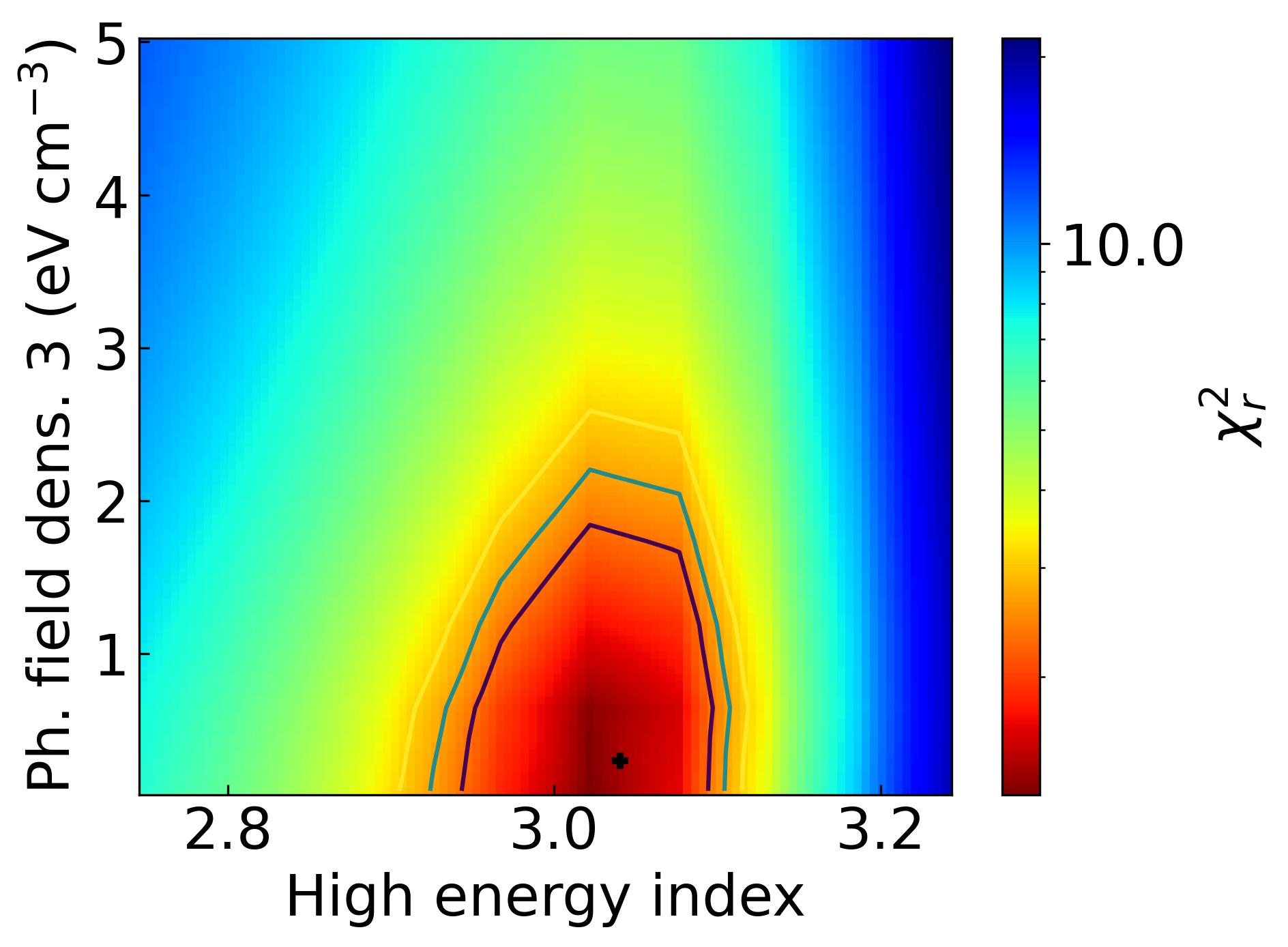}
\\
\includegraphics[width=0.16\textwidth]{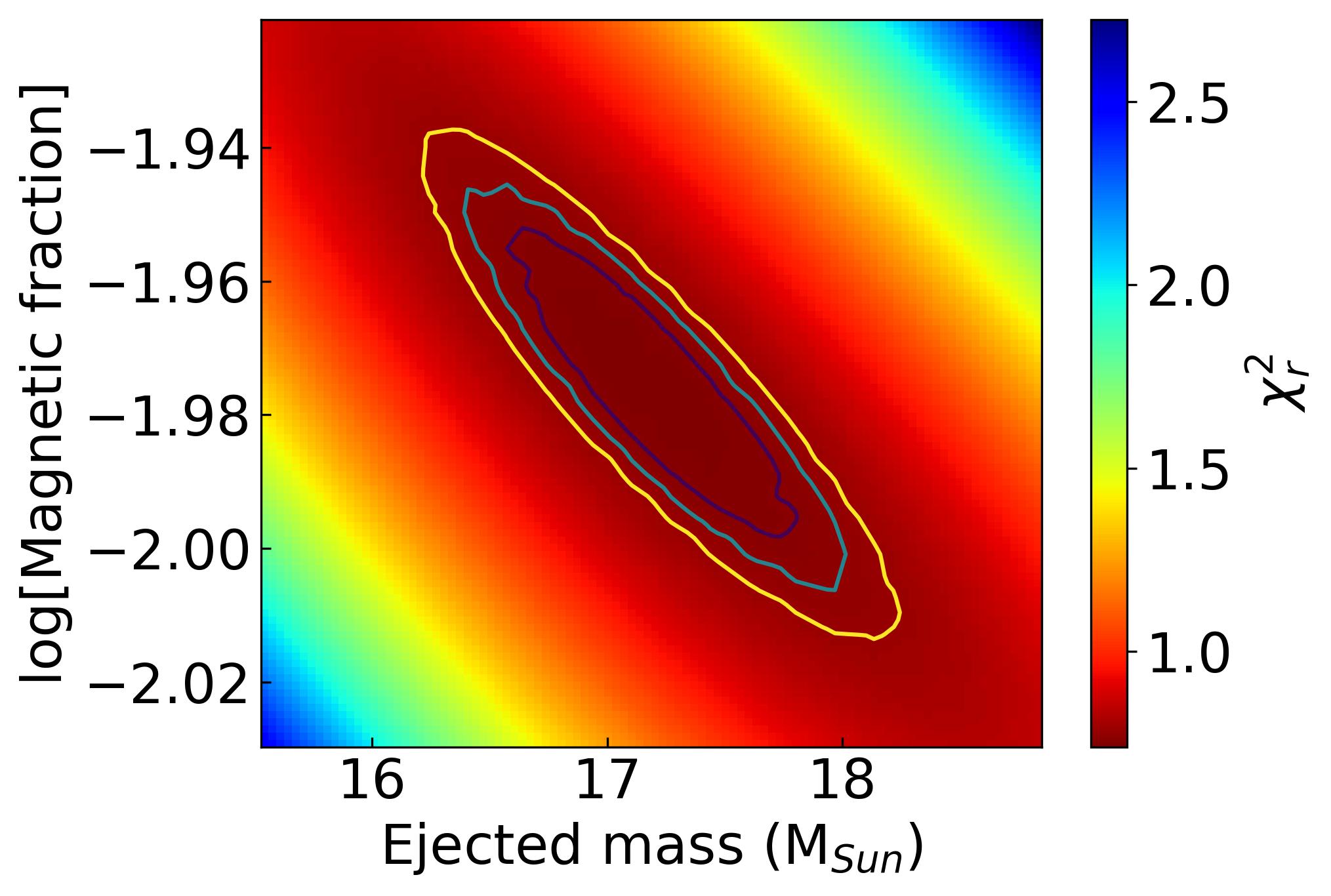}
\includegraphics[width=0.16\textwidth]{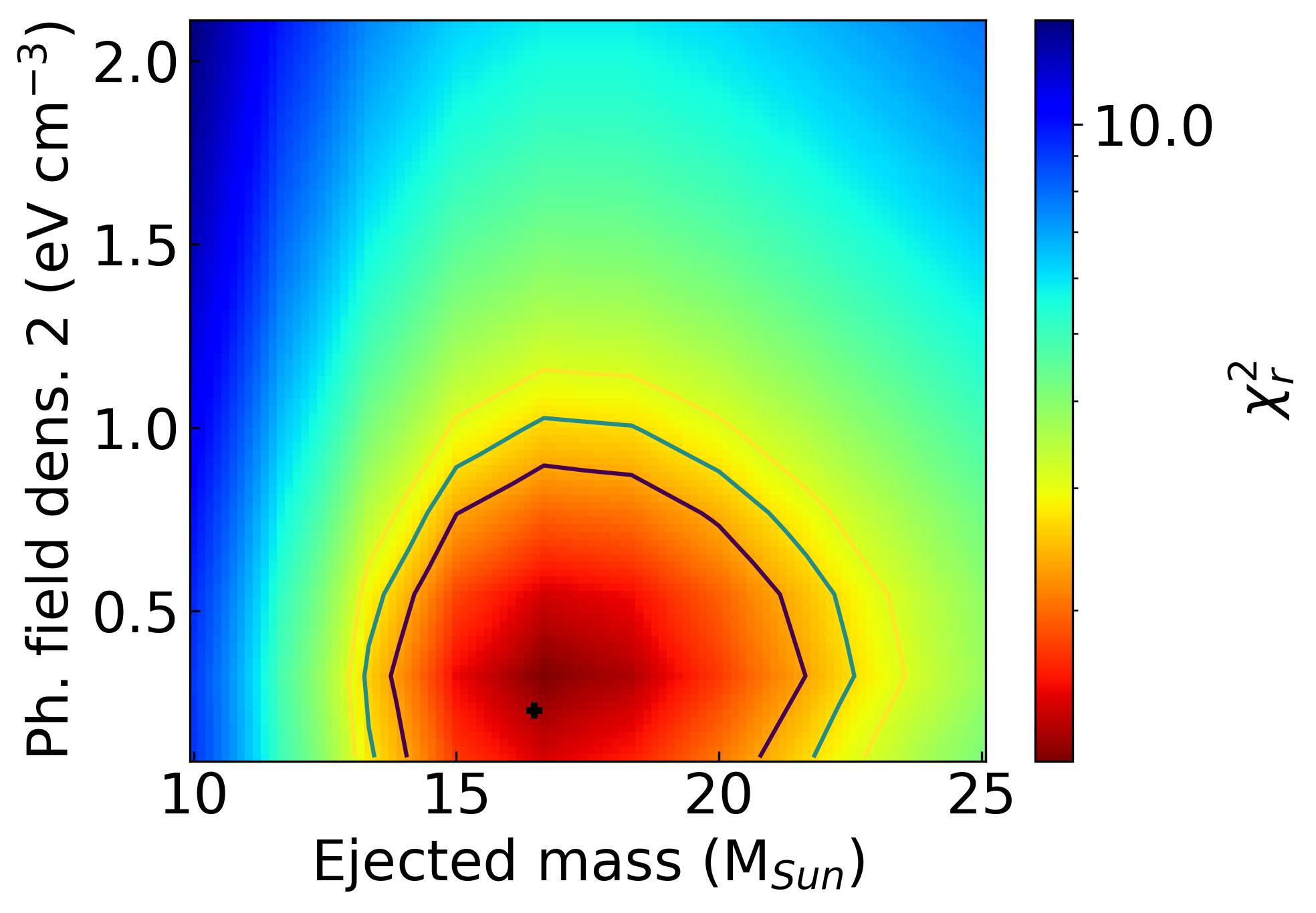}
\includegraphics[width=0.16\textwidth]{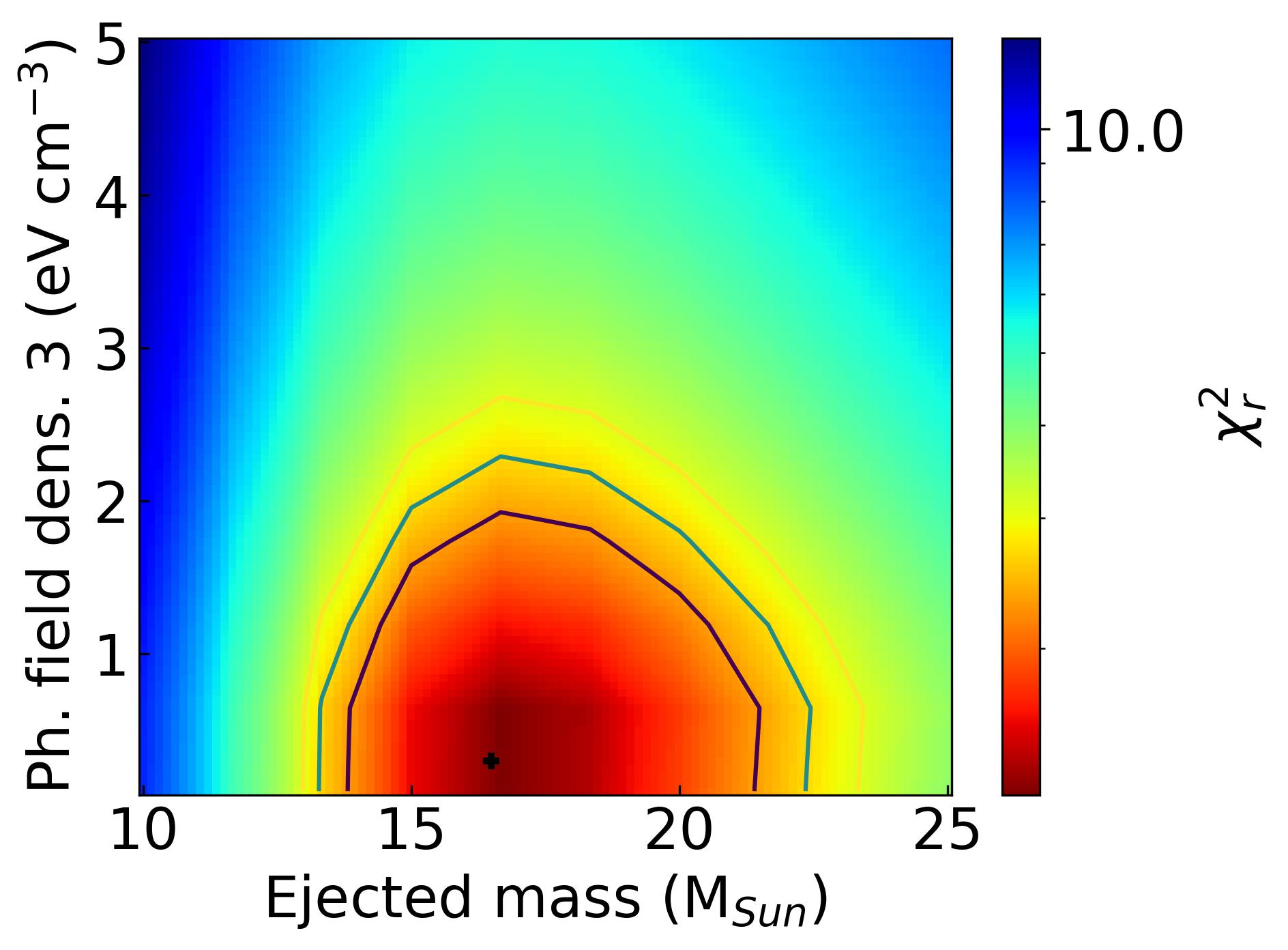}
\\
\includegraphics[width=0.16\textwidth]{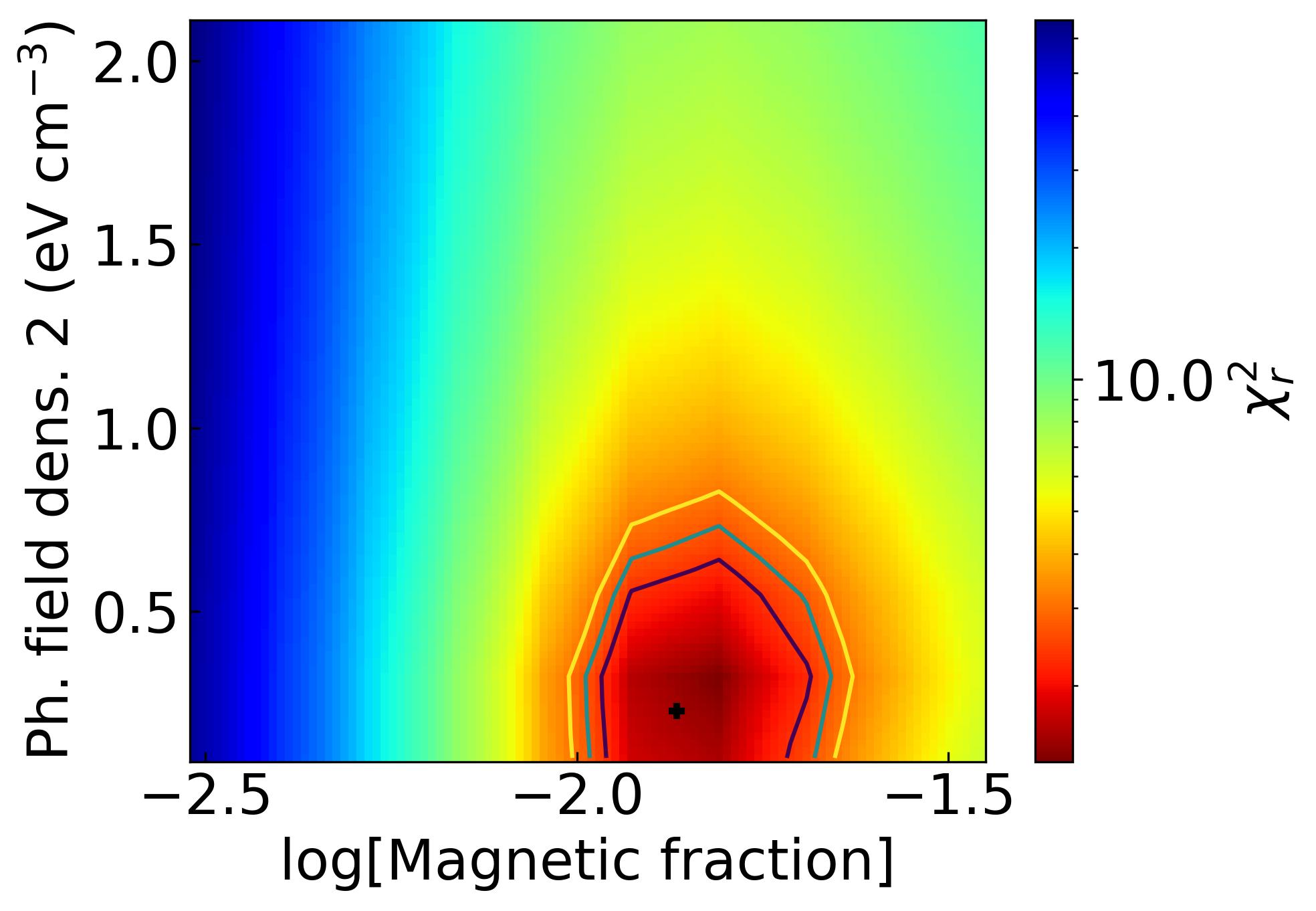}
\includegraphics[width=0.16\textwidth]{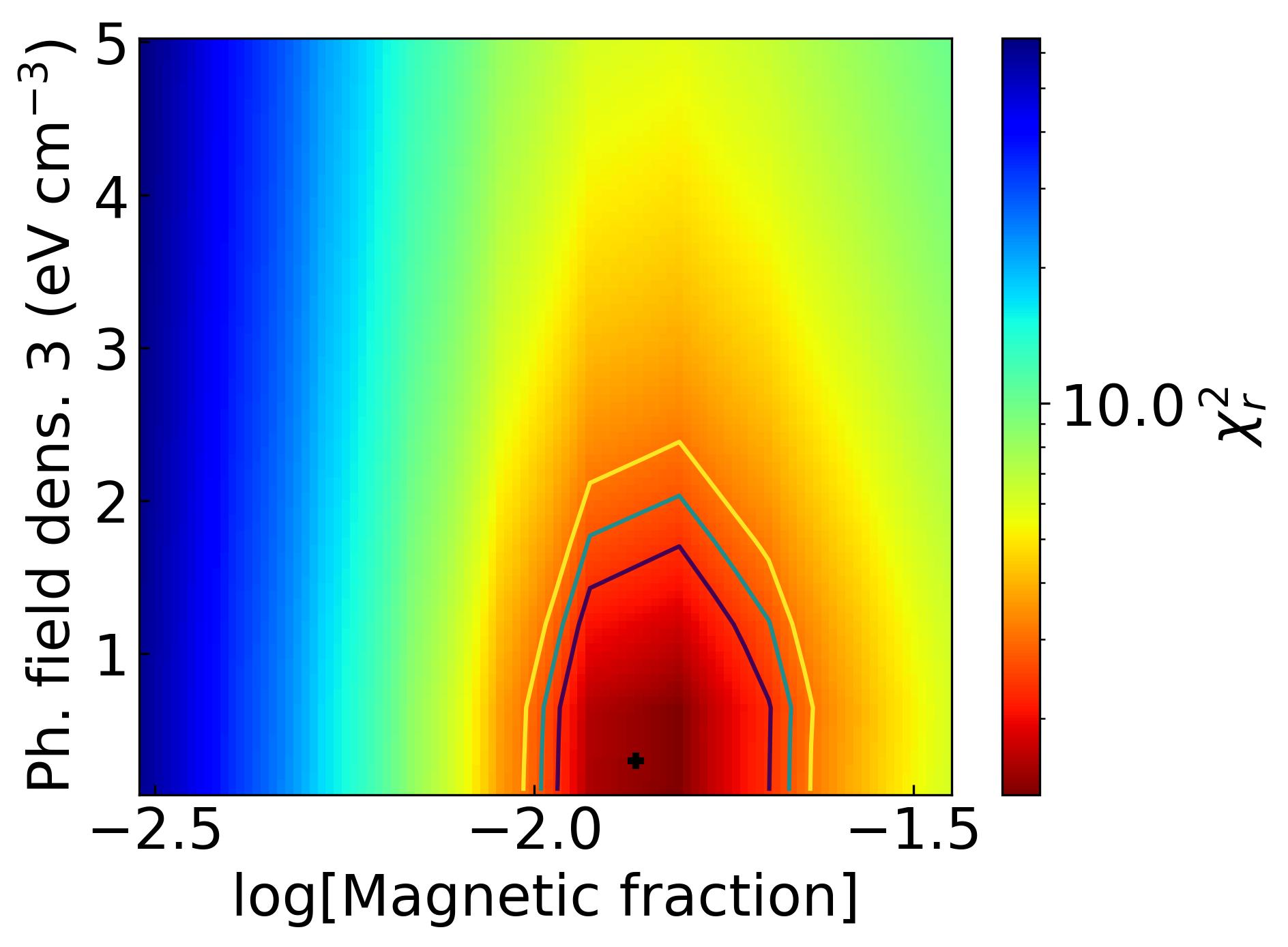}
\\
\includegraphics[width=0.16\textwidth]{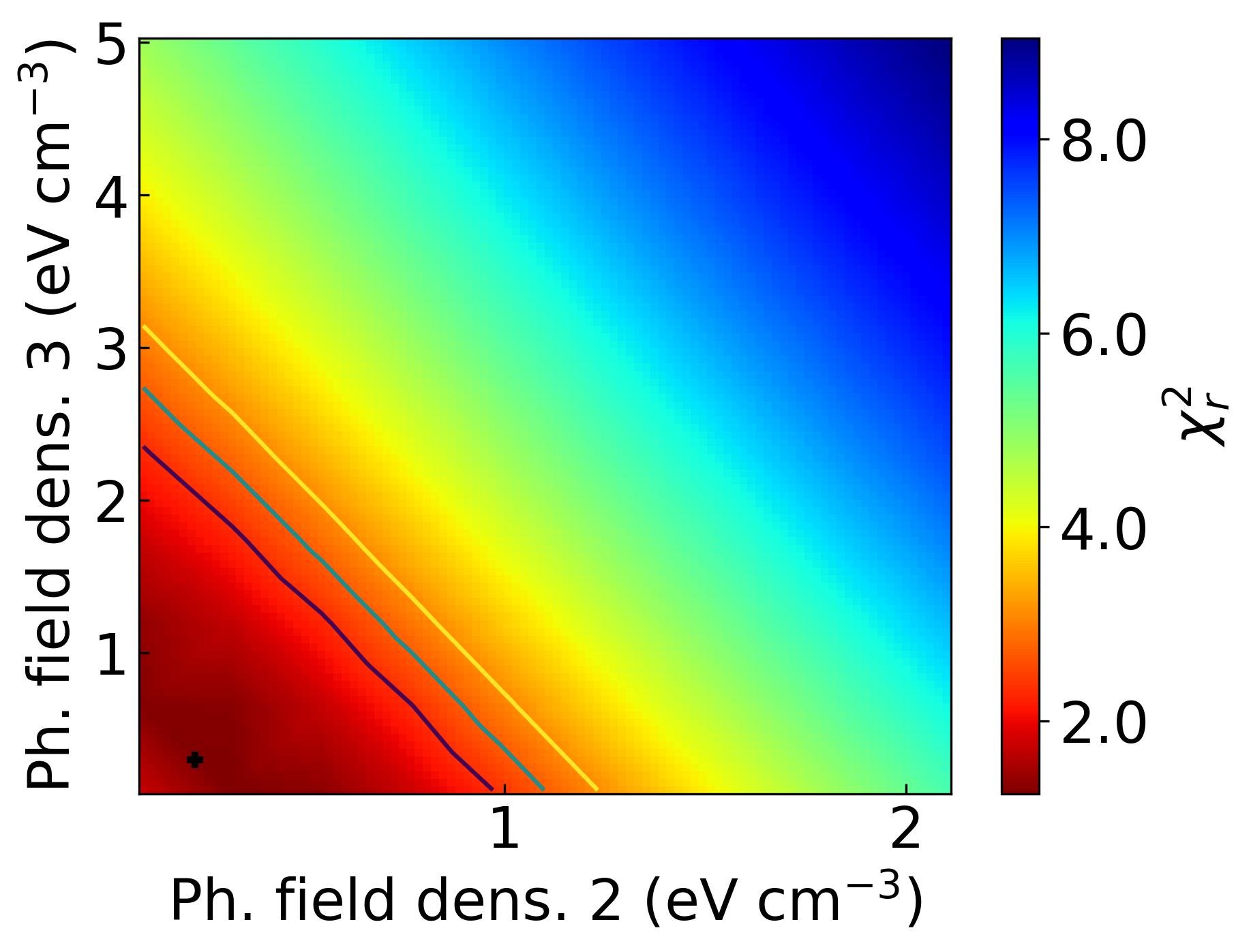}
\caption{Reduced $\chi^2$ maps for 3C 58. The contours have the same correspondence as in Figure \ref{fig:crabchi}.}
\label{fig:3c58chi}
\end{figure*}

We obtain a fit with a reduced chi-squared of 0.74 and a systematic uncertainty of 0.01, which is on the limit considered for this parameter.
In terms of the reduced $\chi^2$, the fit is a bit farther from ideal, but we understand that this may be strongly influenced by the number of data points considered in X-rays (189 of 232 in total) and radio to a lesser extent (24 of the remaining 43 taking out the X-ray points), where the fit is really good. The data outside these ranges have errors larger than an order of magnitude, then their contribution to the reduced $\chi^2$ becomes really low.
Regarding the solution convergence (see Figure \ref{fig:3c58chi}), if we compare with the solution for the Crab Nebula, the ellipsoids in the $\chi^2$ maps seem more elongated. Therefore the relative errors of the parameters are higher, but in terms of degeneracy, the ellipsoids are still closed and the solution is not degenerate.

As in the case of the Crab Nebula,  we also discount from the energy input for particles the pulsed radiation emitted ($1.8 \times 10^{-3}$, see Table~\ref{tab:fixed} and \citealt{li2018}), with no noticeable effect on the spectrum and the parameters. The modelled flux surpasses the data points at energies higher than 100 GeV. For Crab, this situation gives a chance to speculate on possible corrections to the model due to the high quality of the data, but in this case, we have less data points and their dispersion makes this difference no-significant in comparison with the variations of the model in other parts of the IC spectrum.  

Finally, regarding convergence of the solution, and as in the previous case, we have also performed $\sim$150 simulations with random sets of initial parameters converging to the final solution in more than 70\% of the cases. The initial sets of parameters for the non-convergent cases have in general values close to the limit range for the energy break $\gamma_\text{b}$ and magnetic fraction $\eta_\text{B}$. Thus, a very similar scenario and results as in the case of the Crab Nebula appears.

\subsubsection{The fit in the context of other models}

There are not many models of 3C 58 apart from \cite{tanaka2013} and 
the studies we did earlier with TIDE in the framework of different observations
\cite{torres2013,li2018}.
As expected from the outcome of Crab, the differences with the previous fit published in \cite{li2018} are also few, and are mainly explained by the improvement on the expansion model and the mathematical fitting used in the version presented in this paper, as commented before for the case of Crab.

Regarding the model of \cite{tanaka2013}, we recall that the nebula was detected one year later at gamma-rays and VHE \cite{abdo2013,ackermann2013,aleksic2014} and thus these data has not been used then. Despite of this, there is a general good agreement with our obtained slopes for the low and high energy injection power-laws (1.0 \& 3.0, respectively) and the location of the energy break in the two alternative models that they present ($5-9 \times 10^4$).
The expansion model is a ballistic expansion of the nebula where the parameter to be fitted is the expansion velocity.
In this case, the two models are indeed different, from 780 to 2000 km s$^{-1}$.
In our model, the expansion velocity at 2.5 kyr is 1056 km s$^{-1}$, which is roughly in agreement with mentioned models too, but in our case we have to take into account that it is not constant and changes with time like $\sim t^{1/5}$. 
Finally, the magnetic field obtained by us is 15.6 $\mu$G, which is close but lower than the minimum value given by \cite{tanaka2013} where the magnetic field lies between 17 and 40 $\mu$G. This small deviation in the magnetic field can be produced by the disagreement of the models on the considered radius of the PWN. While \cite{tanaka2013} consider a radius of 2 pc for 3C 58, our model predicts a slightly larger radius of 2.31, which is closer to what is observed. The magnetic fraction in \cite{tanaka2013} is also a factor 3 lower than what we obtain from the fit. 
However it is very difficult to compare both models directly since the evolution in time is computed differently.
In fact, we remark that comparing models based on different theoretical priors provides not more than a qualitative feeling of the stability of the solution with respect to the assumptions.

\begin{table*}
\begin{threeparttable}
\center
\scriptsize
\caption{Parameters obtained for the Crab Nebula and 3C 58 from TIDEfit. The parameters where a searching range is defined are those used to do the fit. The quantities inside the parenthesis are the confidence interval at 95\%.}
\label{tab:results}
\begin{tabular}{llccccc}
\hline
Parameter & Symbol & Range & \multicolumn{2}{c}{Crab Nebula} & \multicolumn{2}{c}{3C 58}\\
 & & & Age fixed & Age free & Age fixed & Age free\\
\hline
\hline
Free parameters \\
\hline
Age (yr) & $t_\text{age}$ & [300,1500 / 2000,5000] & 968 & 914 (901,931) & 2500 & 2108 (2075,2140)\\
Magnetic fraction ($\times 10^{-2}$) & $\eta_\text{B}$ & [0.01,50] & 2.048 (2.027,2.053) & 2.2 (2.1,2.4) & 1.06 (1.03,1.09) & 1.77 (1.70,1.84)\\
Ejected mass ($M_\odot$) & $M_\text{ej}$ & [7,12 / 12,25] & 7.9 (7.6,8.1) & 7.0 (7,7.2) & 17.2 (16.8,17.6) & 10.8 (10.5,11.0)\\
Energy break ($\times 10^5$) & $\gamma_{b}$ & [0.1,100] & 4.9 (4.7,5.3) & 5.8 (5.4,6.2) & 0.88 (0.86,0.90) & 0.76 (0.74,0.78)\\
Low energy index & $\alpha_{l}$ & [1,4] & 1.46 (1.44,1.48) & 1.51 (1.49,1.53) & 1.00 (1,1.03) & 1.00 (1,1.04)\\
High energy index & $\alpha_{h}$ & [1,4] & 2.470 (2.468,2.473) & 2.478 (2.476,2.481) & 3.012 (3.009,3.015) & 3.012 (3.009,3.015)\\
FIR energy density (eV cm$^{-3}$) & $w_\text{fir}$ & [0.1,5] & 0.10 (0.1,0.15) & 0.10 (0.1,0.18) & 0.22 (0.1,0.42) & 0.37 (0.1,0.66)\\
NIR energy density (eV cm$^{-3}$) & $w_\text{nir}$ & [0.1,5] & 0.1 (0.1,1.38) & 0.3 (0.1,2.4) & 0.41 (0.1,0.99) & 0.75 (0.1,1.56)\\
Systematic uncertainty & f & [0.01,0.5] & 0.106 & 0.103 & 0.010 & 0.010\\
\hline
Derived values\\
\hline
Magnetic field ($\mu$G) & $B$ & \ldots &  86.7 (84.4,89.1) & 88.5 (86.6,91.4) & 15.6 (15.3,15.9) & 18.0 (17.6,18.4)\\
Maximum energy at injection ($\times 10^9$) & $\gamma_{max}(t)$ & \ldots & 9.25 (9.21,9.26) & 9.6 (9.4,10.1) & 2.70 (2.66,2.74) & 3.51 (3.44,3.58)\\
PWN Radius (pc) & $R_\text{pwn}$ & \ldots & 1.82 (1.79,1.85) & 1.78 (1.73,1.83) & 2.31 (2.28,2.33) & 2.30 (2.25,2.35)\\
\hline 
Statistics\\
\hline
Degrees of freedom\tnote{$^\dagger$} & DOF & \ldots & 580 & 579 & 456 & 455\\
Reduced chi-squared & $\chi^2_\text{r}$ & \ldots & 1.013 & 0.987 & 0.740 & 0.718\\
\hline
\end{tabular}
\begin{tablenotes}
\item [$\dagger$] Age free models have one more free parameter, but also one more data point, the spin-down luminosity. It is necessary to fix correctly the initial spin-down luminosity and the initial spin-down age.
\end{tablenotes}
\end{threeparttable}
\end{table*}

\section{Using detailed observations and PWN models to determine the system's age}

The quantity of data available and its completeness in the entire electromagnetic spectrum in both cases allow us to explore using 
TIDEfit to estimate the system's age directly from the fit, adding it to the rest of the free parameters.
In Table \ref{tab:results}, we show the results obtained for both nebulae when such approach is adopted.
In terms of reduced $\chi^2$, the quality of the fits are quite similar when comparing with those where the age is fixed, and the age obtained is very close to the expected.
In the case of Crab, the true age falls inside the 2$\sigma$ range obtained from the model.
For 3C 58, for which a real age is not known, we find an age of 2059 yr, which implies a $\sim$440 yr younger PWN than assumed in the previous model.
The change in the age has concurrently changed the values of the magnetic fraction, the ejected mass, and the FIR and NIR energy densities.
The rest of parameters remain vary similar, included derived parameters as the magnetic field, the radius and the maximum energy at injection.
A difference of only 440 years is in fact already pointing towards the larger values in the age dichotomoy stated earlier (where the 
discussion was between 871 and 2500 years).
This is the first time that a radiative code is used to fit the age of PWNe.
Of course, it is crucial that a complete set of data, from radio to VHE exists,  to avoid 
possible degeneracies when the age is fitted.

\section{Concluding remarks}
\label{sec:conc}

\begin{itemize}

\item We have considered how to take into account that the multi-wavelength data for a generic PWN come from different
detectors, and from different observation times (sometimes separated decades
apart). Using the prescriptions given in \citet{hogg2010}, we proposed that a cross-calibration
uncertainty must be taken into account when fitting models, and posed that introducing an extra parameter to deal with it is the best option. 

\item We have successfully incorporated a fitting technique into a fully-time dependent numerical model for PWN. 
The appendix~\ref{sec:appa} accounts for the physical components of the model.
They are treated in much the same way as in earlier incarnations of TIDE, prior to automatic fitting,  
including some improvements like those related to the dynamics, or the accounting of the energy reservoir spent in non-PWN related emission, and others mentioned. 
The difference with former versions of the code, and in general with other codes, is 
that all of these components are now included into an automatic fitting technique. This is a result of an improved single-model computational cost (see Appendix~\ref{sec:imple})
as well as the introduction of a relatively fast algorithm technique that avoids the use of derivatives. 
The algorithm introduced in the code TIDEfit is based on the Nelder-Mead search for solutions, which was also used recently in 
the search of best models for magnetospheric emission from pulsars (see \cite{Iniguez2022}).
TIDEfit is able to cope with variations and minimization of a significant number of parameters (see Table \ref{tab:free-para}).

\item Our model maximizes the likelihood  $\chi^2$ function to search for a solution  instead of minimizing the typical expression for $\chi^2$ function. This is done to take into account systematic uncertainties that appear due to the diverse origin of the data, which is generally ignored in the literature regarding radiative models for PWNe. Our model fits the multi-frequency data once the free parameters of the model
are selected. This is done in a modular way, such that a different number of parameters (and which are those) can be chosen according to the problem at hand
and the number and quality of the data points at our disposal. 

\item This represents an improvement over other PWNe models, where the `fits' are in fact qualitatively obtained without a quality assessment. 
In addition, it allows us to tackle whether there is any significant degeneracy where very different sets of parameters would lead to equally good fits.
We performed convergence tests with random sets of initial conditions for the cases studied in this paper (Crab Nebula and 3C 58). The model converges to the final solution in more than 70\% of the cases for both nebulae independently of the initial set of parameters.
When not, parameters selected are on the verge of the search ranges and the fits are non-compliant.
There is no significant degeneracy since we find that neither Crab nor 3C58 data can be encompasses by any other set of parameters leading to an equally good solution.

\item Another application we introduced was the determination of the system's age by assuming it as a free parameter of the model. 
We find that when a significant dataset is considered, the age can be fixed from the spectrum at a relatively high accuracy, for instance, $<50$ years in the case of Crab.
We have used this possibility to actually provide a new estimation of the age of 3C58, for which a discussion is yet ongoing. We obtain an estimation 
of its age at $2108_{-33}^{+32}$ years, within 95\% confidence level.

\end{itemize}

\section*{Data Availability}

No new observational data is herein presented. The dataset used in all figures is publicly available, and references are given correspondingly.
Any additional theoretical detail required is available from the authors on reasonable request.

\section*{Acknowledgements}

We thank 
Elena Amato,
Rino Bandiera,
Niccolo Bucciantini,
Agnibha de Sarkar, 
Barbara Olmi,
and
Wei Zhang
for comments, discussions, encouragement, and tests. 
This work has been supported by the grants INAF grants MAINSTREAM 2018, SKA-CTA, PRIN-INAF 2019, by ASI-INAF n.2017-14-H.O, and the Spanish grant
PID2021-124581OB-I00 and  
by the program Unidad de Excelencia María de
Maeztu CEX2020-001058-M.

\bibliographystyle{elsarticle-harv} 
  \bibliography{references-jheap.bib}

 \appendix
 \section*{Appendix A: Physical components incorporated in the automatic fitting tool }
\label{sec:appa}

Here, we briefly described the physical components incorporated in the fitting model.

\subsection*{Injection}

The last term in the right-hand side of Eq. (\ref{eq:difflos}), $Q(\gamma,t)$, represents the injection of particles into the PWN.
Whereas TIDE allows for any functional dependency to be incorporated easily (see Appendix \ref{sec:imple}), the default injection is described as a broken power-law whose normalization is provided by the pulsar power
\begin{equation}
\label{injection}
Q(\gamma,t)=Q_0(t)\left \{
\begin{array}{ll}
\left(\frac{\gamma}{\gamma_b} \right)^{-\alpha_l}  & \text{for \;\; }\gamma \le \gamma_b,\\
 \left(\frac{\gamma}{\gamma_b} \right)^{-\alpha_h} \exp(-\gamma / \gamma_\text{max}) & \text{for \;\; }\gamma > \gamma_b ,
\end{array}  \right .
\end{equation}
where $\gamma_b$ is the break energy, $\alpha_l$  and $\alpha_h$ are the spectral indices and $\gamma_\text{max}$ is the exponential cut-off, which is calculated by one or the most restrictive criteria explained in Section \ref{sec:maxene}.
It is interesting to note that including an exponential cut-off into the injection function, but there is no significant difference in the parameters obtained in the fits, since the exponential decay at high energies is generated in a natural way when the energy losses are taken into account during the evolution of the pair spectrum.
The normalization constant $Q_0(t)$ is determined via the pulsar power $L(t)$, as \cite{gelfand2009,Bucciantini2011,martin2012}
\begin{equation}
\eta_\text{p} L(t)=\int_{\gamma_{min}}^{\gamma_{max}} \gamma m_e c^2 Q(\gamma,t) \mathrm{d}\gamma.
\label{eqeta}
\end{equation}
We recall that the spin-down power is the energy reservoir for the whole multi-messenger phenomenology of a PSR/PWN system.
The total spin down power is thus considered to be divided into particle injection ($\eta_\text{p}$), magnetic field powering ($\eta_\text{B}$) and multi-frequency/multi-messenger emission ($\eta_\text{other}$) produced elsewhere from the PWN (e.g., radiation from the magnetosphere, which usually amounts to a few percent of the pulsar power). Thus, the following condition holds
\begin{equation}
\eta_\text{p} + \eta_\text{B} + \eta_\text{other} = 1.
\end{equation}
For the spin-down power $L(t)$, we have
\begin{equation}
\label{edot}
L(t)=4\pi^2 I \frac{\dot{P}}{P^3}
=L_0 \left(1+\frac{t}{\tau_0} \right)^{-\frac{n+1}{n-1}}
\end{equation}
where $P$ and $\dot{P}$ are the observational-determined period and period derivative, $I$ is the pulsar moment of inertia assumed to be $I \sim 10^{45}$ g cm$^2$, $n$ is the breaking index, $\tau_0$ is the initial spin-down timescale, and $L_0$ is the initial luminosity.
If enough information is available, we can derive the parameters $L_0$ and $\tau_0$. Adopting the magnetic dipole model for pulsars, the initial spin-down luminosity $L_0$ is given by
\begin{equation}
L_0 = \frac{2 \pi^2 I}{\tau_c P_0^2} \left(\frac{P}{P_0} \right)^{n-1}; \hspace{0.5cm} 
{\rm with} \hspace{0,5cm} \tau_c=\frac{P}{2\dot{P}}.
\end{equation}
and the initial spin-down timescale of the pulsar
\begin{equation}
\label{spindownage}
\tau_0=\frac{P_0}{(n-1)\dot{P}_0}=\frac{2\tau_c}{n-1}-t_{age}
\end{equation}
where $P_0$ and $\dot{P}_0$ are the initial period and its first derivative and $\tau_c$ is the characteristic age of the pulsar.

\subsubsection*{Burst injection}

The model can also encompass the simulation additional injection of particles due to bursting phenomena of the central engine during a certain amount of time (which can be defined by the user), with a given analytical spectrum and energy.
Such bursts can be treated differently, e.g., we can define a burst magnetic fraction to dedicate part of the injected burst energy to power the magnetic field of the PWN.
These new features has been preliminarily used to explore the possible effects of magnetar bursts in PWNe \citep{martin2020} and in particular for the case of PWN J1119-6127  \cite{blumer2021}.

\subsection*{PWN dynamical evolution:  SNR shock positions}
\label{sec:shockpos}

TIDE integrates the Euler equations as described in Section 2 of \cite{bandiera2020}, which provides an accurate evolution for the radius during the free expansion phase.
Such equations depend on the SNR ejecta density, velocity and pressure profiles.
Their value change depending on the location of the PWN, in particular, if the location of the PWN shell is inside unshocked or shocked ejecta.
Taking this into account, we need to know at every time the position of the SNR forward and reverse shocks. We use the updated trajectories computed in \cite{bandiera2021}. This work improves the accuracy of the trajectories published in \cite{truelove1999} in a 10\% or more depending on the initial density profile.
However, as discussed in \cite{bandiera2021}, this percentage is relevant.
An accurate calculation of these trajectories is necessary to obtain a precise estimation of the time of the beginning of the reverberation, since it characterizes the amount of compression which is especially sensitive for PWNe powered by low spin-down luminosity pulsars \citep{bandiera2020}.
TIDE already incorporates the equations given in \citealt{bandiera2021} as a default.
The initial profiles assumed for the SNR are $P_\text{ej}(r,t) = 0$ for the pressure, $v_\text{ej}(r,t) = r / t$ for the velocity and, for the ejecta density,
\begin{equation}
\rho_\text{ej} = \left \{
\begin{array}{ll}
A (v_\text{t} / r)^\delta / t^{3 - \delta}  & \text{for }r < v_\text{t} t,\\
A (v_\text{t} / r)^\omega / t^{\omega - 3} & \text{for }v_\text{t} t \le r < R_\text{snr} ,
\end{array}  \right .
\end{equation}
As an example for the most usual case, we show the trajectories for $\delta = 0$ below, which are the ones we are using in our calculations in this paper. TIDE can cope with any value of $\delta$. For a complete description, we refer to the original paper. The parameterized formula for the reverse shock is
\begin{multline}
R_\text{rs} = \frac{x^{1.5548} (1 - x)^{0.6824}}{0.01964 + 0.5095 x + 0.1871 x^2} \times \\
\left(1 + \frac{0.02171 \{1 / [0.3338 (\omega - 5)] - 1\}}{1 + 1 / [0.3338 (\omega - 5)]^{2.778}} \right),
\end{multline}
where $\omega$ is the SNR envelope density index, which has to be strictly greater than 5 and $x = t / t_\text{implo}$ being
\begin{eqnarray}
&& t_\text{implo} = 2.399 + \nonumber \\
&& \sqrt{\left(\frac{0.1006}{\omega - 5} \right)^2 + \left(\frac{-0.06494 + 0.7063 / (\omega - 5)}{1 + (\omega - 5)^2} \right)^2}.
\end{eqnarray}
For the forward shock, the formula is simply
\begin{equation}
R_\text{fs} = \frac{1.15169 (t + 1.94)^{2 / 5}}{1 + 0.672 / t + 0.00373 t^2}.
\end{equation}

\subsection*{PWN dynamical evolution: SNR ejecta profiles}

The reverse shock divides two regions in the SNR.
From the reverse shock inwards, we find the unshocked medium where the pressure, velocity and density profiles are the original (or initial) ones of the SNR.
The shape of them has been described in Section \ref{sec:shockpos}.
From the reverse shock outwards, there is the shocked medium. Here the profiles change completely and depending on the total energy injected by the pulsar ($L_0 \tau_0$), we find multiple shock structures which add such a level of complexity that makes impossible or really difficult to build an analytical prescription.
Thus, at the beginning of the reverberation phase more care should be put.
\cite{bandiera2020} is the first of a series of papers that study in depth the evolution of PWNe during the reverberation phase.
\cite{bandiera2022} describes the internal physics involved and give some prescriptions to solve the PWNe evolution at least until the first compression of the nebula.
However, these prescriptions are for the moment only valid for non-radiative situations. 
Future works will adopt a new numerical approach in order to be able to trace the ejecta pressure 
case where synchrotron and other losses are at play too.

Currently, the adopted scheme to solve the dynamic equations is similar to \cite{gelfand2009}.
but taken into account that during the compression there is no new material accreted onto the shell \cite{bandiera2020}. 
Then,
\begin{equation}
M(t) \frac{d v(t)}{d t} = 4 \pi R^2(t) \left[P_\text{pwn}(t) - P_\text{ej}(R,t) \right],
\end{equation}

The current pressure, density and velocity profiles for the SNR shocked and unshocked ejecta are thus still the ones provided in \cite{bandiera1984} and \cite{blondin2001}. We emphasize  that this is in fact a caveat. The main point of note is that 
the outer pressure  is not equal the Sedov solution (or a constant fraction of it) along the entire evolution. In fact it is not constant at all -as assumed in all PWN models so far, and
the thin-shell assumption itself fails along the process, being no longer thin in comparison with the size of the PWN, see \cite{bandiera2022} for extensive discussion of these points.
Whereas incorporating such a detailed physics is not part of the current effort, it will be done in future versions of the code.

\subsection*{Magnetic field}

A fraction of the rotational energy powers the magnetic field at each instant of the evolution, $\eta_\text{B} L(t)$, which is subject to adiabatic losses. Numerically, it is obtained by solving 
\begin{equation}
\frac{d W_B(t)}{dt}=\eta_\text{B} L(t)-\frac{W_B(t)}{R(t)} \frac{d R(t)}{dt},
\end{equation}
where $W_B=B^2 R^3/6$ is the magnetic energy. The solution is 
\begin{equation}
\label{bfield}
B(t)=\frac{1}{R^2(t)}\sqrt{6 \eta_\text{B} \int_0^t L(t') R(t') \mathrm{d}t'}
\end{equation}
where $\eta_\text{B}$ is taken out of the integral only if it is assumed as a 
constant along the whole evolution, what can be relaxed if there are reasons to.

\subsection*{Maximal energy}
\label{sec:maxene}

\begin{figure}
\includegraphics[width=0.5\textwidth]{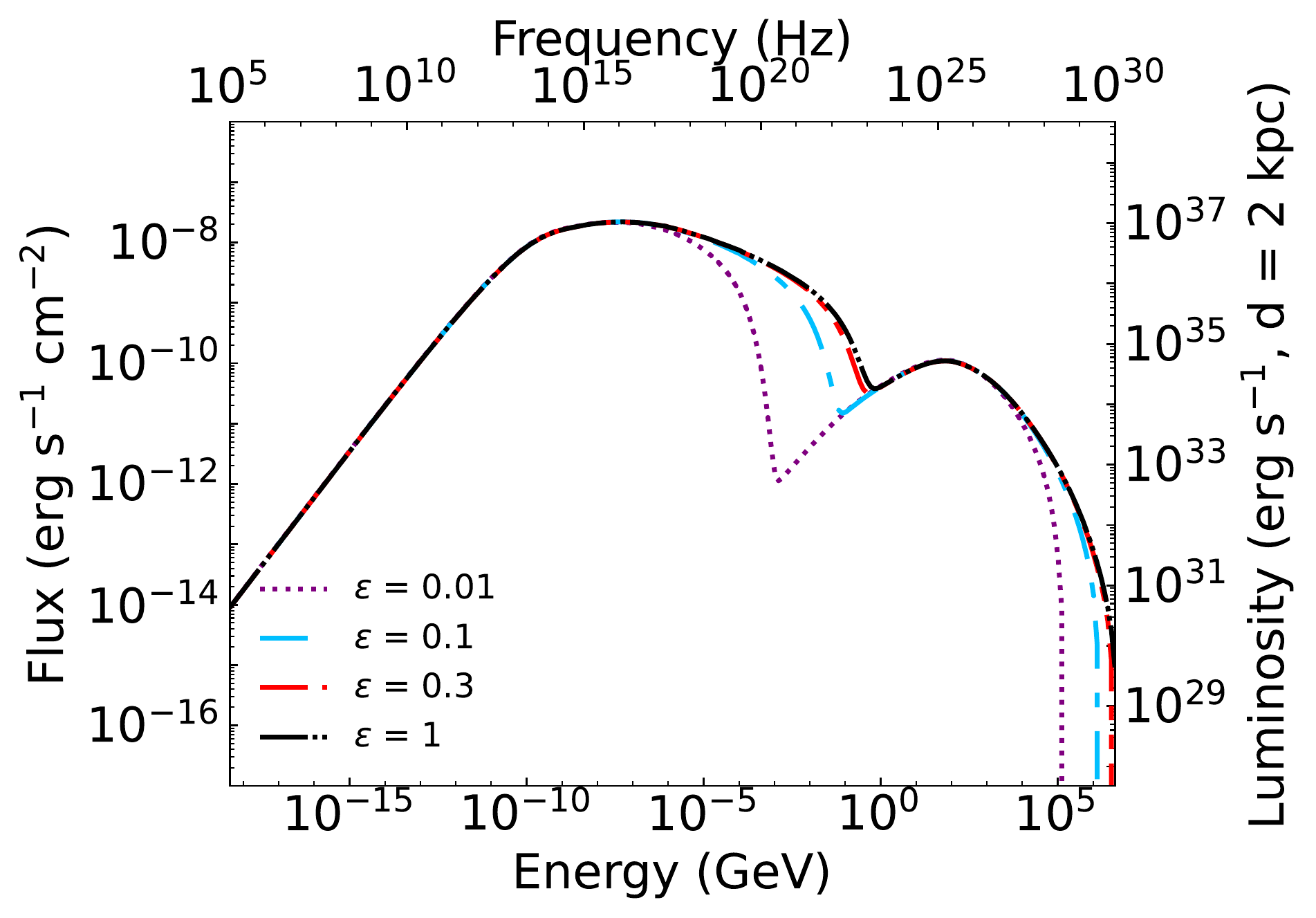}
\caption{Variation of the Crab Nebula spectrum with the containment factor.}
\label{fig:containment}
\end{figure}

The particles' energy is either limited by the balance between the energy lost by particles (mostly due to synchrotron radiation) and the energy gain  \citep{dejager2009}
\begin{equation}
\label{gsync}
\gamma^{sync}_{max}(t)=\frac{3 m_e c^2}{4 e} \sqrt{\frac{\pi}{e B(t)}},
\end{equation}
where $e$ is the electron charge and $B$ the mean magnetic field of the PWN; or by the particles necessary confinement inside the termination shock for acceleration to proceed \cite{dejager2009},
\begin{equation}
\label{eq:gyro}
\gamma^\text{gyro}_\text{max}(t)=\frac{\epsilon e \kappa}{m_e c^2} \sqrt{\frac{\eta_\text{B} L(t)}{c}},
\end{equation}
here $\kappa$ is the magnetic compression ratio (we assume it as 3) and $\epsilon$ is the containment fraction ($\epsilon < 1$),
which is the ratio between the Larmor radius of the pairs and the termination shock radius.
TIDE uses the most restrictive criteria of the two and it is specified in the output at each time-step.
The latter approach is equivalent at the one shown in \citet{hillas1984,venter2007,deona2022} --fixing $\epsilon = 0.5$ in the latter reference.
From \citealt{hillas1984}, it is shown that to confine particles inside an acceleration region with size $L$, the Larmor radius $r_\text{L}$ must accomplish that (see also \cite{deona2022})
\begin{equation}
L > 2 r_\text{L}.
\end{equation}
From the definition of the Larmor radius,  $\gamma_\text{max} \propto e B_\text{TS} r_\text{L} / (m_\text{e} c^2)$. 
\citet{deona2022} calculate the magnetic field at the termination shock using the magnetic field density
\begin{equation}
\frac{B_\text{TS}^2}{8 \pi} = \frac{\eta_\text{B} L(t)}{4 \pi c R_\text{TS}^2} \quad \implies \quad B_\text{TS} = \frac{1}{R_\text{TS}} \sqrt{\frac{2 \eta_\text{B} L(t)}{c}}.
\end{equation}
This is just a different way to write the acceleration conditions. 
Substituting in the expression of $\gamma_\text{max}$, we get
\begin{equation}
\gamma_\text{max}(t)=\frac{\sqrt{2} e \eta_\text{e} }{m_e c^2} \sqrt{\frac{\eta_\text{B} L(t)}{c}},
\end{equation}
where $\eta_\text{e}$ is a proportionality constant that must be lower than 1. Note that the physical dependencies are exactly the same as in Equation \ref{eq:gyro} and that the factor $\varepsilon \kappa$ is their $\sqrt{2} \eta_\text{e}$.
Figure \ref{fig:containment} shows the effect of varying the containment factor in the spectrum of the Crab Nebula.
It changes the maximum energy at injection, thus the synchrotron and IC decay, without changing the magnetic field. 

%

\subsection*{Radiative losses}

The first term in the right-hand side of Eq. \ref{eq:difflos}
accounts for changes on the distribution function due to energy losses, $\dot {\gamma}$. 
The radiative energy losses in $\dot{\gamma}$ include the main radiation mechanisms (see e.g., \cite{Blumenthal1970}).

Synchrotron losses are described by
\begin{equation}
 \label{synclosses}
\dot{\gamma}_{syn}(\gamma,t)=-\frac{4}{3}\frac{\sigma_T}{m_e c}U_B(t)\gamma^2 ,
\end{equation}
where $\sigma_T=(8\pi/3)r^2_0$ is the Thomson cross section, $r_0$ is the electron classical radius, and $U_B(t)=B^2(t)/8\pi$ is the energy density of the magnetic field. 

Inverse Compton losses are described with 
the Klein-Nishina cross section,
\begin{multline}
\label{iclosses}
\dot{\gamma}_{IC}(\gamma)=-\frac{3}{4}\frac{\sigma_T h}{m_e c}\frac{1}{\gamma^2}\int_0^\infty \nu_f \mathrm{d}\nu_f \\ \times
\int_0^\infty \frac{n(\nu_i)}{\nu_i} f(q,\Gamma_{\varepsilon}) \Theta(1-q) \Theta \left(q-\frac{1}{4\gamma^2} \right) \mathrm{d}\nu_i,
\end{multline}
where $h$ is the Planck constant, $\nu_{i,f}$  are the initial and final frequencies of the scattered photons, $\Theta$ is the Heaviside step function, $n$ is
the photon background distribution and 
\begin{equation}
\label{f}
f(q,\Gamma_{\varepsilon})=2q\ln q+(1+2q)(1-q)+\frac{1}{2}\frac{(\Gamma_{\varepsilon}q)^2}{1+\Gamma_{\varepsilon}q},
\end{equation}
with $
\Gamma_{\varepsilon}={4 \gamma h \nu_i}/{m_e c^2},
$
and $
q={h \nu_f} /({\Gamma_{\varepsilon}(\gamma m_e c^2-h \nu_f)}).
$
When $\Gamma_{\varepsilon} \ll 1$ ($\gg 1$), the scatter happens in the Thomson limit (extreme Klein-Nishina limit).

The electron-atom bremsstrahlung (interaction of electrons with the electromagnetic field produced
by the ionized nuclei of the ISM), which is usually sub-dominant at high energies, is described as in \citep{Haug2004}.

\subsection*{Adiabatic losses/gains}

Adiabatic losses (or gains, when contracting) due to the expansion of the nebula are also taken into account in
the first term in the right-hand side of Eq. (\ref{eq:difflos}):
\begin{equation}
\dot{\gamma}_\text{ad}(\gamma,t)=- ({v_{PWN}(t)}/{R_{PWN}(t)})\gamma.
\end{equation}
The latter equation requires knowledge of the dynamical state of the PWN.

\subsection*{Escape and diffusion}

The second term term in the right-hand of Eq. (\ref{eq:difflos}) accounts for the lost particles due to escaping processes with a characteristic timescale $\tau(\gamma,t)$.
TIDE considers Bohm diffusion timescale as default, as given by (see e.g., \citep{vorster2013})
\begin{equation}
\tau_\text{Bohm} = \frac{e B(t) R_\text{pwn}^2}{2 \gamma m_e c^3}.
\end{equation}
The latter equation is the default configuration of the code regarding the escape timescales considered, but the power indices over the energy (e.g., Kolmogorov diffusion index) and/or the magnetic field van be changed easily by the user, taking into account that the normalization and the fundamental constants defining the timescale will vary.

\subsection*{Luminosities}

The final spectrum is computed by summing up the contributions (see \cite{Blumenthal1970})
generated by the particle population 
$N(\gamma,t)$, as calculated from the diffusion-loss equation.
\begin{itemize}
    \item 
The synchrotron luminosity is 
\begin{equation}
\label{synclum}
L_{syn}(\nu,t)=\int_{0}^{\infty} N(\gamma,t)P_{syn}(\nu,\gamma,B(t)) \mathrm{d}\gamma,
\end{equation}
where $P_{syn}(\nu,\gamma,B(t))$ is the power emitted by one electron spiraling in a magnetic field
 \begin{equation}
P_{syn}(\nu,\gamma,B(t))=\frac{\sqrt{3}e^3 B(t)}{m_e c^2}F \left(\frac{\nu}{\nu_c(\gamma,B(t))} \right),
\end{equation}
where $\nu_c$ is the critical frequency, $F(x)=x \int_x^\infty K_{5/3}(y) \mathrm{d}y$, and  $K_{5/3}(y)$ is the modified Bessel function of order $5/3$. 

\item The inverse Compton luminosity is
\begin{equation}
\label{iclum}
L_{IC}(\nu,t)=\frac{3}{4}\sigma_T c h \nu \int_0^\infty \frac{N(\gamma,t)}{\gamma^2} \mathrm{d}\gamma \int_0^\infty \frac{n(\nu_i)}{\nu_i}f(q,\Gamma_{\varepsilon}) \mathrm{d}\nu_i.
\end{equation}
where $\nu_i$ is the initial frequency of the scattered photon, $\nu$ is the final frequency of the photon after scattering, and $n(\nu_i)$ is the photon target field distribution.

Apart from the CMB (temperature of 2.73 K and energy density of 0.25 eV cm$^{-3}$) that is considered always, the photon target can be obtained (in the form of tables) directly from e.g., GALPROP\footnote{\url{https://galprop.stanford.edu/webrun/}} for which TIDE provides an extraction tool. However, it was noticed already that localized environments can have photon densities that significantly deviate from these estimation. To tackle this, we usually consider two photon densities at FIR and NIR energies 
(with typical temperatures of 20-100 and 1000-5000 K) in the computation.
Whereas both the temperature and the energy densities can be taken as free parameters,
the temperature range does not provide significant changes in the output and is usually fixed.
In order to fix the temperatures for FIR and NIR, one possibility --offered by TIDE via a special routine-- is to fit a black body distribution to the spectrum obtained from GALPROP, where the temperature is a free parameter and the black body distribution is normalized to the energy density of the integrated spectrum.
In the case of the FIR, it could be possible to obtain information about it by observing the forming filaments around the PWN shell, but this is really only possible in very high resolution observations (like they do, for instance, in \citealt{meyer2010} for the Crab Nebula).

\item 
The bremsstrahlung luminosity is computed as 
\begin{multline}
\label{bremslum}
L_{Brems}(\nu,t)=\frac{3}{2 \pi} \alpha \sigma_T h c S \int_0^\infty \frac{N(\gamma_i)}{\gamma_i^2} \\
\times \left(\gamma_i^2+\gamma_f^2-\frac{2}{3} \gamma_i \gamma_f \right) \left(\ln \frac{2 \gamma_i \gamma_f m c^2}{h \nu}-\frac{1}{2} \right) \mathrm{d} \gamma_i,
\end{multline}
where $S$ is given in \cite{Haug2004,martin2012}
and $\gamma_{i,f}$ are the  initial and final Lorentz factors of the electron.
The kinematic condition $\gamma_i-\gamma_f=h \nu / m_e c^2$ fix the final energy of the electron in the integral given an initial energy $\gamma_i$ and the energy
of the photon produced $h \nu$.

\item Self-synchrotron luminosity is computed using Eq. (\ref{iclum}) with the synchrotron photon density (as in \cite{Atoyan1989,Atoyan1996})
\begin{equation}
\label{nssc}
n_{SSC}(\nu,R_{syn}(t),t)=\frac{L_{syn}(\nu,t)}{4\pi R_{syn}^2(t) c} \frac{\bar{U}}{h \nu}
\end{equation}
where $R_{syn}/R_{PWN}=1$, $R_{syn}$ is the radius of the volume where the synchrotron radiation is assumed to be produced,
and 
\begin{eqnarray}
\bar{U}&=&\frac{\int_0^\frac{R_{syn}(t)}{R_{PWN}(t)} x^2 U(x) \mathrm{d}x}{\int_0^\frac{R_{syn}(t)}{R_{PWN}(t)} x^2 \mathrm{d}x} \nonumber \\ 
&=& 3 \frac{R_{PWN}^3(t)}{R_{syn}^3(t)} \int_0^\frac{R_{syn}(t)}{R_{PWN}(t)} x^2 U(x) \mathrm{d}x,
\end{eqnarray}
where 
\begin{equation}
U(x)=\frac{3}{2} \int_0^\frac{R_{syn}(t)}{R_{PWN}(t)} \frac{y}{x} \ln \frac{x+y}{|x-y|} \mathrm{d}y.
\end{equation}

\end{itemize}

\subsection*{Polarization}

 Inspired by previous works in polarization determination (e.g., \citealt{bucciantini2017}), TIDE also offers a tool to calculate the degree of circular polarization using the approximated formula given in \cite{linden2015} based on the derivation of \cite{wilson1997}
\begin{equation}
\left(\frac{V}{I} \right) = \frac{4}{\sqrt{3}} \frac{b(\alpha)}{a(\alpha)} \cot \theta \sqrt{\frac{e B \sin \theta}{2 \pi m_e c \nu}},
\end{equation}
where $\theta$ is the pitch angle of the electrons-positron pairs and $\nu$ the emission frequency.
In the derivation of this expression, it is assumed that the pair distribution function is a power-law with an index $\alpha$ for the electrons emitting at frequency $\nu$, then TIDE computes the index of the power-law for such particles assuming monochromatic emission for the pairs and uses it to calculate the $a(\alpha)$ and $b(\alpha)$ coefficients, which have the form \cite{legg1968}
\begin{equation}
a(\alpha) = \Gamma \left(\frac{3 \alpha - 1}{12} \right) \Gamma \left(\frac{3 \alpha + 7}{12} \right) \frac{\alpha + 7 / 3}{\alpha + 1}
\end{equation}
\begin{equation}
b(\alpha) = \Gamma \left(\frac{3 \alpha + 4}{12} \right) \Gamma \left(\frac{3 \alpha + 8}{12} \right) \frac{\alpha + 2}{\alpha}
\end{equation}
Note that this monochromatic approximation is not applied in TIDE for the calculation of the energy losses nor for the calculation of the luminosity. The pitch angle $\theta$ is assumed to be $\pi / 4$ rad and the electron population is assumed to be a power-law.

\section*{Appendix B: A note on implementation and performance}
\label{sec:imple}

TIDE is built using a modular structure which makes possible for advanced users the substitution and/or the implementation of new pieces of code in order to change the important features and add new physics. For example, to include analytic functions for the particle distribution function, change the shape of the injection function, the evolution of the magnetic field or even the entire dynamics of the nebula by only changing the SNR expansion and profiles module.

The routines that imply intense computation are written mostly in Fortran and we use level 3 optimization flags and MPI libraries in order to maximize its efficiency in computational time expense.
The MPI libraries are implemented for the integration of the diffusion-loss equation, then the number of dedicated CPUs will depend on the resolution of the energy grid where we want to integrate the particle distribution function.
In any case, following the results of the performance tests,
we find that 
$<10$ CPUs are enough, since there is no significant improvement for this problem as treated here (only with the 
electron equation being parallelized) with this approach.
The orchestration of the modules and routines are written in Python.

For a Crab-like pulsar, running a single simulation with a given set of parameters for 1 kyr with a time-step of 0.1 yr and an energy grid of 150 points, in 4 CPUs takes about $\sim$4.5 s of computation time. 
This happens because the amount of time needed to exchange information between the CPUs gets larger when we increase the number of CPUs and the execution time of the routine is not significant anymore in comparison with the communication consuming time. Then, if we continue increasing the number of CPUs, the computing time will be even larger due to CPU communication issues.
A fitting simulation of the same PWN at the same age with 7 free parameters take about $\sim$5 hours.
The latter cost is illustrative, because this time can be reduced/increased depending on how close/far is our initial solution from the final one and the computational power of the CPUs used.

\end{document}